\documentclass[twocolumn,english,showpacs,superscriptaddress,prx]{revtex4-1}
\usepackage[latin9]{inputenc}
\usepackage{amsmath}
\usepackage{amsfonts}
\usepackage{babel}
\usepackage{color}
\usepackage{dcolumn}
\usepackage{graphics}
\usepackage{graphicx}
\usepackage{helvet}
\usepackage{longtable}
\usepackage{soul}
\usepackage{bm}

\usepackage{subfigure}
\usepackage{xspace}
\DeclareGraphicsExtensions{.pdf, .png, .jpg, .eps}

\begin{document}

\title{Proximate deconfined quantum critical point in SrCu$_2$(BO$_3$)$_2$}

\author{Yi Cui}
\thanks{These authors contributed equally to this study.}
\affiliation{Department of Physics and Beijing Key Laboratory of
Opto-electronic Functional Materials $\&$ Micro-nano Devices, Renmin
University of China, Beijing, 100872, China}

\author{Lu Liu }
\thanks{These authors contributed equally to this study.}
\affiliation{Beijing National Laboratory for Condensed Matter Physics and
Institute of Physics, Chinese Academy of Sciences, Beijing, 100190, China}
\affiliation{School of Physics, Beijing Institute of Technology, Beijing 100081, China}

\author{Huihang Lin }
\thanks{These authors contributed equally to this study.}
\affiliation{Department of Physics and Beijing Key Laboratory of
Opto-electronic Functional Materials $\&$ Micro-nano Devices, Renmin
University of China, Beijing, 100872, China}

\author{Kai-Hsin Wu}
\affiliation{Department of Physics, Boston University, 590 Commonwealth
Avenue, Boston, Massachusetts 02215, USA}

\author{Wenshan Hong}
\affiliation{Beijing National Laboratory for Condensed Matter Physics and
Institute of Physics, Chinese Academy of Sciences, Beijing, 100190, China}

\author{Xuefei Liu }
\affiliation{Department of Physics and Beijing Key Laboratory of
Opto-electronic Functional Materials $\&$ Micro-nano Devices, Renmin
University of China, Beijing, 100872, China}

\author{Cong Li}
\affiliation{Department of Physics and Beijing Key Laboratory of
Opto-electronic Functional Materials $\&$ Micro-nano Devices, Renmin
University of China, Beijing, 100872, China}

\author{Ze Hu}
\affiliation{Department of Physics and Beijing Key Laboratory of
Opto-electronic Functional Materials $\&$ Micro-nano Devices, Renmin
University of China, Beijing, 100872, China}

\author{Ning Xi}
\affiliation{Department of Physics and Beijing Key Laboratory of
Opto-electronic Functional Materials $\&$ Micro-nano Devices, Renmin
University of China, Beijing, 100872, China}

\author{Shiliang Li}
\affiliation{Beijing National Laboratory for Condensed Matter Physics and
Institute of Physics, Chinese Academy of Sciences, Beijing, 100190, China}
\affiliation{School of Physical Sciences, Graduate University of the
Chinese Academy of Sciences, Beijing, 100190, China}
\affiliation{Songshan Lake Materials Laboratory, Dongguan, Guangdong 523808,
China}

\author{Rong Yu}
\email{rong.yu@ruc.edu.cn}
\affiliation{Department of Physics and Beijing Key Laboratory of
Opto-electronic Functional Materials $\&$ Micro-nano Devices, Renmin
University of China, Beijing, 100872, China}

\author{Anders W. Sandvik}
\email{sandvik@bu.edu}
\affiliation{Department of Physics, Boston University, 590 Commonwealth
Avenue, Boston, Massachusetts 02215, USA}
\affiliation{Beijing National Laboratory for Condensed Matter Physics and
Institute of Physics, Chinese Academy of Sciences, Beijing, 100190, China}

\author{Weiqiang Yu}
\email{wqyu\_phy@ruc.edu.cn}
\affiliation{Department of Physics and Beijing Key Laboratory of
Opto-electronic Functional Materials $\&$ Micro-nano Devices, Renmin
University of China, Beijing, 100872, China}

\date{\today}

\begin{abstract}
{\bf The deconfined quantum critical point (DQCP) represents a paradigm shift in  quantum
matter studies, presenting a ``beyond Landau'' scenario for order--order transitions.
Its experimental realization, however, has remained elusive. Using high-pressure
$^{11}$B NMR measurements on the quantum magnet SrCu$_2$(BO$_3$)$_2$, we here demonstrate a
magnetic-field induced plaquette-singlet to antiferromagnetic transition above 1.8 GPa
at a remarkably low temperature, $T_{\rm c}\simeq 0.07$ K. First-order signatures of the
transition weaken with increasing pressure, and we observe quantum critical scaling at the highest
pressure, 2.4 GPa. Supported by model calculations, we suggest that these observations
can be explained by a proximate DQCP inducing critical quantum fluctuations and emergent O(3) symmetry
of the order parameters. Our findings take the DQCP from a theoretical
concept to a concrete experimental platform.}
\end{abstract}

\maketitle

{\bf Introduction.}---The theoretically proposed deconfined quantum critical
point (DQCP) \cite{Senthil_Science_2004} connects two different ordered ground
states of quantum matter by a continuous quantum phase transition (QPT). This
type of criticality, which has been explored primarily in the context of two-dimensional (2D)
quantum magnets \cite{JQ_PRL_2007}, lies beyond the conventional paradigm of
discontinuous (first-order) transitions between ordered phases with unrelated
symmetries. The DQCP is associated with unconventional phenomena
including fractional spinon excitations and deconfined gauge fluctuations
\cite{Senthil_PRB_2004,Shao_Science_2016,Ma_PRB_2018}, and further intensive
investigations have introduced emergent symmetries
\cite{Nahum_PRL_2015,Zhao_NP_2019,Serna_PRB_2019,Sreejith_PRL_2019,Takahashi_PRR_2020,Xi_arxiv_2021}
and exotic first-order transitions \cite{Nahum_PRX_2015,Wang_PRX_2017}. In a very
recent extended scenario, the DQCP is a multi-critical point \cite{Zhao_PRL_2020,Lu_PRB_2021}
connected to a gapless quantum spin liquid (QSL)
\cite{Liu_arxiv_2020,Yang_PRB_2022,Liu_arxiv_2021,Keles_PRB_2022,Shackleton_PRB_2021}.

\begin{figure*}[t]
\begin{center}
\includegraphics[width=11cm]{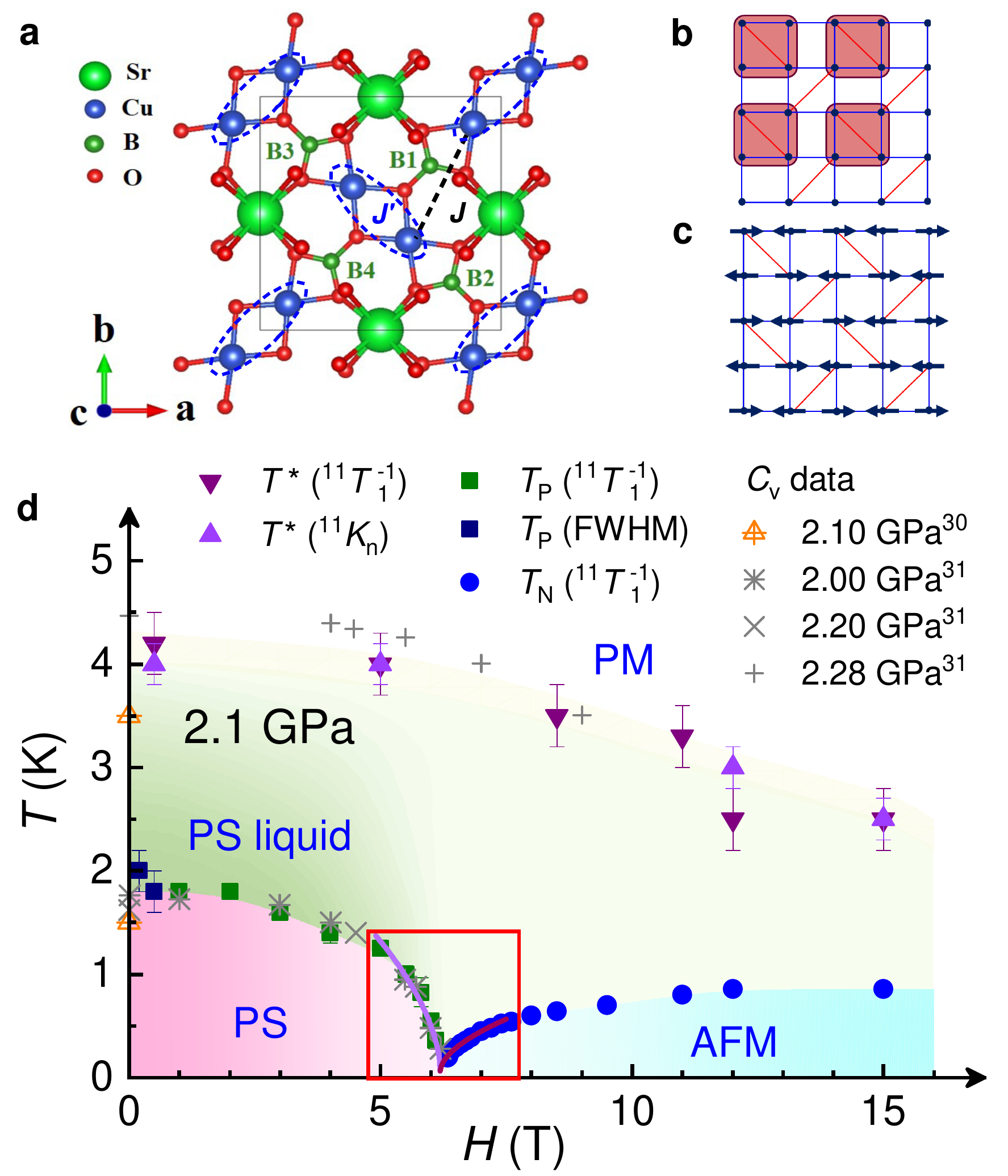}
\end{center}
\caption{\label{fig1} {\bf Experimental overview.}
{\bf a} Atomic structure of a SrCu$_2$(BO$_3$)$_2$ plane. Pairs of Cu$^{2+}$
ions form spin dimers (ellipses) with Heisenberg intra-dimer ($J'$) and inter-dimer
($J$) interactions (black dashed lines). Each unit cell contains four B ions,
whose NMR response we investigate. {\bf b} The PS phase in the equivalent
square lattice of $J$ (blue) and $J'$ bonds (red). In SrCu$_2$(BO$_3$)$_2$,
the singlets (shading) form on the full ($J'$) plaquettes, in one of two
symmetry-equivalent patterns, while in the SSM the singlets form on the empty
plaquettes. {\bf c} The AFM phase, which breaks O(3) symmetry when $H = 0$
and O(2) when $H \not= 0$. For SrCu$_2$(BO$_3$)$_2$ in a $c$-axis field, we find
that the moments order along the $a$ or $b$ axis. {\bf d} Field-temperature phase
diagram at $2.1$~GPa, showing the paramagnetic (PM), PS liquid, ordered PS, and AFM
phases resolved by our NMR measurements (Figs.~\ref{fig3}--\ref{fig5}). The transition
temperatures $T_{\rm P}$ and $T_{\rm N}$, and the crossover temperature $T^*$, are
compared with specific-heat measurements \cite{Guo_PRL_2020,Larrea_Nature_2021}.
The red box marks the regime analyzed in Fig.~\ref{fig5}f.}
\end{figure*}

Although DQCP phenomena are broadly relevant in quantum materials
\cite{Zhang_PRR_2020}, there has been no positive experimental identification
in any system. Quantum magnets in which the interactions can be varied over
a wide enough range to realize two phases bordering a DQCP are rare.
An exception is the layered material SrCu$_2$(BO$_3$)$_2$
\cite{Kageyama_PRL_1999,Miyahara_PRL_1999,Miyahara_2003}, where
antiferromagnetic (AFM) Heisenberg interactions between the $S = 1/2$ Cu$^{2+}$
spins (Fig.~\ref{fig1}a) provide a remarkably faithful realization of the
2D Shastry-Sutherland model (SSM) \cite{Shastry_1981}, in which three different $T=0$
phases are well established versus the ratio $g = J/J'$ of the inter-
to intra-dimer couplings \cite{Koga_PRL_2000,Corboz_PRB_2013}: an exact
dimer-singlet phase (DS, with singlets on the $J'$ bonds), a two-fold degenerate
plaquette-singlet (PS) phase (Fig.~\ref{fig1}b), and a N\'eel AFM phase
(Fig.~\ref{fig1}c). At ambient pressure, SrCu$_2$(BO$_3$)$_2$ is well described
by the $g \simeq 0.63$ SSM with DS ground state \cite{Miyahara_2003}.
An applied pressure increases $g$, driving the system
into a PS phase at $P \simeq 1.8$ GPa \cite{Haravifard_NC_2016,Zayed_NP_2017},
which persists with transition temperature $T_{\rm P} \simeq 2$ K up to
$P \simeq 2.6$ GPa \cite{Guo_PRL_2020,Larrea_Nature_2021}. An AFM phase with
$T_{\rm N}$ from $2.5$ to $4$ K has been detected between $3.2$ and $4$ GPa
\cite{Guo_PRL_2020}.

Here we report a $^{11}$B NMR study of SrCu$_2$(BO$_3$)$_2$ in a magnetic field $H$ up to
$15$ T and pressures up to $2.4$ GPa, with the main goal to characterize the field-driven PS--AFM transition.
As Fig.~\ref{fig1}d shows at $2.1$ GPa, PS and AFM transitions are resolved using their NMR signatures and merge
at $H_{\rm c} \simeq 6$ T and $T_{\rm c} \simeq 0.07$ K. Such a low $T_{\rm c}$ in relation to $T_{\rm P}$ and $T_{\rm N}$
further away from $H_{\rm c}$ indicates proximity to a $T_{\rm c} = 0$ QPT. First-order discontinuities
at $(H_{\rm c},T_{\rm c})$ weaken with increasing pressure, and we observe quantum-critical scaling of the spin-lattice
relaxation at $2.4$ GPa for $T > T_{\rm c}$.

\begin{figure}[t]
\begin{center}
\includegraphics[width=6.5cm]{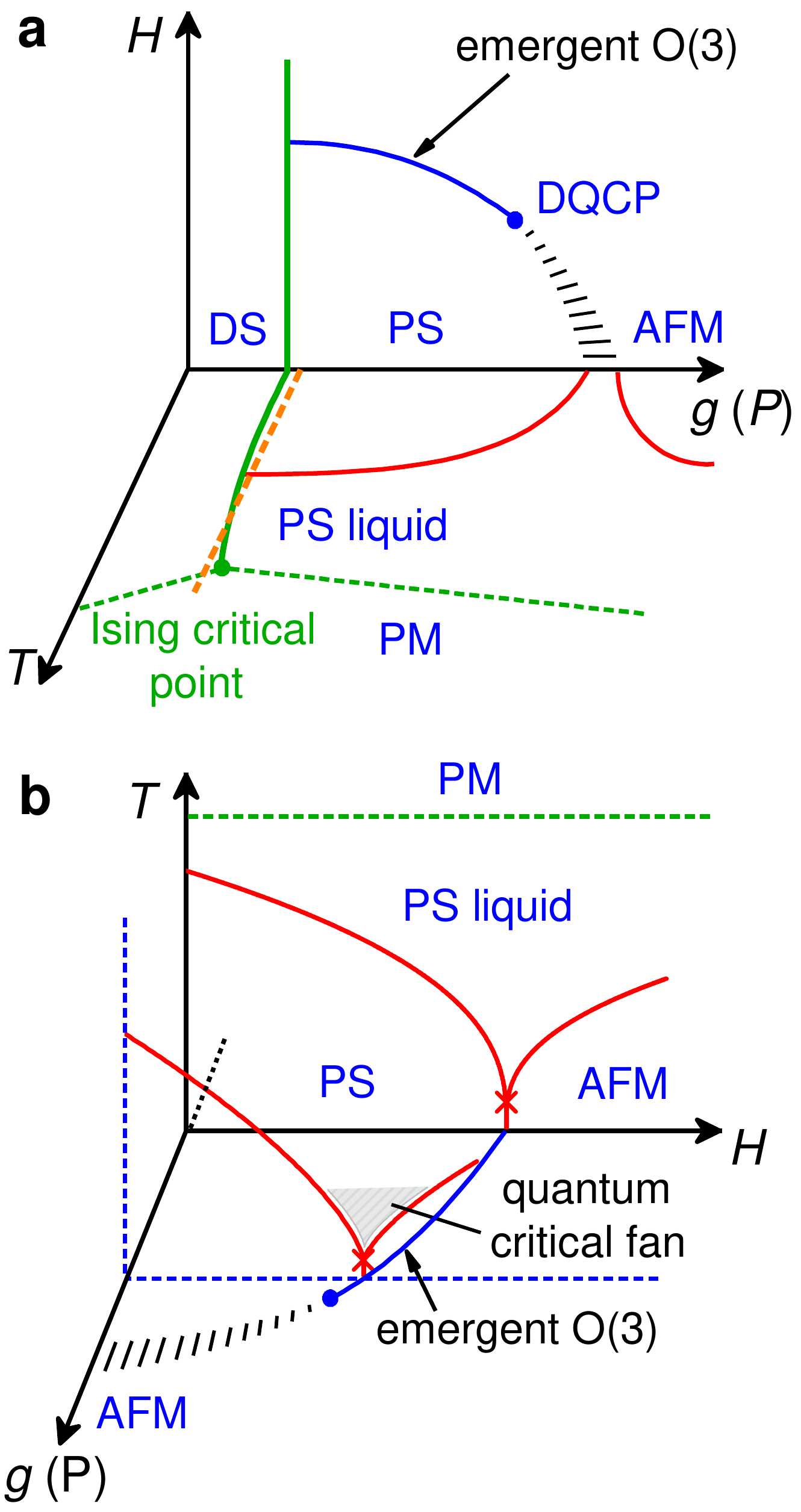}
\end{center}
\caption{\label{fig2} {\bf Schematic phase diagram and DQCP scenario.}
{\bf a} Phases in the space of coupling [$g=J/J'$ in the SSM, $P$ in SrCu$_2$(BO$_3$)$_2$], temperature, and magnetic field.
A multi-critical DQCP separates a line of first-order QPTs and either a QSL phase \cite{Yang_PRB_2022} or a line of generic DQCPs
\cite{Lee_PRX_2019}; the region marked with dashed lines represents this undetermined feature. The first-order DS transition (solid
green line) terminates at an Ising critical point (green circle) \cite{Larrea_Nature_2021}.
The green dashed lines indicate crossovers at $T^*(g)$ into the PM phase. The dashed orange line shows how the slightly curved
first-order DS transition line can be crossed vs $T$ at fixed $P$. The ordered AFM phase at $T > 0$ requires
inter-layer couplings, as in SrCu$_2$(BO$_3$)$_2$. The magnetization plateau states at larger $H$
\cite{Matsuda_PRL_2013,Haravifard_NC_2016} are not shown.
{\bf b} Phase diagram drawn to highlight $(H,T)$ planes exemplified by Fig.~\ref{fig1}d. Red crosses indicate
$T_{\rm c}>0$ caused by weak 3D effects and violations of O(3) symmetry. The shading
represents the ``fan'' in which quantum critical scaling is expected. The blue dashed lines indicate
the plane of highest-pressure ($2.4$ GPa) measurements.}
\end{figure}

Our results support the existence of a multi-critical DQCP controlling the quantum fluctuations at $2.4$ GPa, with
$T_{\rm c}$ on the associated first-order line suppressed by an emergent O($3$) symmetry of the combined scalar PS and
O($2$) AFM order parameters \cite{Zhao_NP_2019,Serna_PRB_2019}. By synthesizing past and present experiments on
SrCu$_2$(BO$_3$)$_2$ and model calculations, we arrive at the global phase diagram depicted in Fig.~\ref{fig2}.
Before further discussing the DQCP scenario, we present our NMR detection of the various phases and transitions.

{\bf NMR identification of phases.}---
We performed $^{11}$B NMR measurements on SrCu$_2$(BO$_3$)$_2$ single crystals
at pressures up to $2.4$ GPa in fields between $0.2$ and $15$~T and temperatures down to $0.07$ K.  Experimental details
are provided in Methods \cite{SI}. We first discuss NMR line shifts to detect the relevant quantum phases and transitions, followed
by results for the spin-lattice relaxation rate $1/T_1$.

\begin{figure*}[t]
\includegraphics[width=16cm]{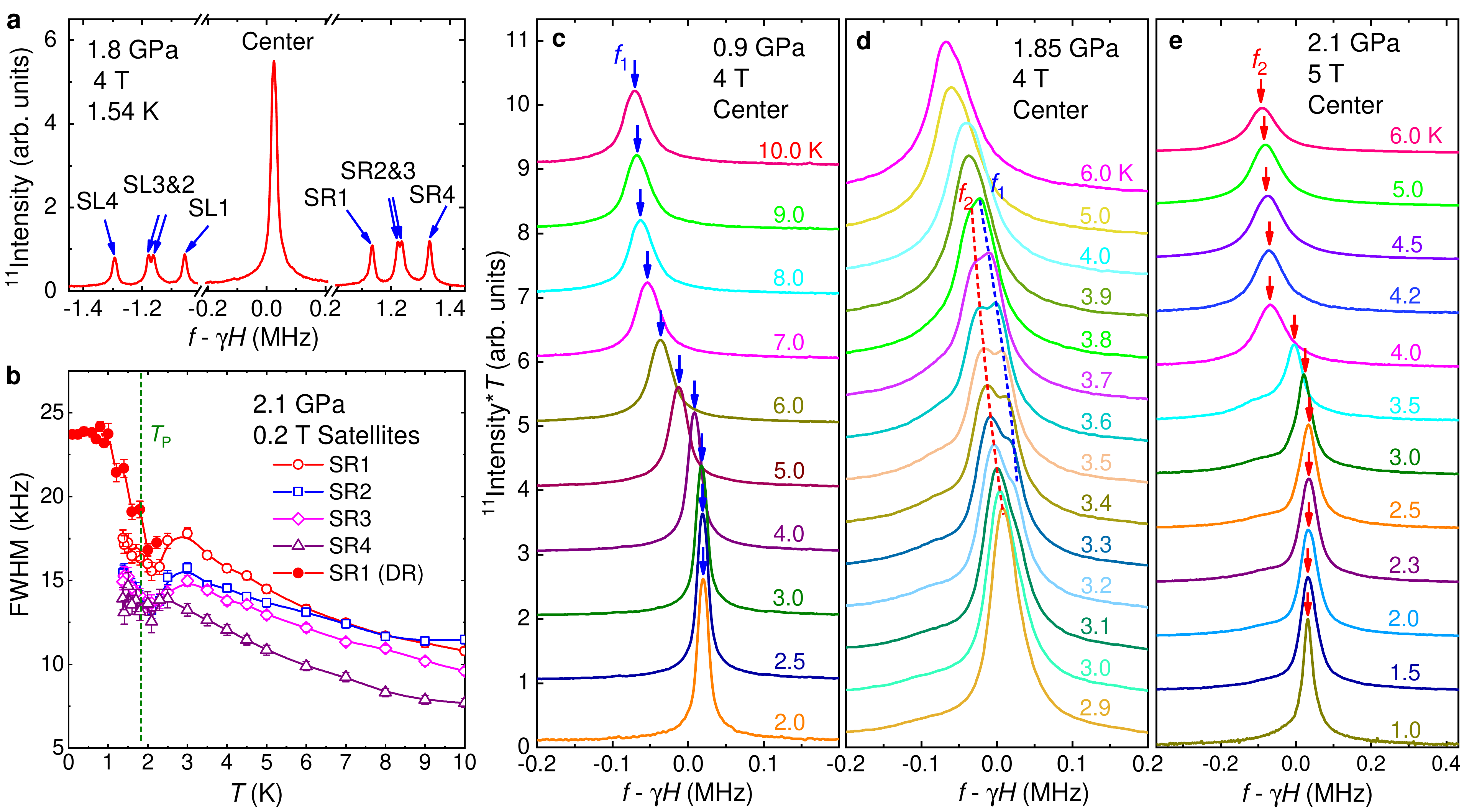}
\caption{\label{fig3} {\bf NMR spectra and line shifts.}
{\bf a} NMR $^{11}$B spectrum at $H = 4$ T and $P = 1.8$ GPa, with the field
applied at 8.6$^{\circ}$ from the crystalline $c$ axis, showing the center line
and two sets of satellites associated with the four B sites (Fig.~\ref{fig1}a).
{\bf b} FWHM of satellites SR1-SR4 shown as a function of temperature at $P =
2.1$~GPa and $H = 0.2$~T. The line at $1.8$~K marks the onset of an upturn
with further cooling. SR1 was measured in the  dilution refrigerator in
addition to the VTI used for all cases (SI Sec.~S2 \cite{SI}). {\bf c-e} NMR center
line for a range  of temperatures (curves shifted vertically) at ({\bf c}) $P=0.9$ GPa,
$H=4$ T, ({\bf d}) $1.85$~GPa, $4$ T, and ({\bf e}) $2.1$ GPa, $5$ T. The peaks in the
DS phase ({\bf c}) and in and above the PS phase ({\bf e}) are marked $f_1$ and $f_2$,
respectively. The split peak in ({\bf d}) reflects phase coexistence.}
\end{figure*}

A typical $^{11}$B NMR spectrum, shown in Fig.~\ref{fig3}a, has a central peak with four satellite peaks on either side, from
inequivalent sites B1-B4 (Fig.~\ref{fig1}a) due to a small tilt angle between field and crystalline $c$ axis (Methods \cite{SI}).
The satellites are sensitive to changes of the lattice structure because of the local coupling between the nuclear
quadrupole moment and the electric-field gradient (Methods \cite{SI}).
As shown at a low field and $P = 2.1$ GPa in Fig.~\ref{fig3}b, the full-width at
half maximum (FWHM) height of the satellites increases on cooling below $10$~K until
a maximum at $T \simeq 3$~K, reflecting increasing lattice fluctuations when the
spins form fluctuating plaquette singlets above the ordered PS phase \cite{Larrea_Nature_2021}.
This PS liquid crosses over to the trivial PM state at higher temperature.

Below $1.8$~K, the FWHM in Fig.~\ref{fig3}b rises sharply and saturates around $1$~K. As explained
in Supplemental Information (SI) Sec.~S2 \cite{SI}, the rapid broadening
follows from an orthogonal lattice distortion when a full-plaquette (FP) PS state (Fig.~\ref{fig1}b)
forms. The FWHM as a proxy for the PS order parameter is further corroborated by the
consistency of $T_{\rm P} \simeq 1.8$ K at the low field applied in Fig.~\ref{fig3}b
with the location of a sharp specific-heat peak \cite{Guo_PRL_2020,Larrea_Nature_2021},
marked in Fig.~\ref{fig1}d.

\begin{figure*}[t]
\begin{center}
\includegraphics[width=16.3cm]{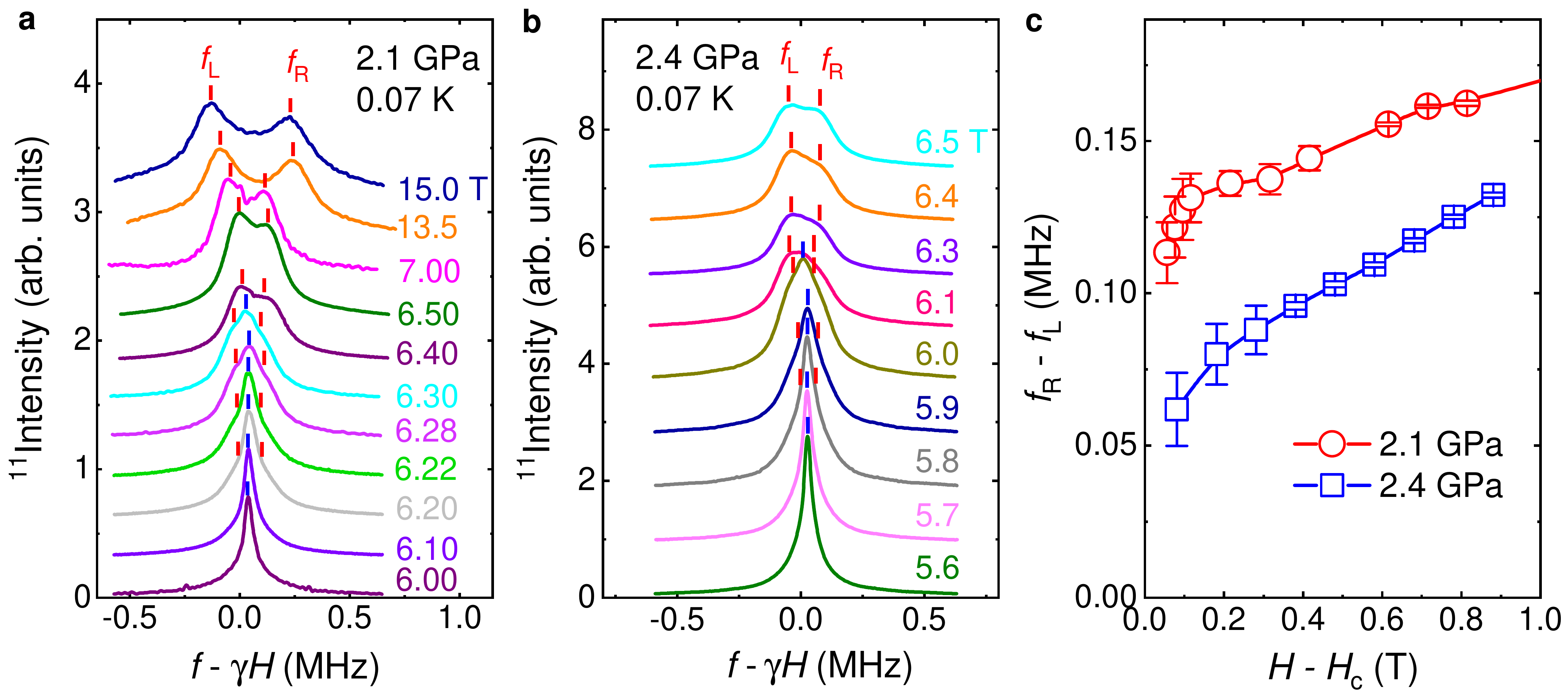}
\end{center}
\caption{\label{fig4} {\bf AFM transition.}
Splitting of the NMR center line with increasing $H$ at $T = 0.07$ K is shown in {\bf a} and {\bf b}
for $P = 2.1$~GPa and $2.4$~GPa, respectively. The two peaks marked $f_L$ and $f_R$ (red bars) indicate AFM order
developing above $H \simeq 6$~T. A center peak (blue bars) remaining at slightly higher fields indicate phase
coexistence. {\bf c} Proxy AFM order parameter $f_R - f_L$ vs $H-H_c$, where $H_c$ is determined using the
spin-lattice relaxation rate (Fig.~\ref{fig5}).}
\end{figure*}

Figure \ref{fig3}c shows the evolution of the central peak with $T$ at
$P = 0.9$~GPa and $H = 4$~T. The negative Knight shift at the higher temperatures reflects
the hyperfine coupling $A_{\rm hf} \simeq -0.259$~T/$\mu_{\rm B}$
(SI, Sec.~S3~\cite{SI}) for ${\vec H} \parallel {\hat c}$ \cite{kodama_JPCM_2002,Waki_JPSJ_2007}.
The shift increases rapidly below $T^* \simeq 7$ K when dimer singlets form in the DS state.
At $2.1$ GPa, Fig.~\ref{fig3}e,  PS order forms below $2$ K but the Knight shift changes rapidly at $T \simeq 4$ K
also in this case when the PS liquid forms.

The first-order transition between the DS phase and the PS or PS liquid phase terminates at an Ising-type critical point,
which at $H=0$ is located at $P \simeq 1.9$ GPa, $T \simeq 3.3$ K \cite{Larrea_Nature_2021}. At low $T$, the DS--PS transition
takes place between $1.7$ and $1.8$ GPa \cite{Guo_PRL_2020}. The first-order DS line must therefore bend slightly, as indicated in
Fig.~\ref{fig2}a, and can be crossed versus $T$ at fixed $P$. Indeed, at $1.85$ GPa, Fig.~\ref{fig3}d, the central peak between $3$
and $4$ K is split, indicating phase coexistence. Previously a different splitting was reported at $2.4$ GPa \cite{Waki_JPSJ_2007,Takigawa_JPSJ_2010},
perhaps caused by pressure inhomogeneity, but we observe the double peak only at $1.85$ and $1.95$ GPa (SI, Sec.~S3~\cite{SI}). Outside
this pressure range, $T^*$ likely only marks a rapid cross-over between the PM and PS liquid, whose associated sharp specific-heat peaks
\cite{Guo_PRL_2020,Larrea_Nature_2021} can be explained by an analogy \cite{Larrea_Nature_2021} with the Widom line away from the gas--liquid
critical point. We have found no NMR signatures of a structural transition here or at higher temperatures (SI Sec.~S3 \cite{SI}).

Above $1.95$~GPa, AFM order emerges at high fields and leads to splitting of the central NMR line by alternating positive and negative
hyperfine fields, as shown at $2.1$~GPa and  $2.4$~GPa in Fig.~\ref{fig4}a and Fig.~\ref{fig4}b, respectively, both at $T=0.07$ K.
The sudden rise with field of the peak-splitting $f_R - f_L$ (a proxy AFM order parameter), shown in Fig.~\ref{fig4}c, signals a
discontinuous onset of AFM order at $H_c(P)$, with the discontinuity much weaker at the higher pressure.

In the AFM state, the uniform magnetization does not exhibit any obvious
discontinuity at $H_{\rm c}$ and remains below $2\%$ of the saturated moment at our highest field of
$15$ T (SI, Sec.~S4 \cite{SI}). A cross-over temperature $T^*$ persists also at high fields, where the PS liquid develops
increasing spin fluctuations (discussed further below).

\begin{figure*}[t]
\includegraphics[width=16.3cm]{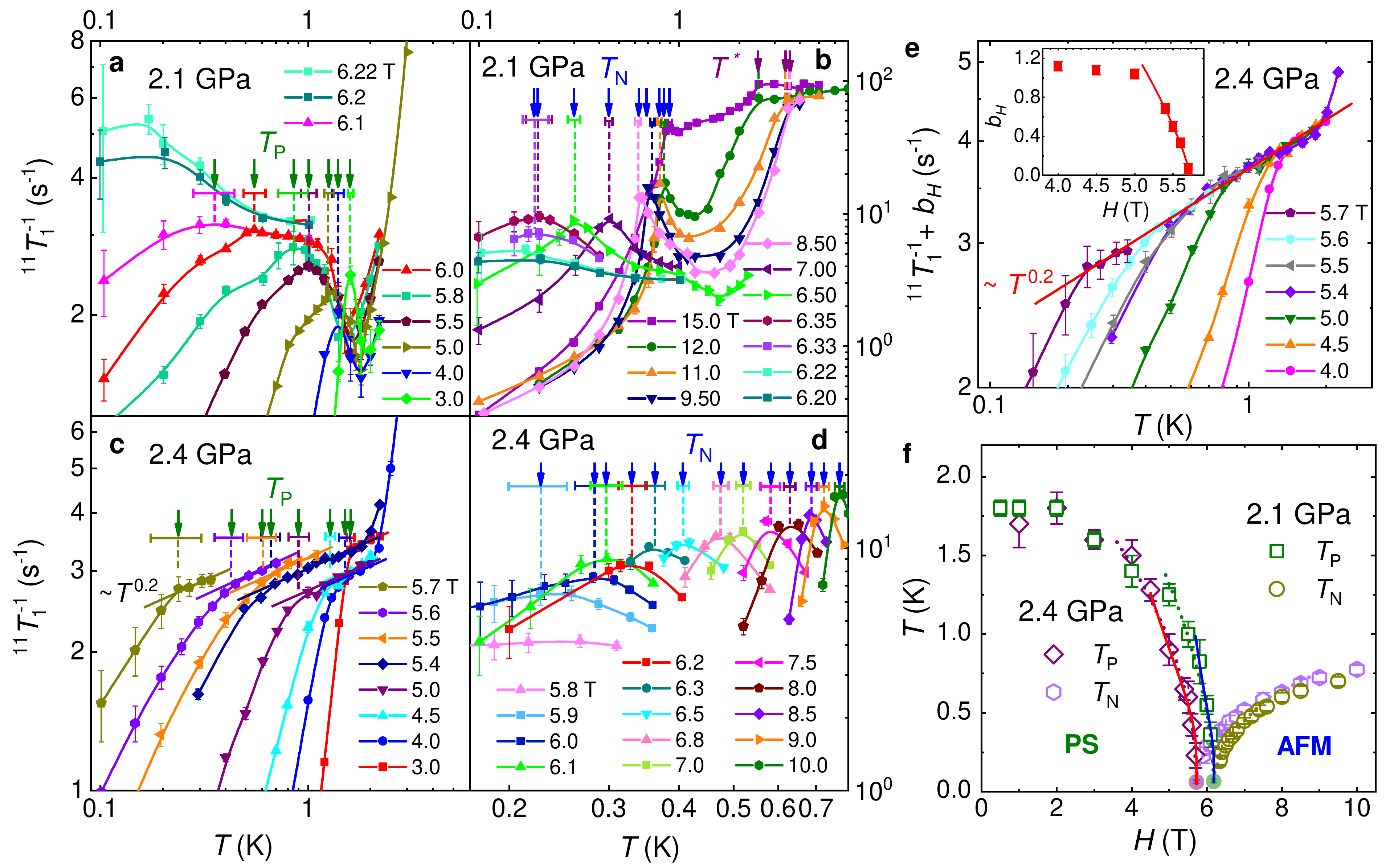}
\caption{\label{fig5}
{\bf Spin-lattice relaxation.}
$1/^{11}T_1 (T)$ measured at $2.1$ ({\bf a,b}) and $2.4$ GPa ({\bf c,d}),
separated to show the PS ({\bf a,c}) and AFM states ({\bf b,d}). The drop
in $1/T_1$ at a $T^* \simeq 4$~K ({\bf a,b}) indicates the sharp crossover
into the PS liquid. The peaks at lower $T$ mark $T_{\rm P}$ and $T_{\rm N}$,
with uncertainties indicated by the horizontal bars. At $2.4$ GPa, no low-$T$ PS
peak is observed ({\bf c}), but $T_{\rm P}$ can be extracted from the sudden
change from thermally activated to quantum critical behavior, $1/T_1 = aT^{\eta}-b_H$.
Power-law scaling of the offset, $b_H \propto (H_c-H)^d$ with $d \approx 0.8$,
close to $H_c$ is shown in the inset of {\bf e}. The common scaling form with constant
$a$ and $\eta \approx 0.2$ is demonstrated in main panel {\bf e}, where $b_H$ has
been added. {\bf f} Low-temperature phase diagrams at $2.1$ and $2.4$~GPa. The solid
and dotted lines indicate the phase boundaries modeled by  respectively, a logarithmic
form of $T_{\rm P}$ and near-critical forms of both $T_{\rm P}$ and $T_{\rm N}$
(SI, Sec.~S6 \cite{SI}). The latter fits give the $T_{\rm c}$ values indicated with circles.}
\end{figure*}

{\bf Spin-lattice relaxation rate.}---
$1/T_1$ is a direct probe of low-energy spin
fluctuations and can detect the PS and AFM transitions
more precisely than (but consistent with) the line shifts. Figures \ref{fig5}a
and \ref{fig5}b show $1/T_1$ at $P = 2.1$~GPa for a wide range of applied fields that we group
below and above $6.2$~T, corresponding respectively to the low-$T$ PS and AFM phases;
Figs.~\ref{fig5}c and \ref{fig5}d show the same at $2.4$~GPa with the
separation at $5.8$~T.

At $2.1$~GPa, Fig.~\ref{fig5}b, we find a sharp drop of
$1/T_1$ at $T^* \simeq 3$--$4$ K and a broad peak or sharper kink
below $2$~K. At low fields in Fig.~\ref{fig5}a, the latter feature extends to $6.1$ T
and clearly marks the opening of a spin gap below $T_{\rm P}$. At $P = 2.4$~GPa, we do not find
a peak at $T_{\rm P}$ (Fig.~\ref{fig5}c) but rather a sharp crossover from a low-$T$ gapped
regime to a window with power-law behavior that is analyzed in Fig.~\ref{fig5}e
and will be further discussed below. At the higher
fields in Fig.~\ref{fig5}b and \ref{fig5}d, the low-$T$ features (below $0.8$ K) are much
sharper and coincide with the NMR peak splitting in Figs.~\ref{fig4}a and \ref{fig4}b.
Thus, we can safely identify these peaks for $H \ge 6.33$ T as $T_{\rm N}$ \cite{Moriya_1963}.
The minimum in $1/T_1$ around $1.5$ K in Fig.~\ref{fig5}b increases with the field, indicating increasing
spin fluctuations in the PS liquid state.

Figure \ref{fig5}f shows very clear field-induced PS--AFM transitions revealed by these
signals at both $P=2.1$ and $2.4$ GPa. The PS and AFM boundaries $T_{\rm P}(H)$ and
$T_{\rm N}(H)$ meet at a remarkably low $T_{\rm c}$. Given phase coexistence (Figs.~\ref{fig4}a,b)
the quantum phase transition at $H_{\rm c}$ is clearly first-order. The proxy AFM order parameter
$f_R-f_L$ in Fig.~\ref{fig4}c is consistent with $H_{\rm c}$ determined from $1/T_1$ at both pressures.
The much smaller first-order discontinuity of $f_R-f_L$ at the higher pressure indicates the approach
toward a continuous QPT.

\begin{figure*}[t]
\begin{center}
\includegraphics[width=12cm]{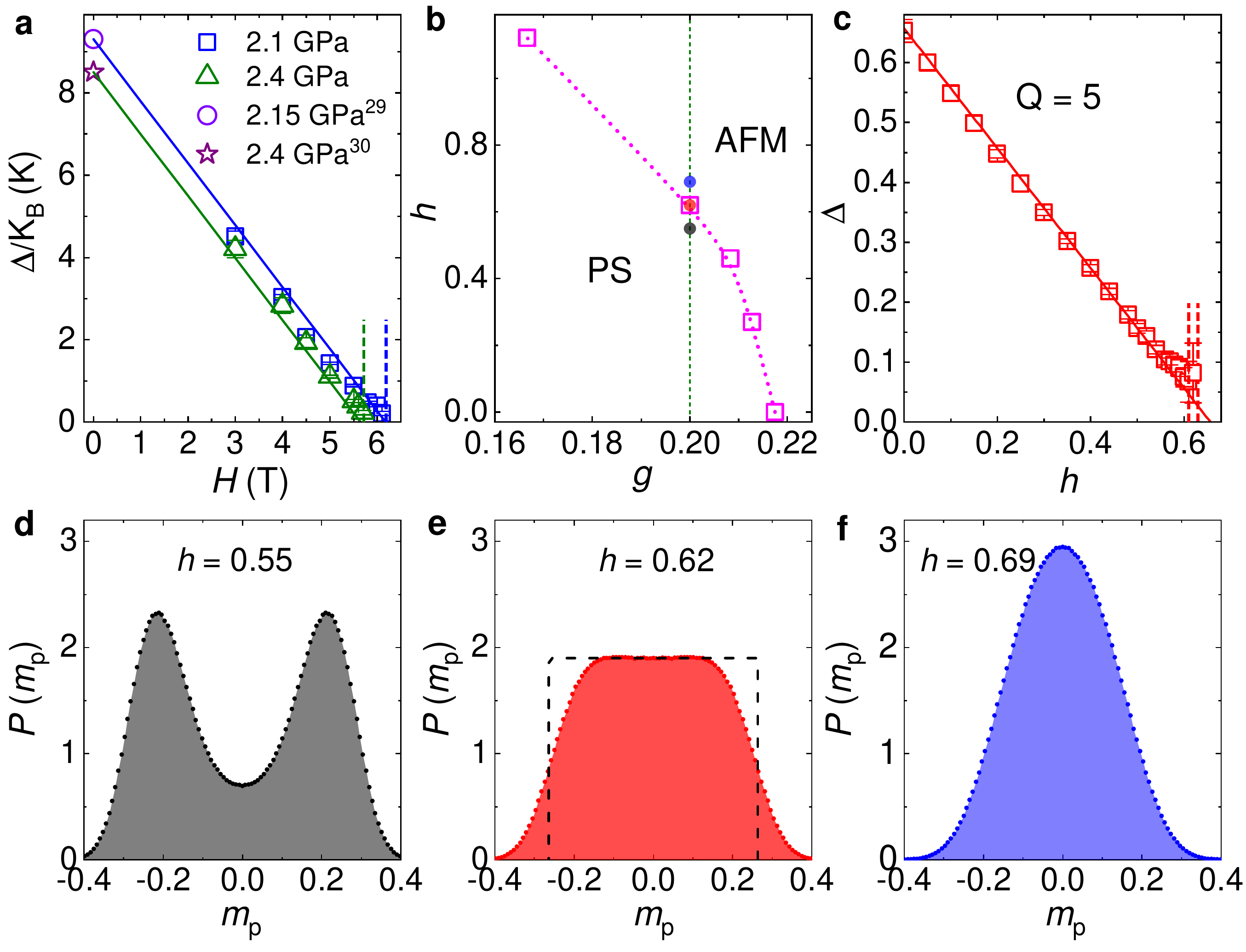}
\end{center}
\caption{\label{fig6} {\bf Spin gap and emergent symmetry.}
  {\bf a} Field dependent gap of SrCu$_2$(BO$_3$)$_2$. The lines show the expected form
  $\Delta(H)=\Delta(0)-\tilde g \mu_BH$ where $\Delta(0)$ are reported zero-field gaps
  \cite{Zayed_NP_2017,Guo_PRL_2020} and $\tilde g=2.28$ is the known $g$-factor (SI, Sec.~S5).
  The vertical dashed lines represent $H_{\rm c}$ values from Fig.~\ref{fig5}f.
  {\bf b} Ground-state phase diagram of the CBJQM vs
  $g = J/Q$ and field $h$. The PS and AFM phases are separated by a
  first-order transition. The vertical line and closely spaced points mark the
  parameters in panels {\bf d}-{\bf f}. {\bf c} Spin gap of the CBJQM
  at $g=0.2$. The dashed vertical line indicates $h_{\rm c}$ and
  the solid line is a fit to $\Delta(h)=\Delta(0)-h$. {\bf d}-{\bf f} Distribution
  of the plaquette order parameter. Double-peak ({\bf d}), plateau ({\bf e}), and
  single-peak ({\bf f}) distributions are found respectively in the PS phase
  ($h = 0.55$), at the transition ($h = 0.62$), and in the AFM phase ($h = 0.69$).}
\end{figure*}

We have extracted the PS spin gap $\Delta$ by fitting $1/T_1$ below $T_{\rm P}$ to a semi-empirical form
$T^{-a}{\rm e}^{-\Delta/k_{B}T}$ with $a\approx 1$ (SI, Sec.~S5 \cite{SI}). As expected, a linear decrease with
$H$ of $\Delta$ at both pressures is revealed in Fig.~\ref{fig6}a, on account of the field-lowering of the $S=1$
($S^z=1$) state above the singlet PS ground state. The results are compatible with previously determined $H=0$ gap
estimates \cite{Zayed_NP_2017,Guo_PRL_2020} and the known $g$-factor.

At a first-order transition into the AFM phase, the PS
gap should jump discontinuously to zero at $H_c$ (given that the AFM state is gapless), but, despite the clear
first-order signals described above (Fig.~\ref{fig4}c), we find $\Delta(H_c)$ values indistinguishable from
zero within statistical errors. We will discuss the anomalously small gap discontinuity in the context of the
proximate DQCP scenario further below.

{\bf Deconfined quantum criticality.}---
The SSM at $H = 0$ has been a candidate for a DQCP separating
its coupling-induced PS and AFM ground states \cite{Zhao_NP_2019,Lee_PRX_2019}. The singlets in the PS phase
of the model occupy the empty plaquettes, in contrast to the FP state in SrCu$_2$(BO$_3$)$_2$
(Fig.~\ref{fig1}b). This aspect of the PS state depends sensitively on other possible weak
interactions beyond the SSM \cite{Boos_PRB_2019,Xi_arxiv_2021}, and the SSM description of the
global phase diagram of SrCu$_2$(BO$_3$)$_2$ should remain valid.

There is mounting evidence that a gapless QSL phase can exist between a PS state (or closely related spontaneously dimerized state) and
the AFM state in frustrated 2D quantum spin systems \cite{Gong_PRL_2014,Wang_PRL_2018,Nomura_PRX_2021,Shackleton_PRB_2021,Liu_arxiv_2020}
and that these QSL phases generically end at multi-critical DQCPs \cite{Yang_PRB_2022,Lu_PRB_2021,Liu_arxiv_2021}. Beyond such a point, the
transition without intervening QSL is expected to be first-order, with the coexistence state at $H=0$ inheriting (and breaking) the emergent O(4)
or SO(5) symmetry (depending on the type of singlet-ordered phase \cite{Nahum_PRX_2015,Wang_PRX_2017,Zhao_NP_2019,Serna_PRB_2019,Takahashi_PRR_2020})
of the DQCP.

In the $H=0$ SSM, early calculations indicated a first-order PS--AFM transition \cite{Corboz_PRB_2013}, and a recent calculation suggested
an O(4) [from the O(3) AFM and scalar PS order parameters] multi-critical DQCP in an extended parameter space \cite{Xi_arxiv_2021}. A generic
O(4) DQCP had previously been proposed \cite{Lee_PRX_2019}. The intervening gapless QSL between the PS and AFM phases was identified
very recently \cite{Yang_PRB_2022,Keles_PRB_2022} and may be explained by an instability of the conventional DQCP \cite{Lu_PRB_2021}.
These theoretical insights along with our NMR results for SrCu$_2$(BO$_3$)$_2$ suggest the scenario in Fig.~\ref{fig2}. Since no
experiment so far (including ours) have explicitly confirmed a QSL phase, the possibility remains that there is instead another
line of PS--AFM transitions. Though the dashed regions in the phase diagrams in Fig.~\ref{fig2} can represent either possibility,
specific heat measurements at $H = 0$ \cite{Guo_PRL_2020,Larrea_Nature_2021} found no phase transition between $2.6$ and $3.2$ GPa,
consistent with a QSL ground state evolving into the $T>0$ PS liquid.

A putative multi-critical DQCP at $H>0$ should evolve from a corresponding $H=0$ DQCP with emergent O($4$) symmetry
\cite{Lee_PRX_2019,Zhao_NP_2019}. While this O(4) point exists only in an extended parameter space outside the ($g,H,T$) cube
in Fig.~\ref{fig2}, the fact that the field-induced magnetization is very small at $H_c$ (SI, Sec.~S4) suggests that the putative
$H>0$ DQCP still hosts an approximate O(4) symmetry, with stronger O(3) character developing on the first-order line. Strictly
speaking, at $H>0$ the DQCP may evolve into a near-critical triple point with first-order signatures at the lowest energy scales.

Closer proximity of SrCu$_2$(BO$_3$)$_2$ to some continuous QPT with increasing pressure is certainly supported by our observation of a weaker
discontinuity of the AFM order parameter at $2.4$ GPa than at $2.1$ GPa (Figs.~\ref{fig4}c). Moreover, at a clearly first-order transition
correspondingly high $T_c$ values would normally be expected. The low $T_c$ at both pressures then point to a mechanism suppressing
long-range order also rather far away from the QPT. The DQCP scenario offers this possibility through its emergent continuous symmetry
inherited (at least up to some large length scale) by the first-order line. An ideal 2D coexistence state with continuous order
parameter symmetry must have $T_{\rm c}=0$, but weak violations of the symmetry (in combination with 3D effects \cite{Sun_CPB_2021})
would imply a low $T_{\rm c} >0$, as observed in SrCu$_2$(BO$_3$)$_2$.

In the scenario of a first-order transition with emergent O($3$) symmetry, the Ising-type PS order can be understood as an uniaxial
deformation of the O(3) order parameter. A logarithmic form of the PS transition temperature is then expected;
$T_{\rm P} \propto \ln^{-1}[a(H_{\rm c}-H)]$, for some value of $a$ \cite{Zhao_NP_2019,Irkhin_PRB_1998}. Fits of the experimental data to
this form (SI, Sec.~S6 \cite{SI}) are shown with solid curves in Fig.~\ref{fig5}f and indeed describe the behavior close to $H_{\rm c}$.

To describe $T_{\rm N}(H)$, we note again that inter-layer interactions are required for $T_{\rm N} > 0$ in a spin-isotropic system. These
couplings also change a continuous QPT ($T_{\rm c} = 0$) into a first-order line extending to a bicritical or triple point at $T_{\rm c} > 0$
\cite{Lee_PRX_2019,Sun_CPB_2021} (red crosses in Fig.~\ref{fig2}b). Given the extremely low $T_{\rm c}$ values in SrCu$_2$(BO$_3$)$_2$, a
modified critical form with the same exponent $\phi$ governing both transitions above $T_{\rm c}$ may be expected from DQCP dualities
\cite{Wang_PRX_2017,Qin_PRX_2017}: $T_{\rm P,N} = T_{\rm c} + a_{\rm P,N} |H - H_{\rm c}|^{\phi}$. Fits with independent exponents $\phi$ for
the PS and AFM transitions (Sec.~S6, SI~\cite{SI}) indeed support a common value and motivate joint fitting with a single $\phi$. Such
fits are shown with the dashed curves in
Fig.~\ref{fig5}, where $T_{\rm c}$ is in the range $0.05$-$0.07$ K at both  pressures. At $2.1$ GPa $H_{\rm c} = 6.184 \pm 0.005$,
$\phi = 0.57 \pm 0.03$, while at $2.4$ GPa $H_{\rm c} = 5.719 \pm 0.009$, $\phi = 0.50 \pm 0.03$. These fits, where $\phi$ is close to estimates
for both SO(5) \cite{Zhao_PRL_2020,Nahum_PRX_2015} and O(4) \cite{Qin_PRX_2017} DQCPs (see further Sec.~S6A), do not rule out the alternative
logarithmic form of $T_{\rm P}$ but further validate the very low $T_{\rm c}$ values and common transition field $H_{\rm c}$ for both
order parameters.

{\bf Quantum-critical relaxation.}---
In Fig.~\ref{fig5}c, $1/T_1$ at $2.4$ GPa exhibits $T^\eta$ scaling with $\eta \approx 0.2$ within a window of temperatures for several fields
close to $H_{\rm c}$ on the PS side. The ensemble of fits is further analyzed in Fig.~\ref{fig5}e using the expected quantum-critical form
$1/T_1 = aT^{\eta}-b_H$ \cite{Chubukov_PRB_1994}, where $a$ is a constant and $b_H>0$ for $H<H_{\rm c}$. The fact that scaling behavior
is not observed at $2.1$~GPa (Fig.~\ref{fig5}a) suggests that only the system at $2.4$ GPa is sufficiently close to a continuous QPT that
it realizes the quantum critical fan \cite{Chubukov_PRB_1994}, depicted in Fig.~\ref{fig2}b, where $T$ is the largest energy scale (but
low enough so that  the correlation length is well above the lattice constant). The value of $\eta$ is compatible with an estimate for
an O(4) DQCP \cite{Qin_PRX_2017}  and slightly below the SO(5) value \cite{JQ_PRL_2007,Nahum_PRX_2015}.

On the AFM side (Fig.~\ref{fig5}d),  $1/T_1$ is dominated by the 3D effects causing $T>0$ AFM order, with the associated peak in $1/T_1$
masking any 2D quantum criticality, unlike the PS side, where the spin correlations and 3D effects are much weaker. We lack 2.4
GPa data above the temperatures in Fig.~\ref{fig5}c. At 2.1 GPa, no scaling is observed between $T_{\rm N}$ and $T^*$ in Fig.~\ref{fig5}b,
where a sharp drop below $T^*$ is immediately followed by strong precursors to AFM ordering.

{\bf Quantum spin model.}---
We now turn to the checkerboard $J$-$Q$ model (CBJQM), in which
four-spin interactions $Q$ replace $J'$ in the SSM. The CBJQM is amenable to quantum Monte Carlo
simulations and hosts PS and AFM phases separated by a first-order transition with emergent O($4$)
symmetry at zero field \cite{Zhao_NP_2019}. We here simulate (Methods \cite{SI})
the same model in a field, defining $g=J/Q$ and $h=H/J$ with $J=1$.

In the phase diagram in Fig.~\ref{fig6}b, the field-driven PS--AFM transition is first-order. The PS gap $\Delta(h)$
obtained from the low-temperature susceptibility (SI, Sec.~S8 \cite{SI}) is shown in Fig.~\ref{fig6}c at $g=0.2$, below
the $h=0$ transition at $g_c(0)\approx 0.217$. The expected linear form $\Delta(h)=\Delta(0)-h$ for an $S^z=1$ excitation
is observed for $h<h_c$, with $h_c$ slightly less than $\Delta(0)$ implying a small gap discontinuity at $h_c$. We also observe
(SI, Sec.~S8 \cite{SI}) a very small magnetization jump, about $0.002$ per spin. These behaviors are reminiscent of the
well-known ``spin-flop'' transitions from Ising to canted XY AFM phases, but with anomalously small magnetization
discontinuity. We argue in SI (Sec.~S8 \cite{SI}) that the small magnetization and gap discontinuities, which
decrease further upon moving closer to $g_c(0)$, reflect an approximate emergent O($3$) symmetry in the
CBJQM at $h>0$.

The emergent symmetry can also be studied directly. At $h = 0$, the O($3$) AFM order parameter
$(m_x,m_y,m_z)$ combines with the scalar PS order parameter $m_p$ into an O($4$) vector $(m_x,m_y,m_z,m_p)$
at the $T=0$ transition \cite{Zhao_NP_2019,Sun_CPB_2021}. To detect the putative O($3$) symmetry of
$(m_x,m_y,m_p)$ at $h > 0$, we study the distribution $P(m_p)$ along the vertical line in Fig.~\ref{fig6}b.
In the PS phase, Fig.~\ref{fig6}d, $P(m_p)$ exhibits the expected double peak, reflecting the Z$_2$ symmetry
that is broken in the thermodynamic limit. In the AFM phase, Fig.~\ref{fig6}f, there is a single central peak,
reflecting the lack of PS order.

At a conventional first-order transition, a three-peak distribution would follow from coexisting PS and
AFM orders. In contrast, the distribution in the coexistence state in Fig.~\ref{fig6}e is nearly uniform
over a range of $m_p$ values (with finite-size rounded edges). The distribution $P(m_p)$ obtained by
integrating an O(3) symmetric $P(m_p,m_x,m_y)$ over $m_x$ and $m_y$ should indeed be uniform for $m_p \in [-R,R]$,
where $R \equiv {\rm max}(|m_p|)$; thus the approximately flat distribution demonstrates
emergent O($3$) symmetry in the presence of finite-size fluctuations of $R$. Though this symmetry cannot be exact,
i.e., it exists up to some finite length scale, it is responsible for suppressing $T_c$ and the gap at $H_c$;
see further SI, Sec.~S8 \cite{SI}, where we also show supporting results for cross correlations between
the PS and AFM order parameters.

We expect the same O($3$) emergent symmetry
at the PS--AFM transition in SrCu$_2$(BO$_3$)$_2$, where the ordered coexistence state breaks
the symmetry. The symmetry should be violated on long length scales, because of the distance
to the DQCP and also by 3D couplings. One of the Goldstone modes associated with the coexistence
state then develops a small gap. Studies of the CBJQM with inter-layer couplings suggest that
the symmetry is surprisingly robust \cite{Sun_CPB_2021}.

Emergent O($3$) symmetry on large length scales in SrCu$_2$(BO$_3$)$_2$ is supported, in particular,
by our results at $2.1$ GPa, where Fig.~\ref{fig4}c shows a large discontinuity in the AFM order
parameter but $T_{\rm c}$ is low and the gap (Fig.~\ref{fig6}a) is very small at $H_{\rm c}$.
Moreover, the uniform magnetization is extremely small and does not exhibit a discernible discontinuity
(SI, Sec.~S4 \cite{SI}). These behaviors are analogous to those in the CBJQM for $g$ close to $g_c(0)$.

{\bf Discussion.}---
Our high-pressure NMR experiments on
SrCu$_2$(BO$_3$)$_2$ in a magnetic field establish the first example of a
quantum magnet realizing DQCP phenomenology, which so far existed only
in the realm of field theories and model studies. We have demonstrated
PS and AFM transitions with $T_{\rm P}(H)$ and $T_{\rm N}(H)$ merging at
$T_{\rm c} \simeq 0.07$ K and $H_{\rm c} \simeq 6$ T. The PS--AFM transition at
$H_{\rm c}$ is first-order, with discontinuity weakening with increasing pressure.

We have argued that the remarkable suppression of $T_{\rm c}$ and absence of significant
PS gap discontinuity are consequences of emergent O($3$) symmetry generated by a nearby
DQCP. At the highest pressure, $2.4$ GPa, $1/T_1$ exhibits critical scaling for
$T$ between $0.2$ and $2$ K, indicating sufficient proximity to the DQCP (which is
likely of the multi-critical type \cite{Zhao_PRL_2020,Lu_PRB_2021,Yang_PRB_2022,Liu_arxiv_2021})
for realizing the characteristic quantum-critical fan \cite{Chubukov_PRB_1994} on
the gapped PS side of the transition. Strong 3D AFM ordering effects on the
gapless side of the transition mask putative quantum-criticality in $1/T_1$ there,
but the AFM ordering temperature $T_{\rm N}$ vanishes in a way very similar to the PS
ordering $T_{\rm P}$, again in support of emergent symmetry of the order parameters.

The $H=0$ AFM phase was previously detected in the specific heat between
$3.2$ and $4$ GPa \cite{Guo_PRL_2020}, with $T_{\rm N}$ from $2$~K to $3.5$~K.
Subsequently, results at $H > 0$ were also reported \cite{Larrea_Nature_2021}.
However, while $T_{\rm P}(H)$ from the specific heat agrees well with our PS transitions
in Fig.~\ref{fig1}d, the heat capacity peak assumed to signal the AFM transition did
not drop below $1$ K \cite{Larrea_Nature_2021}, extending above the PS phase at
fields as low as $3$~T. It may be difficult to detect the small specific-heat peak
signaling the AFM transition \cite{Guo_PRL_2020} in high-field measurements at low
temperatures.

Beyond the highest pressure reached here, a plausible scenario
\cite{Yang_PRB_2022,Lu_PRB_2021} is a QSL between the PS and AFM phases (Fig.~\ref{fig2}).
Our experiments do not directly address the putative QSL, and further investigations
should elucidate the low-$T$, $H = 0$ state between $2.6$ and $3$ GPa (where
no order has been detected \cite{Guo_PRL_2020,Larrea_Nature_2021}) and its
evolution as $H$ approaches $5.7$ T, where our current experiments point to
a DQCP slightly above $2.4$ GPa.

{\bf Acknowledgments.}---
We would like to thank Bruce Normand for his extensive suggestions and constructive criticism.
We also thank Wenan Guo, Fr\'ed\'eric Mila, Masashi Takigawa, Yiming Wang, Zhi-Yuan Xie, and Yi-Zhuang You
for helpful discussions.
This work was supported by the National
Natural Science Foundation of China under Grants No.~12134020, 12104503,
12174441, 51872328, and 11874401, the Ministry of Science and Technology of
China under Grants No.~2016YFA0300504 and 2017YFA0302903, the Simons Foundation
under Simons Investigator Grant No.~511064, the China Postdoctoral Science
Foundation under Grant No.~2020M680797, Beijing Institute of Technology Research
Fund Program for Young Scholars, and the Fundamental Research Funds for
the Central Universities and the Research Funds of Renmin University of China
under Grants No.~21XNLG18 and 18XNLG24. Some of the numerical calculations
were carried out on the Shared Computing Cluster managed by Boston University's
Research Computing Services.

{\bf Author contributions.}--- Y.C. performed NMR measurements and data analysis
with assistance from C.L., Z.H., and W.Y.. W.H. and S.L. provided single
crystals. X.L. and H.L. performed the Bayesian fitting analysis. L.L., H.L., N.X.,
and K.H.W. performed numerical simulations with guidance from R.Y. and A.W.S..
W.Y., A.W.S., and R.Y. guided the project and wrote the manuscript with
input from all the authors.

{\bf Competing interests.}--- The authors declare that they have no competing interests.

{\bf Data and materials availability.}--- All data needed to evaluate the conclusions
in the paper are present in the manuscript or the Supplementary Materials.

\onecolumngrid

\newpage

\begin{center}

{\large Supplemental Information}
\vskip5mm

{\bf\ Proximate deconfined quantum critical point in SrCu$_2$(BO$_3$)$_2$}

\vskip5mm

Yi Cui, Lu Liu, Huihang Lin, Kai-Hsin Wu, Wenshan Hong,
Xuefei Liu, Cong Li, Ze Hu, Ning Xi, Shiliang Li,\\
Rong Yu, Anders W. Sandvik, and Weiqiang Yu

\end{center}

\vskip5mm

\twocolumngrid

\setcounter{equation}{0}
\setcounter{figure}{0}
\setcounter{section}{0}
\setcounter{table}{0}
\setcounter{page}{1}

\renewcommand{\theequation}{S\arabic{equation}}
\renewcommand{\thefigure}{S\arabic{figure}}
\renewcommand{\thetable}{S\arabic{table}}
\renewcommand{\thesection}{S\arabic{section}}
\renewcommand{\thepage}{S\arabic{page}}


\section{Methods}
\label{smethod}

\subsection{Experiments}

Single crystals of SrCu$_2$(BO$_3$)$_2$ were grown by the floating-zone
method. We used a NiCrAl piston cell for the high-pressure NMR measurements
and Daphne 7373 oil as the pressure medium. The pressure was calibrated using
the low-temperature Cu$_2$O NQR frequency~\cite{Reyes_1992}, and the highest
pressure achieved was 2.4 GPa. NMR experiments were performed in two types of
cryostat: a variable-temperature insert (VTI) was used for measurements in
the temperature range from 1.5~K to 50~K and a dilution refrigerator was used
to achieve temperatures ranging from 0.07~K to 2~K.

\subsection{NMR spectrum}

The $^{11}$B spectra were obtained by the standard spin-echo technique. $^{11}$B
nuclei have spin $I = 3/2$, whence the NMR spectra and line widths contain
contributions from both local magnetism and electric-field gradients (EFGs).
The local Hamiltonian for a nucleus with spin $I$ and quadrupole moment
$Q$ can be described by the following form,
\begin{eqnarray}
  {\mathcal H}_{\rm n} & = & {\mathbf I}\cdot{\mathbf B}+ \sum_{i} A_{\rm hf}^{{\alpha}{\beta}}(i){I^{\alpha}}{S_i^{\beta}} \nonumber \\
  && + \frac {e^2qQ}{4I(2I-1)}[3I_z^2-{\mathbf I}^2+\eta(I_x^2-I_y^2)],
  \label{hhh}
\end{eqnarray}
where the three terms describe, respectively, the coupling of the nuclear spin to the external field,
the hyperfine coupling between the nuclear spin and neighboring electronic spins, and the coupling
between the nuclear quadrupole moment and the local EFG. The factor $q$ in the last term is from the
EFG tensor (the component along the principal axis), which is produced by the local electronic
structure of the ions.

To identify the ordered phases and clarify their nature, NMR spectra were measured
over a wide field range from $H=0.2$ T to $15$ T along the crystalline $c$ axis,
with a top-tuning circuit allowing for the corresponding wide range of frequencies.
The full spectrum was obtained by the
frequency-sweep method, which covers one center line and two sets of satellites
on each side of it.
Figure~3a of the main text shows a typical NMR spectrum at
4~T. It contains one center line, with a frequency of approximately zero
relative to $^{11}\gamma H$, where $^{11}\gamma$ is the Zeeman factor, and two
sets of satellites aligned from $\pm 1.0$~MHz to $\pm 1.4$~MHz. We define the
Knight shift as $^{11}K_n = f/{^{11}\gamma}H - 1$, where $f$ is the position of
the center peak in the spectrum.

The NMR magnetic field was oriented primarily along the
$c$ axis, with a $8.6^\circ$ tilting applied to separate the four $^{11}$B sites
producing satellites in the spectrum.
The angle is calibrated by different satellite frequencies $\nu_Q$
of four sets of satellites shown in Fig.~3a,
as reported by the earlier NMR study~\cite{Waki_JPSJ_2007}.
The four satellites arise because each
$^{11}$B site has a different orientation of the principal EFG axis relative
to the (tilted) external field. Each satellite line cannot be assigned to a
specific $^{11}$B site because we did not characterize the exact direction of the
field in the sample $ab$ plane.

\subsection{NMR line widths}

For a system with local inhomogeneity, the FWHM of the NMR center line has two
additive terms, the hyperfine field contribution, which scales linearly with
the applied field, and a second-order EFG correction, which is weak and scales
inversely with the field. In a field of 0.2~T and at temperatures below 5~K,
the FWHM of the center line is of order 2~kHz. By contrast, although the FWHM
of the satellites has a similar hyperfine-field contribution, it has a large
and nearly field-independent first-order EFG contribution, making the FWHM of
each satellite larger than 10~kHz for the same temperature range and field.

\subsection{NMR spin-lattice relaxation rate}

The relaxation time $^{11}T_1$ was measured using the
spin-inversion method by applying a $\pi$ inversion pulse (of duration
4~$\mu$s). In all phases displaying a line-splitting, the $1/^{11}T_1$
data reported in the main text were measured on the higher-frequency peaks,
meaning $f_R$ in Fig.~4a, although $1/^{11}T_1$ measured on $f_L$ was verified
to be consistent in every case with that on $f_R$.

Note that $1/^{11}T_1$ overall increases with the field, as shown Fig.~5 of the main text.
In particular, this increase follows the expectations in the quantum-critical fan on the
PS side of the transition, analyzed in Fig.~5e. This behavior indicates dominant contributions
from the hyperfine field fluctuations originating from the critical spin fluctuations. By contrast,
the EFG contributions, which are caused indirectly by bond and plaquette singlet fluctuations
that couple to the lattice [causing fluctuations of $q$ in Eq.~(\ref{hhh})], are expected to decrease
with the field because the PS fluctuations are suppressed by the field.

The most likely reason for the absence of visible EFG contributions in the $1/^{11}T_1$ data is that
the singlet fluctuations in the frustrated magnet induce bond-length fluctuations of relatively small
amplitude, i.e., the fluctuations of $q$ in Eq.~(\ref{hhh}) are small and generate insignificant
contributions to $1/T_1$ compared to the spin fluctuations mediated directly through the hyperfine coupling.

Evidence for different scales of the effective EFG and hyperfine couplings come from the way we have
detected the PS and AFM order parameters using the NMR line shape. In the case of PS order, in
Fig.~3b of the main text we used the FWHM of the central and satellite lines, because the expected
peak splitting is too small to observe. Thus, the lattice deformation induced by the frozen singlets,
i.e., the modulation of $q$ in Eq.~(\ref{hhh}), is very small. In contrast, the peak splitting is very
substantial in the AFM phase, Fig.~4. Given the small modulation of $q$ in the ordered PS state, the
fluctuations of $q$ above and close to the PS transition temperature, induced indirectly by quantum fluctuations
of bond and plaquette singlets, should also be small and contribute insignificantly to $1/T_1$ compared
to the spin fluctuations mediated directly by the large coupling $A_{\rm hf}$. We are therefore
justified in analyzing $1/T_1$ solely in terms of spin fluctuations detected through the hyperfine
coupling---this assumption is often made in NMR studies of quantum magnets without the
supporting evidence we have here from the weak response in the PS state.

It is also possible, in principle, that $1/^{11}T_1$, as an anisotropic
probe of EFG and magnetic fluctuations, may not significantly sense the anisotropic EFG
fluctuations geometrically by accident \cite{Yogi_JPSJ_2011}. Because of our
direct evidence of insignificant EFG contributions due to the small amplitude of the
lattice fluctuations, there is no need to invoke such a mechanism here, however.

\subsection{Fitting using Bayesian inference}

To obtain the fitting parameters for the functional forms of $T_{\rm N}(H)$
and $T_{\rm P}(H)$, with these phase boundaries coming together at a low common
$T_{\rm c}$ value at $H = H_{\rm c}$, we followed the Bayesian inference procedure
of Ref.~\cite{Allenspach_2021}. Sets of points spanning the multidimensional
parameter space are generated by a Markov-chain Monte Carlo process and the output
is a statistical model for the probability that a given parameter set can describe
the measured data. The errors inherent to the data are then reflected systematically
as the uncertainties in each of the fitting parameters. For further details we
refer to Ref.~\cite{Allenspach_2021}.

The functional forms used to model the transition temperatures are given in the main
text and the maximum-probability results for pressures $2.1$ and $2.4$ GPa are shown in
Fig.~5f. Probability distributions for the system parameters consistent with the
measured data are shown in Sec.~\ref{sstats} of the SI, and these distributions underlie
the error bars on the parameters reported in the main text.

\subsection{QMC simulations of spin models}
\label{methods:cbjq}

The Hamiltonian of the CBJQM, i.e., the $J$-$Q$ model \cite{JQ_PRL_2007} defined on the
checkerboard lattice \cite{Zhao_NP_2019}, subject to an applied magnetic field is given by
\begin{equation}
{\mathcal H} = - J \! \sum_{\langle ij \rangle} \! P_{ij} - Q \!\!\!\!
\sum_{ijkl\in\square_s} \!\!\!\! (P_{ij} P_{kl} + P_{ik} P_{jl})
 - H \! \sum_{i} \! {S^z_i},
\label{hjqh}
\end{equation}
where $\mathbf{S}_i$ is an $S = 1/2$ spin operator at site $i$ and $J > 0 $ is the
nearest-neighbor AFM Heisenberg coupling, with the corresponding operator
defined for convenience as a singlet projector,
\begin{equation}
P_{ij} = 1/4 - \mathbf{S}_i \cdot \mathbf{S}_j.
\end{equation}
The second term describes a partial ring-exchange-type interaction (or
correlated singlet projection) defined on plaquettes $\square_s$ with a
checkerboard arrangement on the square lattice. The third term describes
the external magnetic field of strength $H$.

For the simulation results shown in Figs.~6b-6f of the main text, we took
$J = 1$ as the unit of energy to define the dimensionless parameter $g = J/Q$
and the reduced field $h = H/J$. We studied this model by stochastic series
expansion (SSE) quantum Monte Carlo (QMC) simulations~\cite{Sandvik_PRE_2002}
on systems of sizes $L$$\times$$L$ up to a maximum of $L = 64$ and down to a
minimum temperature of $T = J/L$. The SSE method is exact, in the sense that
it is based on a construction corresponding to a complete representation of
the imaginary-time dimension of a quantum system when mapped into a classical
model (on which the Monte Carlo sampling is performed), and thus the simulation
results are only affected by statistical errors.

To characterize the PS--AFM transition (Sec.~\ref{cbjqma} and \ref{cbjqmb}),
we used an order parameter for the PS phase of the form
\begin{equation}
m_p = \frac{2}{L^2} \sum_{\mathbf{i}\in\square_s} (-1)^{i_x} S^z
(\mathbf{i}) S^z (\mathbf{i}+\hat{x}) S^z (\mathbf{i} + \hat{y})
S^z (\mathbf{i} + \hat{x} + \hat{y}),
\label{mpdef}
\end{equation}
where $\mathbf{i}$ is the position vector of the checkerboard plaquettes and $i_x$
denotes the row index of $\mathbf{i}$. We also calculated the uniform magnetic
susceptibility,
\begin{equation}
\chi = \frac{\partial m}{\partial h},~~~m = \frac{1}{N} \sum_{i=1}^N
\langle S^z_i \rangle,
\label{chidef1}
\end{equation}
where $N = L^2$. In the simulations we obtained the susceptibility in the standard way
from the magnetization fluctuations,
\begin{equation}
\chi = \frac{N}{T}\bigl [\langle m^2\rangle-\langle m \rangle^2 \bigr],
\label{chiqmc}
\end{equation}
where $\langle m\rangle=0$ in the absence of external field.
We extracted the spin excitation gap in the PS phase by
fitting our numerical results to the form $\chi(T) \propto e^{-\Delta/T}$ at
low temperatures for each field.

In addition to the above observables, which can be obtained using diagonal estimators in the
$S^z$ basis used, we also study cross-correlations between the AFM and PS order parameters. In
the case of $h=0$, we can use the diagonal correlator $\langle m_z^2 m_p^2\rangle$, where
$m_z$ is $z$-component of the staggered magnetization, but when
$h>0$ the magnetization operator in $\langle m_x^2 m_p^2\rangle=\langle m_y^2 m_p^2\rangle$
is off-diagonal and a more complicated estimator based on the directed-loop SSE updates is required.
Essentially, in the directed-loop updates used to evolve the configurations \cite{Sandvik_PRE_2002},
the probability distribution of the two open ends of a uncompleted loop (i.e., a string) at the
construction stage is related to the off-diagonal operator $m_x^2 + m_y^2$ (here in the case
where the string ends are located at the same ``time slice'' in the SSE operator space).
We follow the implementation discussed by Dorneich and Troyer \cite{Dorneich_PRE_2001}, with the
modification that each contribution is also multiplied by the diagonal operator $m_p^2$ when
accumulating the cross-correlation function $\langle (m_x^2+m_y^2) m_p^2\rangle$.

To complement the simulations of the CBJQM, we also apply SSE QMC simulations to study
an anisotropic XXZ Heisenberg $S=1/2$ spin model given by the Hamiltonian
\begin{equation}
{\mathcal H} = -J \! \sum_{\langle ij \rangle} \! (P_{ij} - \lambda S^z_iS^z_j) - H \! \sum_{i} \! {S^z_i},
\label{hxxz}
\end{equation}
where $\lambda$ expresses the spin anisotropy as a deviation from
the Heisenberg model.

The XXZ model clearly has an exact O($3$) symmetry at $\lambda = 0$, while perturbations
with $\lambda < 0$ imply an O($2$) and with $\lambda > 0$ a Z$_2$ (Ising) order parameter.
For $\lambda > 0$, the system undergoes a first-order ``spin-flop'' transition versus the
magnetic field; for small $h$ remaining in the Ising phase and transitioning through a
level crossing into a canted XY (planar) AFM. We use this transition as a well understood
benchmark case for comparing and contrasting with the CBJQM results in Sec.~\ref{scbjq} of the SI.

\section{Inequivalent Boron sites in the PS phase}
\label{stib}

\begin{figure}[t]
\includegraphics[width=8.5cm]{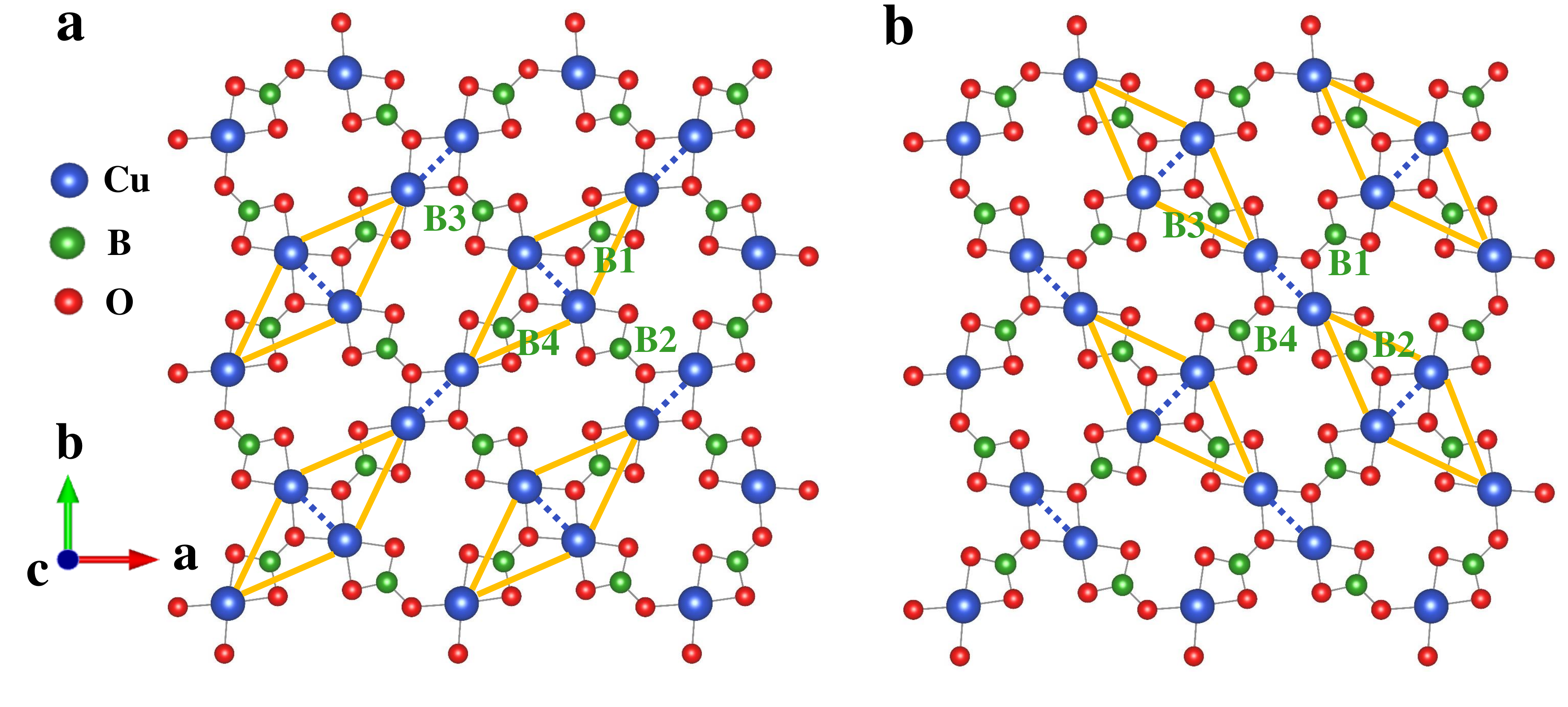}
\caption{\label{fptwinss2}
Illustration of the FP phases in one planar layer of SrCu$_2$(BO$_3$)$_2$,
where each singlet forms on the yellow diamond composed of four Cu
ions, which includes one Cu dimer (short dashed lines). B1 to B4
label the four B sites in each unit cell of SrCu$_2$(BO$_3$)$_2$, which
are inequivalent in a magnetic field tilted away from the $c$-axis.
{\bf a} and  {\bf b} represent the two degenerate FP states, in each of
which only one type of diamond (dimer) orientation is selected.}
\end{figure}

The planar lattice structure of SrCu$_2$(BO$_3$)$_2$ is shown in Fig.~1a of the
main text, and is represented in a more detailed way in Fig.~\ref{fptwinss2}
to illustrate the singlet formation in the full-plaquette (FP) variant of the
PS state. Each unit cell contains four B sites, B1 to B4, which are equivalent
in the DS state at zero field. For applied fields with an in-plane
component, these four sites become inequivalent by field orientation with
respect to the B-O bond directions. The NMR spectrum of $^{11}$B consists of
one center line and two satellite lines (Fig.~3a of the main text); the
satellite frequencies are affected by both the hyperfine field and a
first-order correction from the EFG, whereas the center line is affected by
the hyperfine field and a weaker second-order EFG correction (which is
negligible at fields above 2~T).

\begin{figure}[t]
\includegraphics[width=8.5cm]{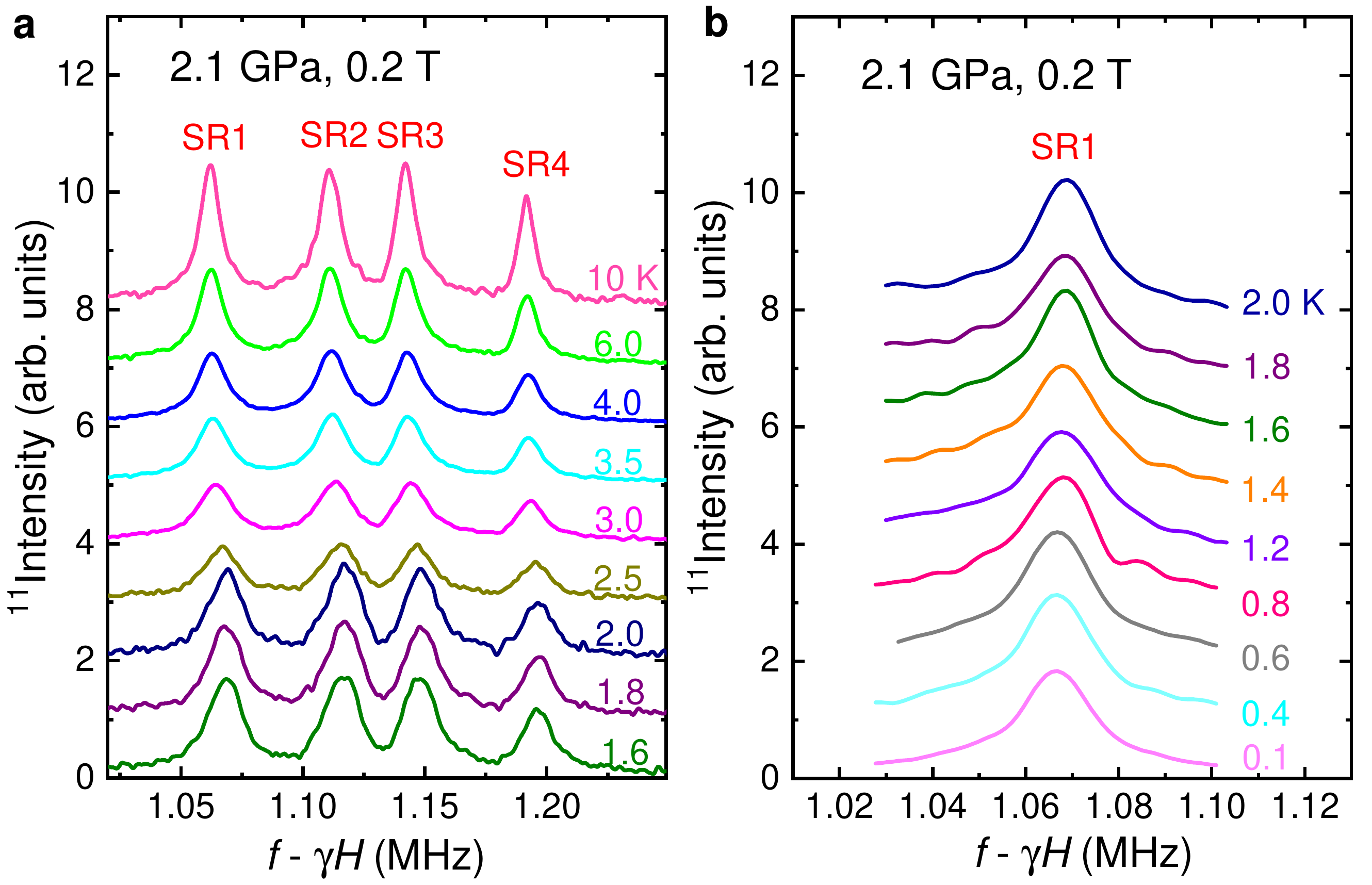}
\caption{\label{fpspecs3}
{\bf a} High-frequency NMR satellite spectra measured in a conventional He
refrigerator under a pressure of 2.1~GPa and a small field of 0.2~T. SR1-SR4
are associated with the four B sites, as labelled in Fig.~\ref{fptwinss2},
although the precise B site corresponding to each NMR line is not determined.
{\bf b} Satellite SR1 measured in the dilution refrigerator at 2.1~GPa and
0.2~T.}
\end{figure}

When the applied field is tilted away from the $c$ axis, failure to discern
four peaks in the center line at low fields simply indicates that the
difference in hyperfine effects on the four sites is too small to be
resolved. By contrast, the four distinguishable satellite peaks shown in
Fig.~\ref{fpspecs3}a are a clear indication for strongly differing EFG
effects (i.e.~principal axes of the local EFG) at sites B1 to B4.

Recent studies of the SSM find a close energetic competition
between the two possible plaquette states \cite{Boos_PRB_2019,Xi_arxiv_2021}; the FP
state in which a dimer is enclosed in the local four-spin singlet (represented in
Fig.~1b of the main text) and the empty-plaquette (EP) phase in which no dimer
is enclosed. The double degeneracy of the FP phase is shown explicitly for the
planar lattice structure of SrCu$_2$(BO$_3$)$_2$ in Figs.~\ref{fptwinss2}a and
\ref{fptwinss2}b; the two diamond (dimer) directions are mutually orthogonal.
Because of the shape of the four-spin units involved in the spontaneous
breaking of the two-fold symmetry when the FP state forms, this phase
transition must be accompanied by an orthorhombic distortion of the lattice.
In a bulk single crystal, when the plaquette phase is entered on cooling, the degeneracy associated
with the distortion in the FP phase is expected to lead to structurally twinned
domains, i.e., macroscopic coexistence of the degenerate ground states by
phase separation.

Figure \ref{fpspecs3} shows NMR satellite spectra measured through the FP
PS transition in both conventional and dilution refrigerators, under very low
field to reduce the magnetic broadening of the line. Each line has a Lorentzian
shape, a result we used to fit their FWHM, which is displayed in Fig.~3b of the
main text. The strong change of FWHM observed below 1.8~K in the NMR satellite,
but not in the center line, is understood from local lattice distortions as
follows: The twinning of the FP phase leaves any given B site (e.g.~B1) either
inside (Fig.~\ref{fptwinss2}a) or outside (Fig.~\ref{fptwinss2}b) the local
diamond-shaped FP~\cite{Waki_JPSJ_2007}. This leads to a splitting of the
$^{11}$B spectra, because the different B1 sites have different hyperfine
fields and local EFGs. The nuclear quadrupole moment on the inside site experiences
a larger EFG due to the lattice contraction, and conversely for the outside site,
causing a further splitting of each satellite.

The large increase observed in the FWHM of the  satellites below 1.8~K in Fig.~3b of the
main text is consistent with such an EFG effect, which is too small to be resolved as
a line splitting but is responsible for the order parameter behavior of the FWHM.
The much weaker increase in the FWHM of the center line excludes a dominant role for
magnetic effects on the NMR line shapes at such a low field, meaning that the explanation
requires local structural distortions.

In contrast to the two inequivalent sites forming in the FP state,
in an EP state all the B sites reside outside the square plaquettes and experience
a reduced EFG, whence a narrowing of the FWHM would be expected. Thus, we have
confirmed that the PS state is of FP type, where the lattice has undergone an
orthogonal distortion \cite{Boos_PRB_2019}.

\section{NMR Knight shift and $1/T_1$ around the DS--PS phase transition}
\label{slp}

\begin{figure}[t]
\includegraphics[width=8.5cm]{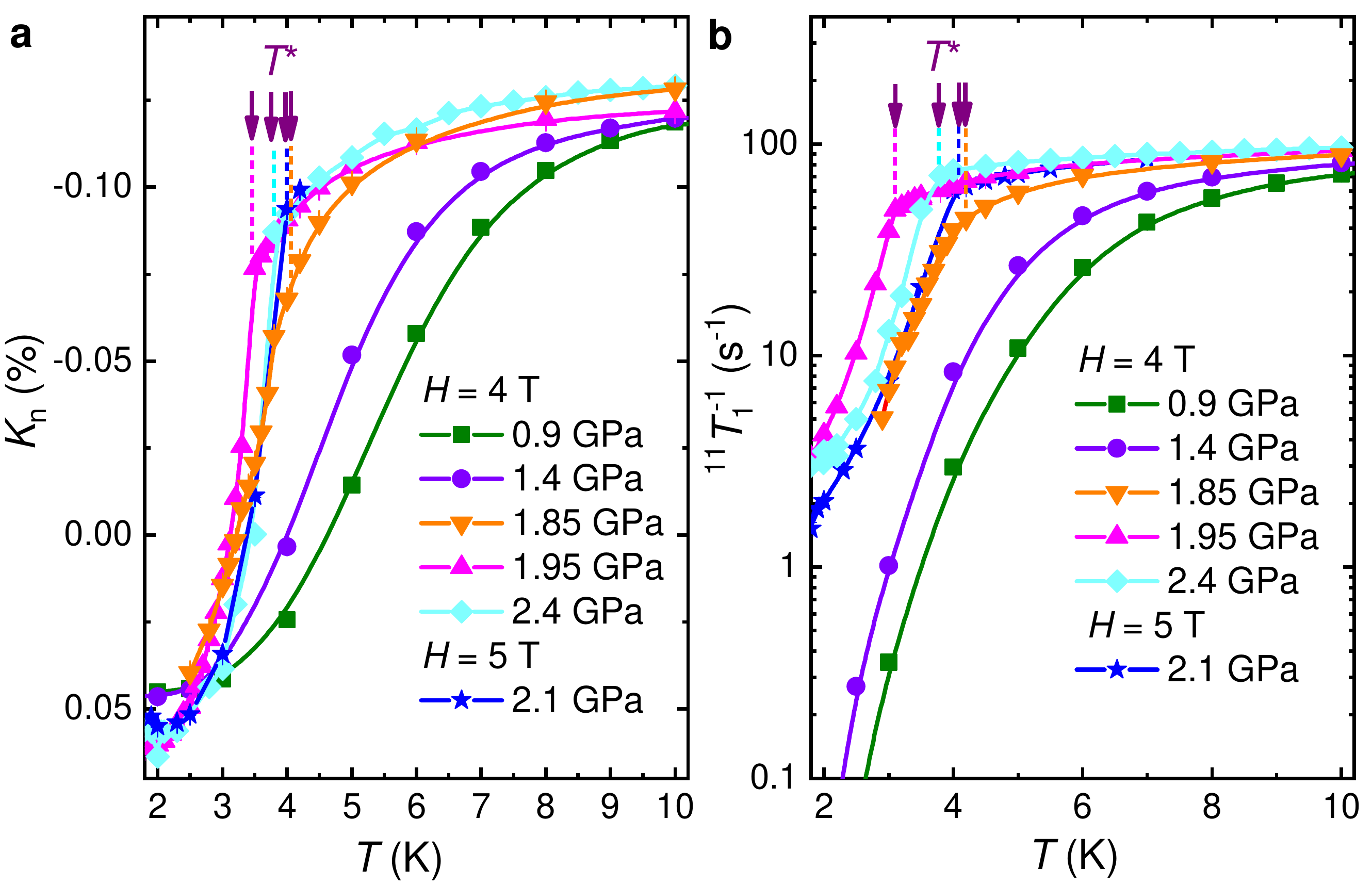}
\caption{\label{ffps1}
{\bf a} Knight shift, $^{11}K_n (T)$, shown for a number of different
pressures and fields. {\bf b} Spin-lattice relaxation rate, $1/^{11}T_1 (T)$,
shown for the same pressures and fields. In both panels, $T^*$ indicates the
temperature below which a dramatic change is observed in $^{11}K_n$ and
$1/^{11}T_1$. Here $^{11}K_n$ and $1/^{11}T_1$ for pressures at and above
1.85~GPa were measured on the $f_2$ peak shown in Fig.~3 of the main text.}
\end{figure}

Figure~3c in the main text shows the evolution of the NMR center line (arrows) for
several temperatures at pressure 0.9~GPa and field 4~T. The spectrum has a
single peak with a negative Knight shift for temperatures from 10~K to 2~K,
consistent with the negative hyperfine coupling, $A_{\rm hf} \simeq -0.259$
T/$\mu_{\rm B}$, for ${\vec H} \parallel {\hat c}$ \cite{kodama_JPCM_2002,
Waki_JPSJ_2007}. The change of the $^{11}$B NMR Knight shift, $^{11}K_n (T)$, is shown
in Fig.~\ref{ffps1}a for two low field values and a wide range of pressures.
The saturation of $^{11}K_n$ below 2~K to approximately $0.05\%$ that is
observed at all pressures is caused by a temperature-independent orbital
contribution, $^{11}K_{\rm orb}$, as a result of which the magnetic contribution
to the hyperfine coupling is then taken as ${^{11}K}_s (T) = {^{11}K}_n (T) -
{^{11}K}_{\rm orb}$. Similarly, $^{11}K_n$ at all pressures decreases slowly when
cooled down to 10~K from high temperatures (data not shown), consistent
with PM behavior given the negative hyperfine coupling.

Returning to the interpretation of Fig.~3c, the change of
$^{11}K_n (T)$ is smooth for pressures below 1.85 GPa, matching the rapid
drop of the susceptibility as the temperature decreased below the spin
gap of the DS state, and consistent with a crossover but no phase transition
(only an increasing density of singlets on the $J'$ bonds as $T$ is lowered).
Figure \ref{ffps1}b shows the analogous behavior in spin-lattice relaxation
rate, $1/^{11}T_1$, for the same pressures and fields.

By contrast, for pressures at and above $1.85$ GPa Fig.~\ref{ffps1} shows more
sudden drops of $^{11}K_n$ and $1/^{11}T_1$ on cooling through an onset temperature that we
label as $T^*$. The reduction of the uniform susceptibility and the low-energy spin fluctuations
here indicate either a sharp crossover or a true phase transition into the PS
liquid phase (where plaquette singlets form but do not yet order) identified
in the main text.

To distinguish between a phase transition and a sharp cross-over, Figs.~3d
in the main text is helpful. It again shows the center NMR line at a series of temperatures,
in this case at 1.85~GPa and 4~T. Above 4~K, the spectrum has a single peak with
a negative Knight shift. The Knight shift again increases rapidly on cooling, and in
this case a line splitting is observed below 4~K with two peaks labeled as $f_1$ and $f_2$
respectively. The spectral weight of the $f_1$ peak first increases on cooling below $4$ K
but then vanishes at $2.9$ K. This line splitting is observed only at 1.85 and 1.95 GPa.

The appearance of two lines over a finite temperature range suggests the phase
coexistence expected when the system crosses the first-order transition between the DS
and PS liquid phases, i.e., at temperatures below the critical point where this line of
transitions terminates, as indicated schematically in Fig.~2a of the main text.
Given that the system at $1.85$ GPa is in the PS phase at low temperatures, the $f_1$ peak, which
is gradually suppressed below 4~K and is absent below 2.9 K, must be associated with the
DS phase. The $f_2$ peak, which remains as a single peak below 2.9 K, corresponds to another
phase. Because of its spectral character, discussed below (see also Sec.~\ref{stib}), we identify
it as arising from regions in the PS liquid phase.

Thus the line splitting is the signal that the process occurring around 4~K
at $1.85$ and $1.95$ GPa is a genuine first-order magnetic transition from
the PM phase into the PS liquid. This is possible when the first-order DS
transition line is not vertical in the ($P,T$) plane, but has a finite slope
toward higher pressures as $T$ increases, as represented in Fig.~2a
of the main text. Within the uncertainties of pressure calibration in different
specific-heat measurements, this is consistent with the identification of
the DS--PS transition between $1.7$ and $1.8$ GPa at temperatures below 2 K in
Ref.~\cite{Guo_PRL_2020}, while the Ising-type critical point was located
at $3.3$ K and approximately $1.9$ GPa in Ref.~\cite{Larrea_Nature_2021}.

It should be noted that both phases are of singlet type and therefore the first-order
line is not sensitive to the magnetic field strength even at the rather high fields,
$4$ and $5$ T, used in Figs.~3c-e. Phase coexistence over a rather wider range of
temperatures, $3 \sim 4$ K, is explained by a very mild slanting of the transition line
in the region where this line is crossed. Thus, the system remains very close to the
transition line for an extended range of temperatures.

The highest pressure at which we have observed this line splitting is $1.95$ GPa, indicating that phase coexistence
is terminated there, in good agreement with $P \simeq 1.9$ GPa for the critical end point determined from the heat capacity
in Ref.~\cite{Larrea_Nature_2021}. Certainly no line splitting is observed at $2.1$ GPa in Fig.~3e of the
main text. This is in contrast to Ref.~\cite{Waki_JPSJ_2007}, where a different line splitting was observed
at $2.4$ GPa \cite{Waki_JPSJ_2007}.

Previous NMR works at 2.4 GPa reported two types of line splitting,
with one occurring below 30~K, and one below 4~K with a large line split
of $\sim$0.2 MHz \cite{Waki_JPSJ_2007,Takigawa_JPSJ_2010}.
These line splitting are different from our observation with
a much smaller line split ($\sim$0.03 MHz), at lower pressures.
The previous reported line splits may be caused uniaxial pressure inhomogeneity
which is enhanced upon cooling and breaks the tetragonal crystal symmetry.
Our sample resolves clearly the PS and the AFM phases and their respective
transition temperatures, indicative of high sample quality and
pressure homogeneity. Therefore, it is unlikely that disorder effects
are significant in our sample and smears out other orderings,
in contrast to a doped Han purple compound \cite{Allenspach_2021}.

Above the small pressure window $1.85 \sim 1.95$ GPa, the single center peak with its very rapid change in
$K_n$ and $1/T_1$ at $3 \sim 4$ K persists at all higher pressure measurements in Fig.~\ref{ffps1}.
Because the change from the high-$T$ PM phase to the PS liquid phase should not be
a true phase transition when the pressure is sufficiently high or low for the first-order
DS line to be avoided, we conclude that $T^*$ for $P > 1.95$ GPa reflects an extremely sharp crossover.
We stress that, in the PS liquid, the plaquette singlets fluctuate between the two orthogonally
directed local plaquette types, so that no local symmetry-breaking occurs in this phase, only
in the ordered PS state below $2$ K. The cross-over is manifestly very sharp close to the Ising-type
critical point, as reflected also in the sharp peaks in the specific heat \cite{Guo_PRL_2020,Larrea_Nature_2021}.
Thus, the NMR signal can easily be mistaken for a true phase transition, which would not be compatible
with the critical-point scenario and other aspects of the phase diagram.

Though all of our observations at $3 \sim 4$ K are consistent with the critical-point scenario \cite{Larrea_Nature_2021}
and sharp cross-overs at $T^*$ from the PM phase, we still offer an alternative scenario for completeness: The critical Ising
point scenario was considered theoretically from the perspective of a 2D system \cite{Larrea_Nature_2021}, but in principle
3D effects could turn the critical point into a triple point (with large fluctuations since the 3D couplings should be very weak);
the nexus of three first-order transitions. Then, the PM phase (which can be interpreted as a gas phase) would turn upon lowering
the temperature into either the PS liquid or the DS phase (which then also should be considered as another liquid-like phase)
through a very weak first-order transition without any symmetry breaking. This alternative scenario, for which we do not have any
evidence, experimental or theoretical, would not affect the other parts of the phase diagram in Fig.~2, including the DQCP scenario.

\section{AFM spectra and phase transitions}
\label{safm}

\subsection{Continuous temperature-driven transition}

\begin{figure}[t]
\includegraphics[width=8.5cm]{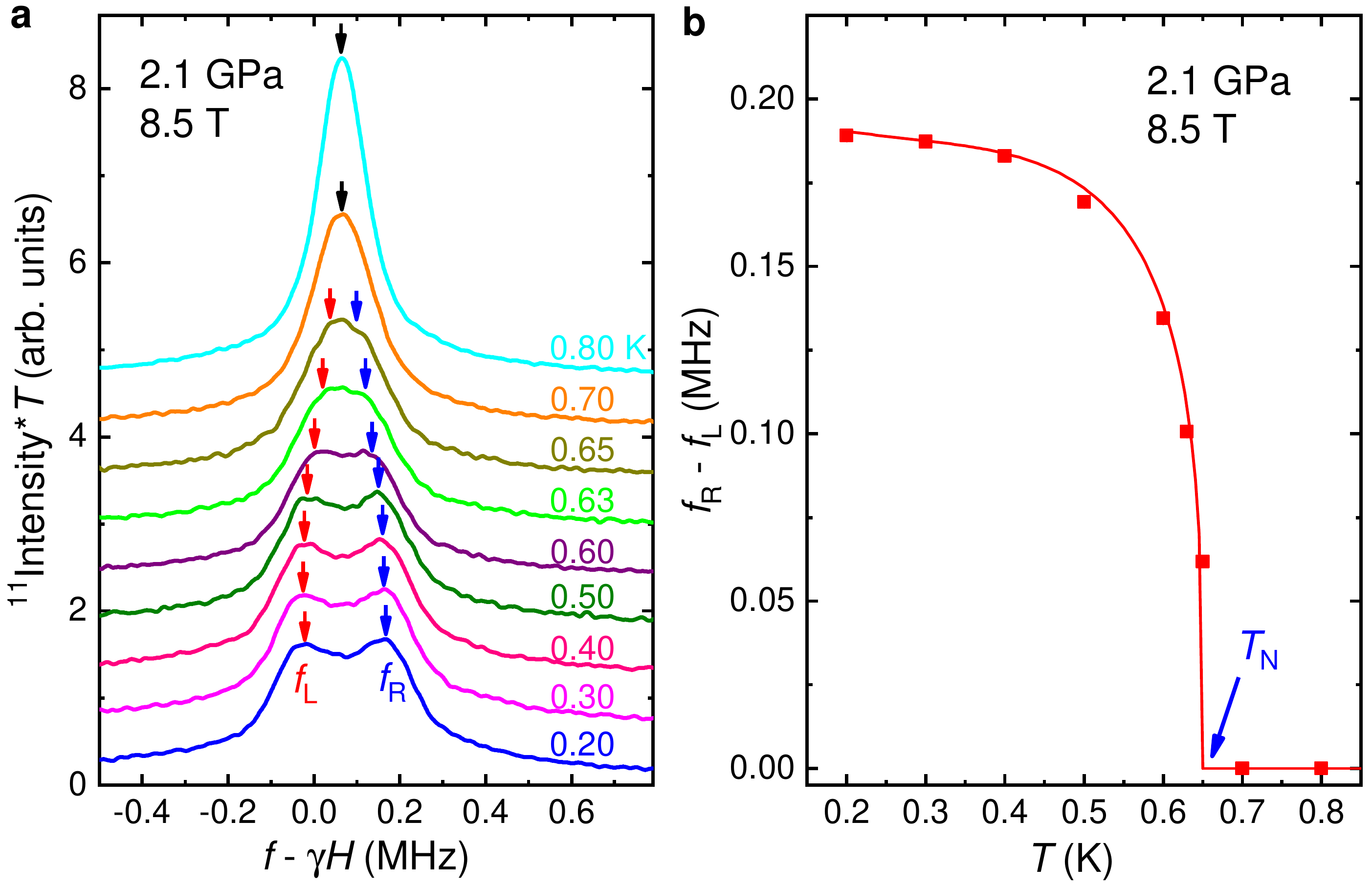}
\caption{\label{afmspecs6}
{\bf a} NMR spectra measured over a sequence of dilution-refrigerator
temperatures at $P=2.1$ GPa and $H=8.5$ T. The data sets are offset vertically for
clarity. Black arrows mark the single-peak locations while the red and blue arrows
mark the locations of the split peaks, at the respective frequencies $f_L$ and $f_R$
determined by the double-Lorentz fitting,
that signal the AFM phase. {\bf b} Peak-splitting $f_R - f_L$ (a proxy AFM order
parameter) shown as a function of temperature. $T_{\rm N}$ marks the transition temperature
into the AFM phase, where $f_R - f_L$ rises from zero. The fitted curve has the
asymptotic form $f_R - f_L \propto (T_{\rm N}-T)^\beta$ with $T_{\rm N}=0.65$~K and
$\beta = 0.305$.}
\end{figure}

In Fig.~4a and 4b of the main text we showed the splitting of the NMR center line as
the system is driven from the PS to the AFM phase by increasing the applied field.
The two field-induced peaks are located symmetrically around the zero of frequency
measured relative to $f_0 = {^{11}\gamma} H$. This symmetrical line splitting indicates
the onset of equal negative and positive hyperfine fields and constitutes direct evidence
for AFM ordering. To complement the results for the field-driven PS--AFM transition, here we discuss
the line splitting observed at constant field as a function of the temperature. The center line of
the NMR spectrum at 2.1~GPa and 8.5~T is shown in Fig.~\ref{afmspecs6}a over a range of
low temperatures. A single line is observed at and above 0.7~K, but below 0.65~K the spectrum
shows two peaks, labeled as $f_L$ and $f_R$.

We again use the frequency difference $f_R - f_L$ as a proxy for the
AFM order parameter, which we show as a function of temperature in Fig.~\ref{afmspecs6}b. We observe
that $f_R - f_L$ rises rapidly on cooling below $T_{\rm N}$, and note that for this
field, $1/^{11}T_1$ displays a clear peak at the same temperature, as shown in
Fig.~5b of the main text.

Because no obvious three-peak signature of phase separation can be observed in the NMR spectra in Fig.~\ref{afmspecs6}a,
our results in this case establish a continuous or very weakly first-order AFM transition on cooling in a constant
field 8.5 T, which is significantly away from the common transition field $H_{\rm c} \approx 6.2$ T.
Close to the first-order triple point that we have established in the main text at $(T_{\rm c},H_{\rm c})$
and $2.1$ GPa, the transition should remain first-order, while sufficiently far away from the triple point one
should expect continuous transitions (as has been observed in model studies \cite{Sun_CPB_2021}).

The curve fitted to
the data points in Fig.~\ref{afmspecs6}b is of the form $f_R - f_L = F(T_{\rm N}-T)(T_{\rm N}-T)^\beta$, where
$F(x)$ is a second-order polynomial in $x=T_{\rm N}-T$ to account for the cross-over from the asymptotic critical form to
almost temperature independent at lower $T$. We do not have enough data for such a fit to produce quantitatively conclusive
results, e.g., to rigorously test the exponent $\beta \approx 0.35$ corresponding to the expected 3D XY universality
class for the $T>0$ XY AFM transition. Nevertheless the behavior is consistent with a continuous or a very weakly first-order
transition, and the exponent $\beta \approx 0.30$ obtained from the fit is reasonably close to the expected value.

We do not  have enough $f_R - f_L$ data at other fields to systematically study the change from first-order to
continuous AFM transitions as the field strength is varied. Similar to the situation at 2.1~GPa, splitting of the
center NMR line is also observed at 1.95 and 2.4~GPa in the same range of temperature.

Technically, because the $^{11}$B sites are located slightly above the Cu
plane and have dipolar coupling to the Cu moment \cite{Waki_JPSJ_2007},
establishing in-plane AFM moments on the Cu sites ought to produce a dipolar
hyperfine field along the $c$ axis at the $^{11}$B sites, which due to the AFM
order is antiparallel on different sites. Our results are then consistent with
the expected planar rather than $c$-axis AFM order, because the latter would produce
a much larger uniform $c$-axis magnetization, which, as we discuss below in Sec.~\ref{uniform},
is not observed at such low fields.

\subsection{First-order field-driven transition}

The low-temperature $^{11}$B NMR spectra at 2.1~GPa and 2.4~GPa and at 0.07~K, presented
in Fig.~4a and 4b of the main text, show that the field-induced AFM
ordering is accompanied by a narrow regime of magnetic fields exhibiting phase coexistence
around the AFM transition detected at our lowest temperature of $0.07$ K.

As shown in Fig.~4a, at $2.1$ GPa, a significant line splitting is detected all the way
down to the field $H_{\rm c}$ determined from the fits illustrated in Fig.~5f, demonstrating
that the transition is rather strongly first-order in this case. In contrast, at $2.4$ GPa, the
line cannot be reliably separated into three different peaks below $5.8$ T, which is still some distance
away from the transition field $H_{\rm c} = 5.72$ T. Thus, while the transition is also first-order at
this higher pressure, the discontinuity has weakened significantly relative to that at $2.1$ GPa, indicating
that the system is approaching a continuous QCP (which we argue is a DQCP) located at only slightly higher
pressure. The transition field also clearly moves down with increasing pressure, and the trends observed
here suggest that $H_{\rm c}$ at the DQCP should be below $5.7$ T.

\subsection{Orientation of ordered moments}
\label{dipolar}

Because the hyperfine coupling between the $^{11}$B nuclear spins and the Cu
moments in SrCu$_2$(BO$_3$)$_2$ is dominated by dipolar interactions
for the off-diagonal tensor elements of the hyperfine
coupling \cite{Kodama_JPCM_2005}, the
hyperfine field at each $^{11}$B nucleus can be estimated in the AFM state
for different ordered configurations of the local moments. In general, the
hyperfine field, ${\bf H_{\rm in}}$ = ($H_{\rm in}^x$, $H_{\rm in}^y$, $H_{\rm in}^z$),
on a $^{11}$B nucleus due to one neighboring Cu spin can be written as
\begin{equation}
{\bf{H}}_{\rm in}= \widetilde{A} {\bf{m}},
\end{equation}
where ${\bf m} = (m^x, m^y, m^z)$ is the local moment of the Cu ion and the
hyperfine coupling tensor has the form
\begin{equation}
\widetilde{A} = \left[ \begin{matrix}
A^{aa} & A^{ab} & A^{ac} \\
A^{ba} & A^{bb} & A^{bc} \\
A^{ca} & A^{cb} & A^{cc} \end{matrix} \right].
\end{equation}
All of the tensor elements can be estimated by using the general form of the
dipolar field, with the relative positions of the $^{11}$B nucleus and the Cu
ion as input.

\begin{table}[t]
\centering
\begin{tabular}{c|ccc}
 (a) [100] \\
\hline
  site & ~~~ $H_{\rm in}^x$ & ~~~ $H_{\rm in}^y$ & ~~~ $H_{\rm in}^z$ ~~ \\
\hline
  B1 & ~~~ $-0.011$ & ~~~ $-0.608$  & ~~~ $-0.168$ ~~ \\
\hline
  B2 & ~~~ 0.011 & ~~~ $-0.608$ & ~~~ 0.168 ~~    \\
\hline
  B3 & ~~~ 0.011 & ~~~ $-0.608$ & ~~~ $-0.168$ ~~  \\
\hline
  B4 & ~~~ $-0.011$ & ~~~ $-0.608$ & ~~~ 0.168 ~~  \\
\botrule
\end{tabular}
\vskip2mm
\begin{tabular}{c|ccc}
   (b) [010] \\
\hline
  site & ~~~ $H_{\rm in}^x$ & ~~~ $H_{\rm in}^y$ & ~~~ $H_{\rm in}^z$ ~~ \\
\hline
  B1 & ~~~ $-0.608$ & ~~~ $-0.011$ & ~~~ $-0.168$ ~~  \\
\hline
  B2 & ~~~ $-0.608$ & ~~~ 0.011 & ~~~ $-0.168$ ~~    \\
\hline
  B3 & ~~~ $-0.608$ & ~~~ 0.011 & ~~~ 0.168 ~~  \\
\hline
  B4 & ~~~ $-0.608$ & ~~~ $-0.011$ & ~~~ 0.168 ~~    \\
\botrule
\end{tabular}
\vskip2mm
\begin{tabular}{c|ccc}
 (c) [110] \\
\hline
  site & ~~~ $H_{\rm in}^x$ & ~~~ $H_{\rm in}^y$ & ~~~ $H_{\rm in}^z$ ~~     \\
\hline
  B1 & ~~~ $-0.619$ & ~~~ $-0.619$ & ~~~ $-0.336$ ~~    \\
\hline
  B2 & ~~~ $-0.597$ & ~~~ $-0.597$ & ~~~ 0 ~~    \\
\hline
  B3 & ~~~ $-0.597$ & ~~~ $-0.597$ & ~~~ 0 ~~   \\
\hline
  B4 & ~~~ $-0.619$ & ~~~ $-0.619$ & ~~~ 0.336 ~~    \\
\botrule
\end{tabular}
\caption{Hyperfine fields at the $^{11}$B nucleus for a system in which the
AFM moments on the Cu$^{2+}$ ions are oriented along the crystalline [100] (a), [010] (b),
and [110] (c) directions. The field strengths are given in arbitrary units.}
\label{alltables}
\end{table}

In the AFM phase induced by a field applied along the $c$ axis, the Cu
moments, which have strong Heisenberg interactions, should orient in the
crystalline $ab$ plane. The AFM order is collinear type, with parallel
moments on each Cu dimer being antiparallel to those on all four
neighboring dimers (Fig.~1c of the main text). If one assumes that the Cu
moments are oriented along the $[100]$ direction, the hyperfine fields at
the four $^{11}$B nuclei shown in Fig.~\ref{fptwinss2} can be calculated by
summing the contributions from all of the neighboring Cu sites, giving the
result shown in Table~\ref{alltables}(a). We note that the real hyperfine
field in a material is usually enhanced by various factors beyond the simple
dipolar-field calculation, but this approximation is sufficient for the
qualitative conclusions we will draw. If one assumes that the Cu moments are
ordered along the $[010]$ or $[110]$ directions, the resulting hyperfine
fields at the four $^{11}$B sites are those shown respectively in Tables
\ref{alltables}(b) and \ref{alltables}(c).

These calculations show that $H_{\rm in}^z$ has one pair of negative and one pair
of positive values for the four B sites when the AFM moments are oriented along
the $[100]$ or $[010]$ directions. Such a pattern of hyperfine field
components create a double NMR line splitting, which is consistent with our
experimental observations (Fig.~\ref{afmspecs6}). By contrast, moments aligned
in the [110] direction will split the NMR line into three, which is not
consistent with our observations. Similarly, moments oriented in other
directions, which are necessarily of lower symmetry, will also create three
or more split NMR lines. Thus we conclude that the ordered moment in the
field-induced AFM phase is aligned in the $[100]$ or the $[010]$ direction.
Note that the small non-zero angle of the sample alignment in the field was not
taken account in the calculations.

Because the hyperfine field components $H_{\rm in}^{x,y}$ are not zero for planar
AFM moment orientations (Table \ref{alltables}), the spin-lattice relaxation
rate $1/^{11}T_1$, measured with $H \parallel c$, should pick up transverse
fluctuations from the AFM phase. This is demonstrated by the peaks appearing
in $1/^{11}T_1$ at the transition temperature, $T_{\rm N}$, in Figs.~5b and 5d
of the main text.

\begin{figure}[t]
\includegraphics[width=8.5cm]{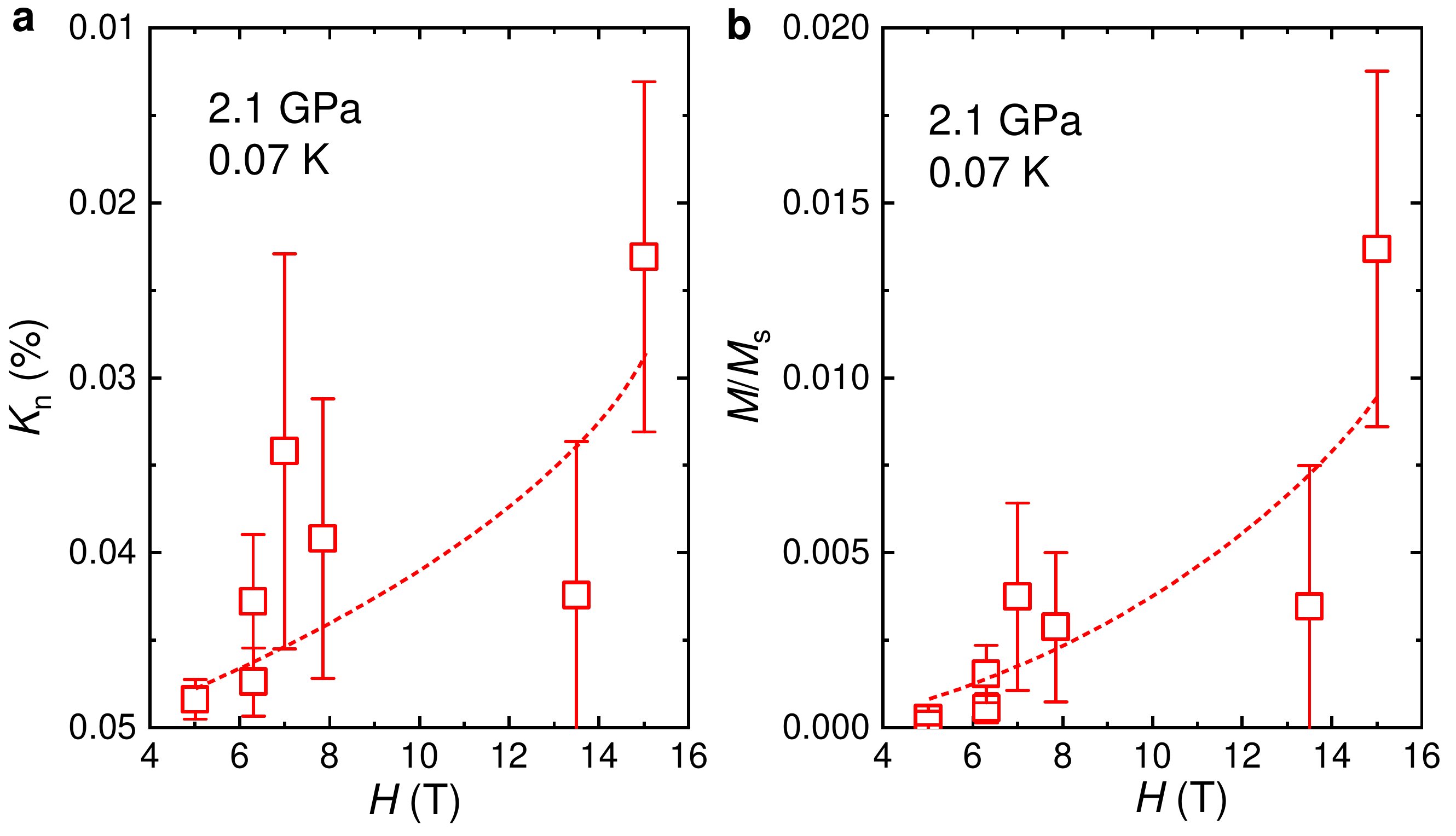}
\caption{\label{afmchis8} {\bf a} Knight shift, $^{11}K_n$, calculated from
the average frequency of the full NMR spectra presented in Fig.~4a of the main
text and shown as a function of field at 2.1~GPa. The dashed line is a guide
to the eye. {\bf b} Uniform magnetization of the system, obtained from the
Knight shift by averaging the frequency of the NMR lines.}
\end{figure}

\subsection{Knight shift and uniform magnetization}
\label{uniform}

To further explore the nature of the AFM phase, we calculated the uniform magnetization,
$M(H)$, at different applied fields from the NMR Knight shift. $^{11}K_n$ for
the AFM phase is calculated from the average frequency, $f$, of the center line
of each spectrum and the uniform magnetization is obtained from its magnetic
part, $K_s$, using the expression $M = {^{11}K}_s H /^{11}A_{\rm hf}$. For a
pressure of 2.1~GPa, $^{11}K_{n}$ at our base temperature of 0.07~K is shown
as a function of field in Fig.~\ref{afmchis8}a. The resulting $M(H)$ is shown
normalized to its saturation value, $M_s$, in Fig.~\ref{afmchis8}b, for which
we used a $g$-factor of $g_c = 2.28$~\cite{kageyama_JPSJ_1998}.

With ${\vec H}\parallel {\hat c}$, we find that the uniform magnetization never exceeds $2\%$ of its
saturation value at $15$ T. Such a high field is about $9$ T above $H_{\rm c}$, and therefore offers an
energy scale of $14$ K to partially polarize the system. This energy scale is not very small compared
to the microscopic AFM interactions  ($J' \approx 57$ K and $J \approx 39$ K at $2.1$ GPa
\cite{Guo_PRL_2020}), and because the couplings are frustrated the effective interaction scale may
be even smaller. Experimentally, however, the $M$ values are still very low as shown above, indicating
that the effective AFM interactions still lock the spins strongly to the XY plane with very little canting

An important aspect of the magnetization is the absence of significant (not clearly detectable)
discontinuity at the transition field; $H_{\rm c} \approx 6.2$ T at $2.1$ GPa in Fig.~\ref{afmchis8}b.
We do expect some discontinuity, similar to spin-flop transitions in uniaxially anisotropic quantum magnets,
but, as we discuss further in Sec.~\ref{scbjq}, an anomalously small discontinuity observed here at the
PS--AFM transition is likely a consequence of the emergent O($3$) symmetry associated with the nearby DQCP.

The small induced magnetization also makes it very unlikely that a magnetization plateau could exist up
to the highest field strength reached here. Thus, our data do not support a supersolid phase at these
fields and pressures. Different studies suggest that a much larger uniform magnetization is
required to stabilize such a phase~\cite{Haravifard_NC_2016,Shi_arxiv_2021}.

\section{Gap analysis in the PS phase}
\label{sgap}

\begin{figure}[t]
\includegraphics[width=8.5cm]{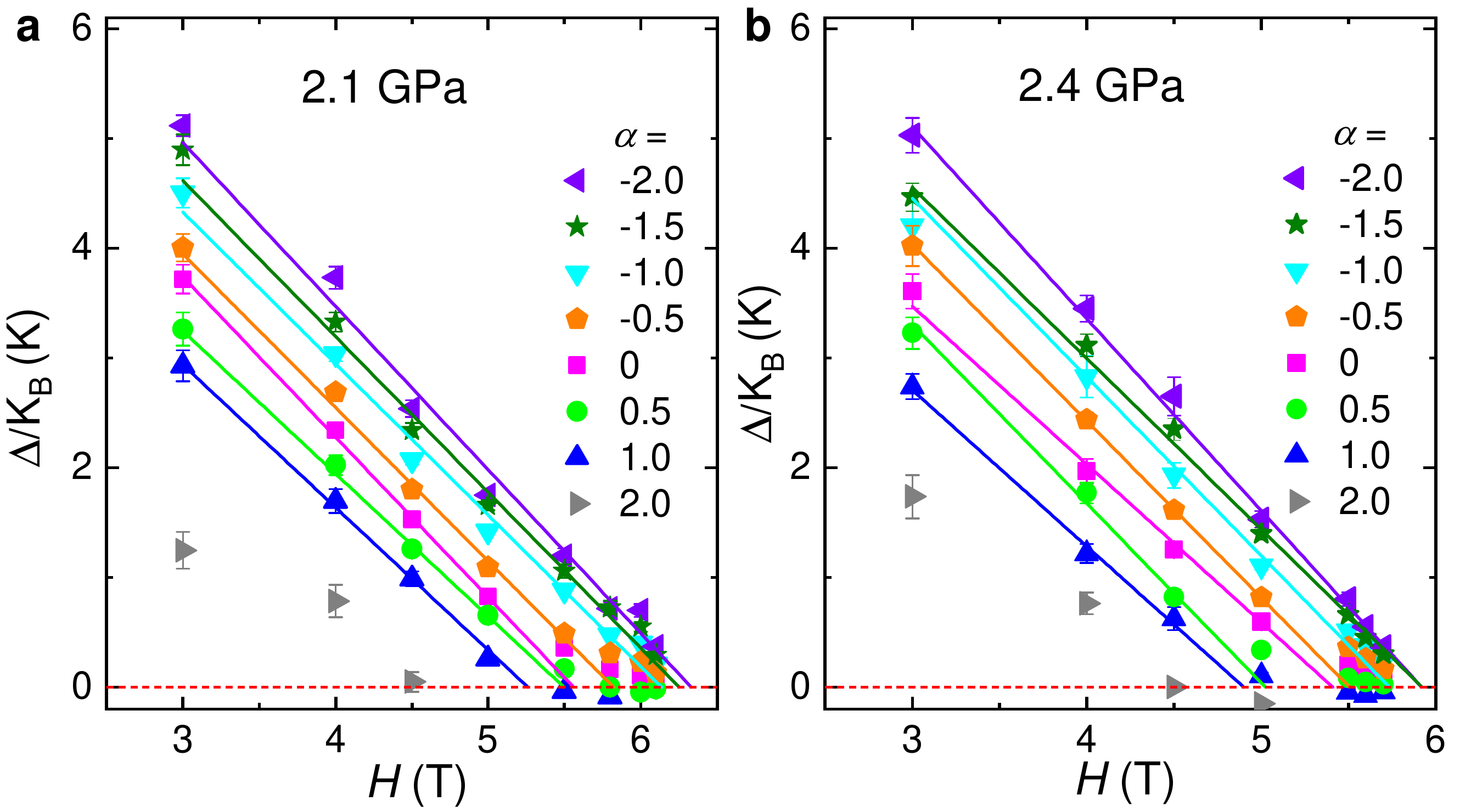}
\caption{\label{gapvsh}
    Spin excitation gaps obtained by fitting $1/T_1$ to the form Eq.~(\ref{t1fitform})
    with a series of values of $\alpha$, plotted as functions of the field at pressure $2.1$ GPa in panel {\bf a} and
    $2.4$ GPa in panel {\bf b}. The straight lines correspond to the fitted form $\Delta/{\rm K_B}=g(H_0-H)$ with
    fitting parameters $g$ and $H_0$.}
\end{figure}

\begin{figure}[t]
\includegraphics[width=6cm]{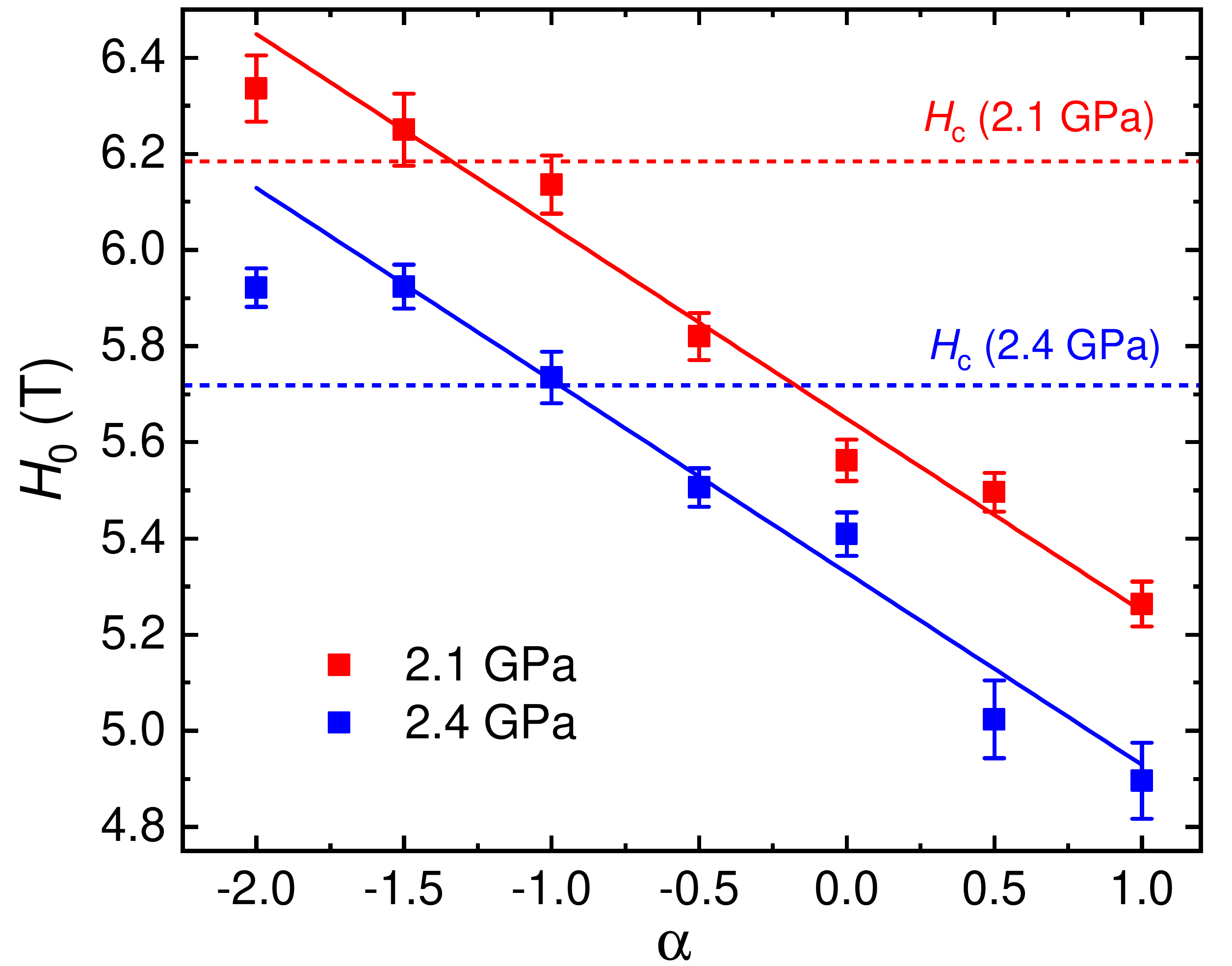}
\caption{\label{hcvsalpha}
    Gap closing fields $H_{\rm 0}$ from the fits in Fig.~\ref{gapvsh} graphed versus $\alpha$
    at both pressures, 2.1~GPa and 2.4~GPa. The red (blue) dashed horizontal line is
    the value of $H_c$ obtained at 2.1~GPa (2.4~GPa) from the analysis of the phase boundaries
    as discussed in Sec.~\ref{sstats} (results shown in Fig.~5f in the main text). The solid
    lines are linear fits, with the $\alpha=-2$ points left out because the value at 2.4 GPa deviates
    significantly from such a form.}
\end{figure}

The spin gap $\Delta$ of the lowest triplet excitations in the PS phase under applied field
can be obtained by fitting the low-temperature spin-lattice relaxation rate $1/^{11}T_1$
to the form
\begin{equation}
1/T_1 \propto T^{\alpha}e^{-{\Delta}/k_BT},
\label{t1fitform}
\end{equation}
where $\alpha$ is a parameter
to be determined. Theoretically, the $T^\alpha$ prefactor arises from the density of states
and matrix element effects. In two dimensions, for a quadratic dispersion of triplet excitations
above the gap, $\alpha=1$ if the matrix elements are constant. However, the matrix elements
typically exhibit divergent singularities at the gap edge, which can lead to negative values of
$\alpha$.

Here we regard $\alpha$ in Eq.~(\ref{t1fitform}) as an empirical parameter to be determined from the
experimental data. Given the rather small amount of data, optimizing $\alpha$ just based on the goodness
of the fit is not possible, as a wide range of values result in statistically acceptable fits. Figs.~\ref{gapvsh}a and
\ref{gapvsh}b show results for the field dependent gaps obtained from fits of $1/T_1$ at
$2.1$~GPa and $2.4$~GPa, respectively, to a series of values of $\alpha$.

To determine the best value of $\alpha$ in Eq.~(\ref{t1fitform}), we next fit the gap values by a linear function
$\Delta \propto(H_{\rm 0}-H)$, with the zero-gap field $H_{\rm 0}$ and the overall factor as fitting parameters.
We use the data points for which the gap is not very small (roughly for $\Delta/{\rm K_B} \agt 0.5$ K), and fitting
is then possible only for $\alpha \le 1$. Here we do not take a discontinuity at $H_c$ into account, but for
consistency with $H_c$ we know that $H_0$ resulting from our procedure must be above the critical field;
$H_0 > H_c$. Knowing $H_c$ from our other measurements, we can then also extract a bound for $\alpha$.

The $H_0$ values obtained by gap fitting for different $\alpha$ (using the range of points roughly consistent
with linearity in each case) are plotted against $\alpha$ in Fig.~\ref{hcvsalpha}. Here we can see that $H_0 > H_c$
only for $\alpha \le -1$, thus establishing the upper bound on $\alpha$ (further supported by the linearity being poor
in the small-gap regime for $\alpha$ larger than this value). Moreover, for $\alpha = -2$, the form of the gap versus
$H$ in Fig.~\ref{gapvsh} appears to deviate more from a straight line than it does for $\alpha=-1.5$. We also note
that $\alpha=-2$ in Eq.~(\ref{t1fitform}) would imply an unusually strong divergence of the spectral function at the
gap edge, thus making such a large negative $\alpha$ value unlikely also from a fundamental perspective.
Overall, this analysis suggests that $\alpha$ should be in the range $-1$ to $-1.5$.

\begin{figure}[t]
\includegraphics[width=8.5cm]{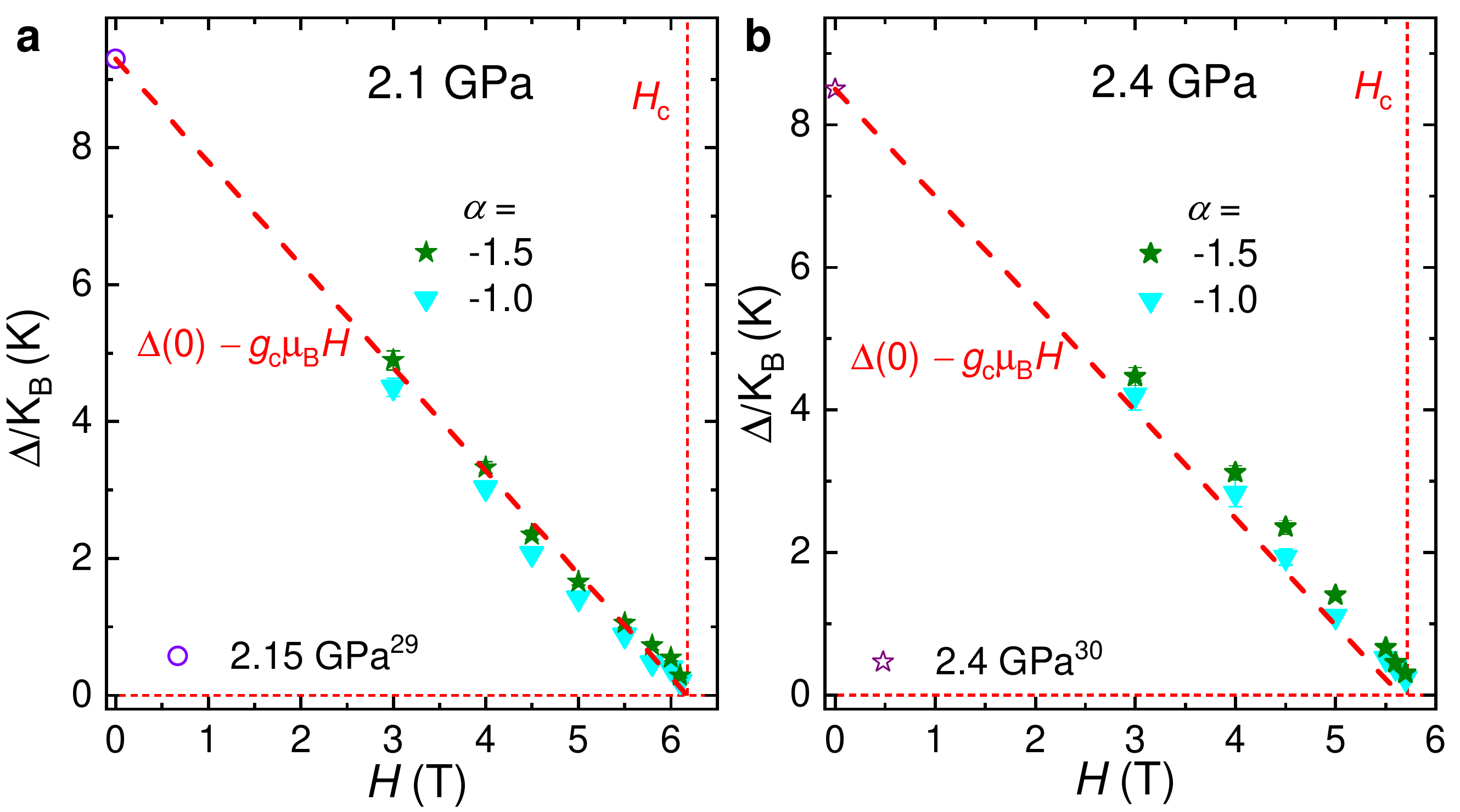}
\caption{\label{gapvsh2}
    Gaps obtained by fits of $1/T_1$ to the form Eq.~(\ref{t1fitform})
    with $\alpha=-1$ (triangles) and $\alpha=-1.5$ (stars) at $2.1$ GPa in {\bf a} and
    $2.4$ GPa in {\bf b}. The dashed lines are the predictions based on the previously extracted
    $H=0$ gaps from inelastic neutron scattering measurements at $2.15$~GPa~\cite{Zayed_NP_2017} (only
    slightly above $2.1$ GPa in our experiments) in {\bf a} and from the specific heat measurements at
    $2.4$~GPa~\cite{Guo_PRL_2020}, in {\bf b}. The vertical dashed lines mark the critical field
    at each pressure, determined by phase boundary analysis in Sec.~\ref{sstats}.}
\end{figure}

In practice, the gaps versus $H$ for $\alpha=-1$ and $-1.5$ look very similar, and we cannot determine which
of these values is better. In Fig.~\ref{gapvsh2} we show our results for the two $\alpha$ values along with the linear
$H>0$ gap predictions based solely on the previously measured $H=0$ gaps \cite{Zayed_NP_2017,Guo_PRL_2020},
$\Delta(H) = \Delta(0)-g_c\mu_BH$, applicable for an $S=1$ excitation, with the previously determined $g_c= 2.28$
\cite{kageyama_JPSJ_1998}. Though there is of course some scatter among our data points, the overall agreement
is remarkably good. We judge that overall $\alpha=-1$ is slightly better than $-1.5$ and show results
for the former in Fig.~6a in the main text. A likely very small discontinuous jump of the gap at $H_{\rm c}$ is
barely distinguishable from zero within the error bars both in Figs.~\ref{gapvsh2}a and \ref{gapvsh2}b.
In Sec.~\ref{cbjqmc} we discuss the small gap discontinuity further in the context of modeling the PS--AFM
transition with the CBJQM.

\begin{figure}[t]
\includegraphics[width=8.5cm]{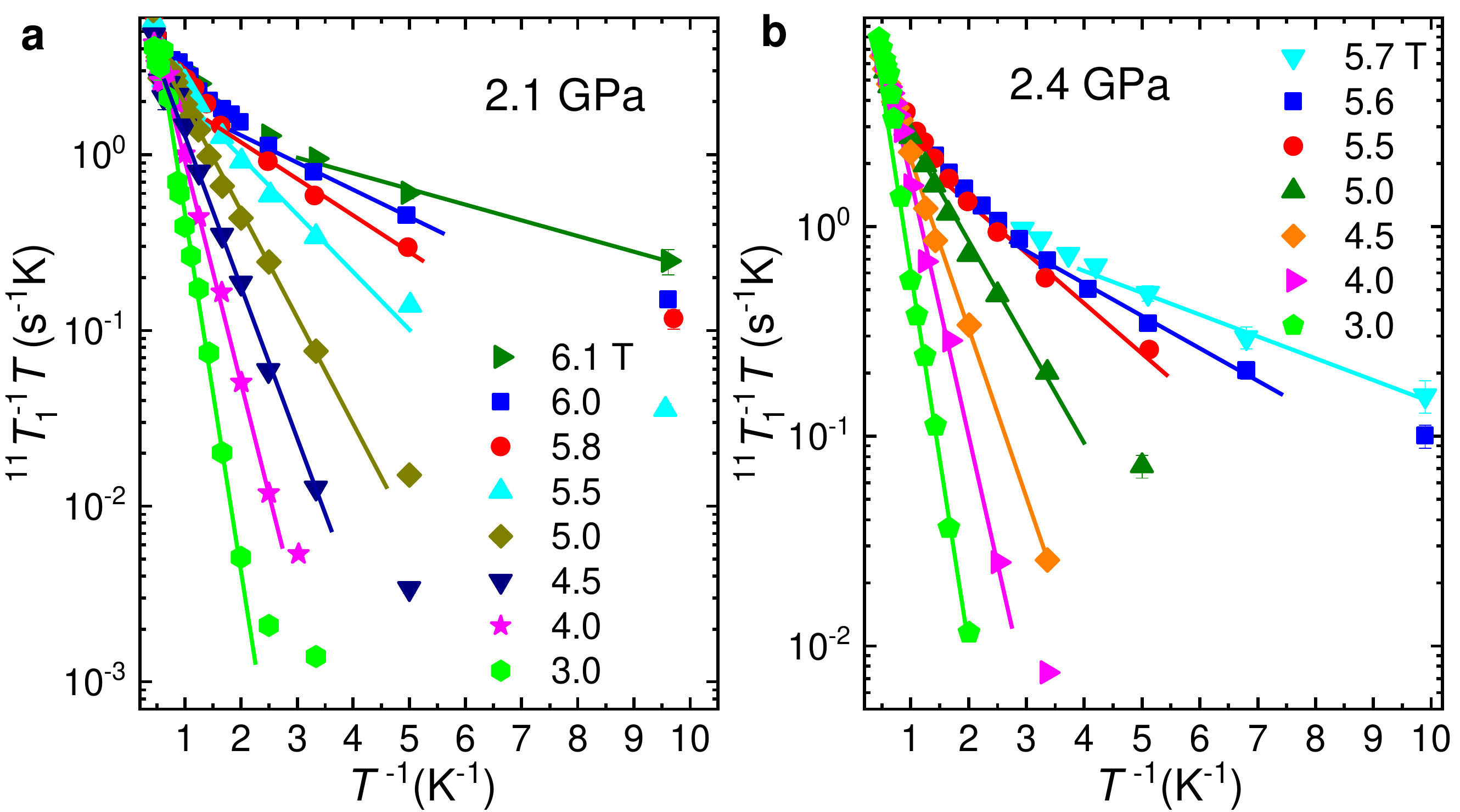}
\caption{\label{gapfit}
  $T/^{11}T_1$ data sets for different field strengths corresponding to the PS phase at low temperatures,
  shown on logarithmic axes as functions of $1/T$ at pressure $2.1$ GPa in {\bf a} and $2.4$ GPa in {\bf b}.
  The solid straight lines are fits to the gap function $T/T_1 \propto e^{-\Delta/k_BT}$ with spin gap $\Delta$.}
\end{figure}

We finally show the  $1/^{11}T_1$ data along with the $\alpha=-1$ fits
in Figs.~\ref{gapfit}a and \ref{gapfit}b, where $T/^{11}T_1$ at each field is plotted
as a function of $1/T$ on semi-log axes. At the low-temperature (high $1/T$) side, the
data fall on a straight line for each field, which indicates a gapped behavior with
the assumed power-law correction with $\alpha=-1$. At very low values of $1/T_1$ in
Fig.~\ref{gapfit} deviate significantly from the fitted lines, likely because of weak disorder effects.
We have not included those points in the fits.

\section{Analysis of experimental phase boundaries}
\label{sstats}

In the main text, we have fitted to two different functional forms of the PS and
AFM transition temperatures,  both of which are shown in Fig.~5f. We here further
motivate these forms and explain the details of our fitting
procedures; first for the modified critical forms in Sec.~\ref{sstats1} and
then in Sec.~\ref{sstats2} for the logarithmic form expected for the PS ordering
temperature at a first-order transition with emergent continuous order-parameter
symmetry.

We stress that both fitting forms should have their ranges of validity in
terms of the distance to a DQCP. Under the conditions of our experiments,
these ranges of validity may both be marginal and partially overlapping. Thus,
while we cannot determined which type of fit is the best, they both lend support
to a simultaneous transition of both order parameters at a point $(H_{\rm c},T_{\rm c})$
with very low $T_{\rm c}$ (in relation to the microscopic energy scales $J$ and
$J'$ of the interacting spins and also to the transition temperatures away
from $H_{\rm c}$) and the direct-transition field $H_{\rm c}$ close to $6$ T (weakly
decreasing with increasing pressure). Below $T_{\rm c}$, the direct field-driven  PS--AFM
transition is first-order for the range of pressures we have reached, less strongly
at $2.4$ GPa than at $2.1$ GPa.

\subsection{Critical point with emergent symmetry}
\label{sstats1}

It is tempting on the basis of Fig.~5f of the main text to fit
our data for $T_{\rm P,N}$ directly to the form expected at a QCP, i.e.,
$T_{\rm P,N} \propto |H - H_{\rm c}|^{\nu z}$. At a DQCP, the dynamical exponent
is $z = 1$ and, for the most well studied case of the transition from an O($3$)
AFM to a four-fold degenerate dimerized state, the emergent symmetry is SO($5$)
and the correlation-length exponent $\nu$ has been estimated by QMC simulations
to $\nu \simeq 0.46$ \cite{Zhao_PRL_2020,Nahum_PRX_2015,Sandvik_CPL_2020} (which
applies to both the PS and the AFM order parameter). At a PS--AFM DQCP in zero
field the emergent symmetry should instead be O($4$) on account of the PS order
parameter being a scalar instead of the two-component dimer order parameter.
In this case a similar value of $\nu$ as above was obtained \cite{Qin_PRX_2017}.

\begin{figure}[t]
\includegraphics[width=8.4cm]{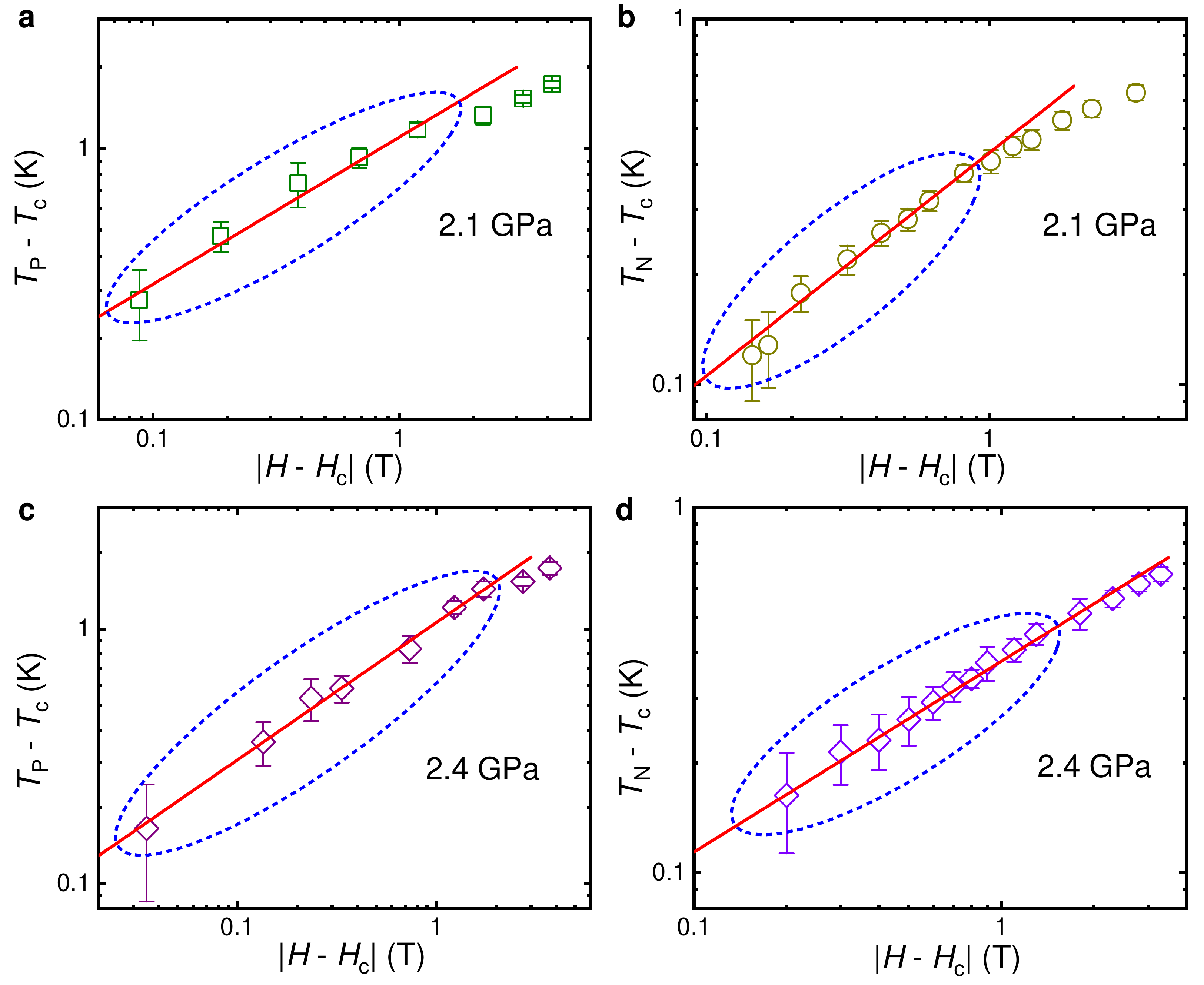}
\caption{\label{scaling0} {\bf Fits of the transition temperatures.}
Power-law scaling of $T_{\rm P,N} - T_{\rm c}$ with $|H - H_{\rm c}|$
obtained by four-parameter fits to each set of data using the near-critical form
of Eq.~(\ref{ebcp0}), shown for the experimental pressures $2.1$ GPa ({\bf a, b})
and $2.4$ GPa ({\bf c, d}). Each fit (straight line) applies to the data (open symbols)
enclosed within the dashed ellipses, in a range about 1~T away from the critical fields.}
\end{figure}

\begin{figure}[t]
\includegraphics[width=8.4cm]{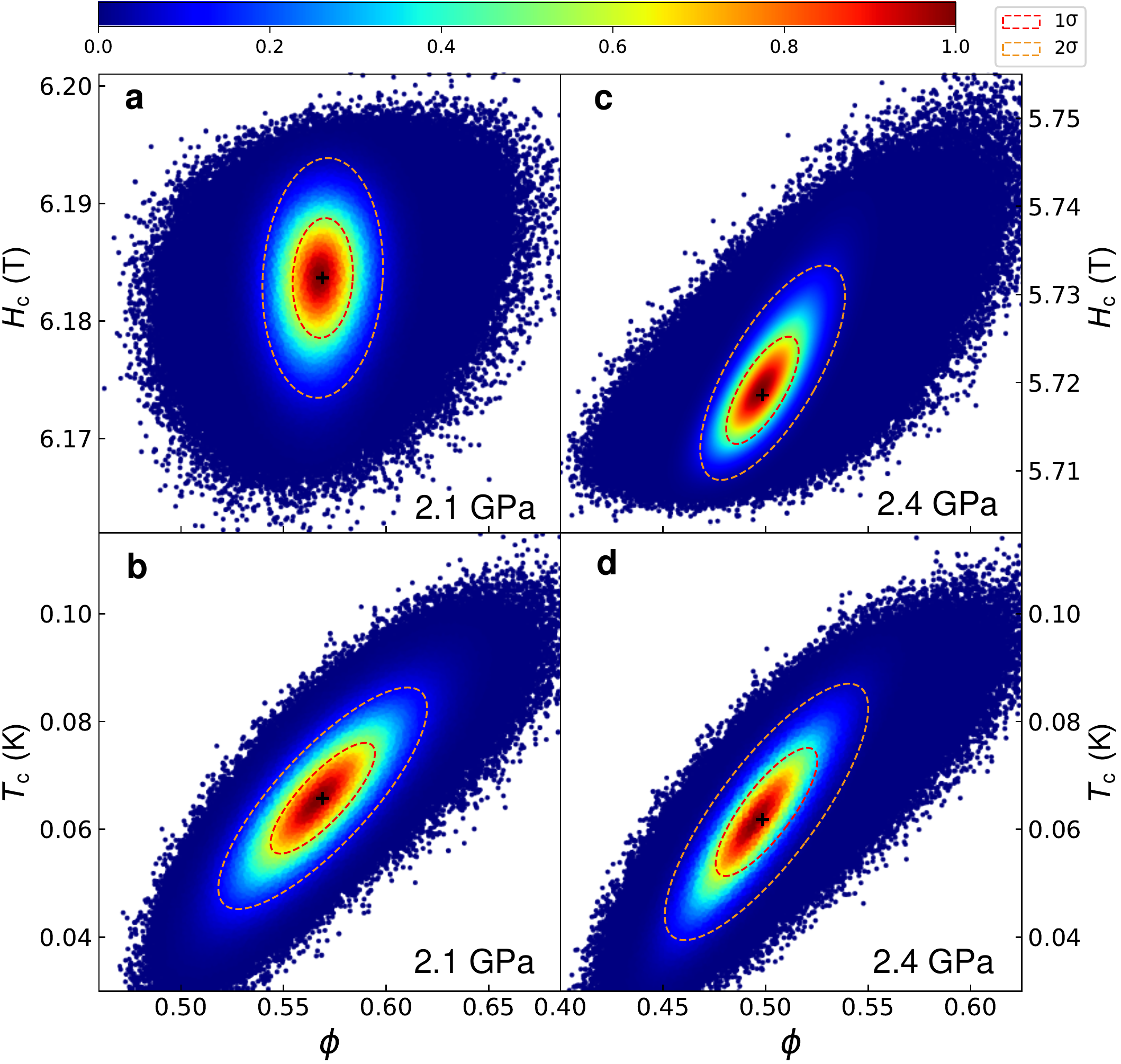}
\caption{\label{contour} Contour maps of the fitting probability function
projected on the space of the parameter pairs ($H_{\rm c},\phi$) ({\bf a,c}) and
($T_{\rm c},\phi$) ({\bf b,d}) at 2.1~GPa ({\bf a,b}) and 2.4 GPa ({\bf c,d}).
Crosses mark the best fits and dashed lines the 1$\sigma$ ($68\%$ CI) and
2$\sigma$ ($95\%$ CI) intervals. These probability functions are based on
the analysis of $N = 12$ data points at each pressure.}
\end{figure}

In the presence of an external magnetic field the symmetry is further reduced to O($3$)
at a putative DQCP separating phases with scalar PS and O($2$) AFM order parameters.
For this DQCP, $\nu$ has not been determined, but, in analogy with the slowly
evolving exponents of the conventional O($N$) transitions (where there is only
one order parameter), one can expect that $\nu$ would remain close to its SO($5$)
and O($4$) values. It is also very possible that the O($3$) DQCP does not strictly
exist but is a weakly first-order triple point, at which scaling properties may
still be governed by the O($4$) DQCP as long as the external field is not very
large, i.e., the zero-field system is close to the DQCP. This closeness to O($4$)
symmetry is supported by the very small field-induced magnetization at $H_c$
in Fig.~\ref{afmchis8}, i.e., the AFM order parameter is still O($3$) here for
all practical purposes and can combine with the scalar PS order to form an
effectively O($4$) emergent symmetry.

An argument against the above identical treatment of the PS and AFM transition
temperatures is that $T_{\rm N}=0$ in an ideal 2D system, i.e., the system orders
only exactly at $T=0$ on account of the continuous order parameter symmetry (which
precludes $T>0$ ordering by the Mermin-Wagner theorem). In our case of O($3$) AFM
symmetry reduced to O($2$) by a magnetic field there is a Kosterlitz-Thouless (KT)
transition into a critical phase existing down to $T=0^+$, which we do not consider
further.

In SrCu$_2$(BO$_3$)$_2$,
the weak 3D couplings allow for order also at $T>0$, as we have found experimentally
with $T_{\rm N}$ overall about half of $T_{\rm P}$ at comparable distances from $H_{\rm c}$
(Fig.~2d in the main paper). The 3D couplings also imply $T_{\rm c}>0$, and, therefore,
the standard critical forms of $T_{\rm P}$ and $T_{\rm N}$ have to be modified.
Phenomenologically, the form
\begin{equation}
T_{\rm P,N} = T_{\rm c} + a |H - H_{\rm c}|^{\phi},
\label{ebcp0}
\end{equation}
where we have defined $\phi=z\nu$, can be used.

We first fit $T_{\rm P}$ and $T_{\rm N}$ independently, each with four parameters $T_{\rm c}$, $H_{\rm c}$, $a$ and
$\phi$, following the Bayesian inference procedure \cite{Allenspach_2021}
briefly described in the Methods section, with data errors included.
Data in about 1~T range from the critical fields are
used to obtain reasonable power-law fitting.
The quality of the fits are depicted in Fig.~\ref{scaling0},
with $H_{\rm c}$ and $T_{\rm c}$ as below.
With statistics within the 95\% credible interval (equivalent to 2$\sigma$),
at 2.1 GPa,
the fit to $T_{\rm P}$ gives $H_{\rm c} = 6.189 \pm 0.017$~T, $T_{\rm c} = 0.074 \pm 0.026$~K, $\phi = 0.542 \pm 0.053$, and
the fit to $T_{\rm N}$ gives $H_{\rm c} = 6.185 \pm 0.017$~T, $T_{\rm c} = 0.070 \pm 0.025$~K, $\phi = 0.609 \pm 0.107$;
at 2.4 GPa,
the fit to $T_{\rm P}$ gives $H_{\rm c} = 5.731 \pm 0.016$~T, $T_{\rm c} = 0.065 \pm 0.025$~K, $\phi = 0.533 \pm 0.060$, and
the fit to $T_{\rm N}$ gives $H_{\rm c} = 5.700 \pm 0.017$~T, $T_{\rm c} = 0.067 \pm 0.024$~K, $\phi = 0.522 \pm 0.103$.

\begin{figure*}[t]
\includegraphics[width=14cm]{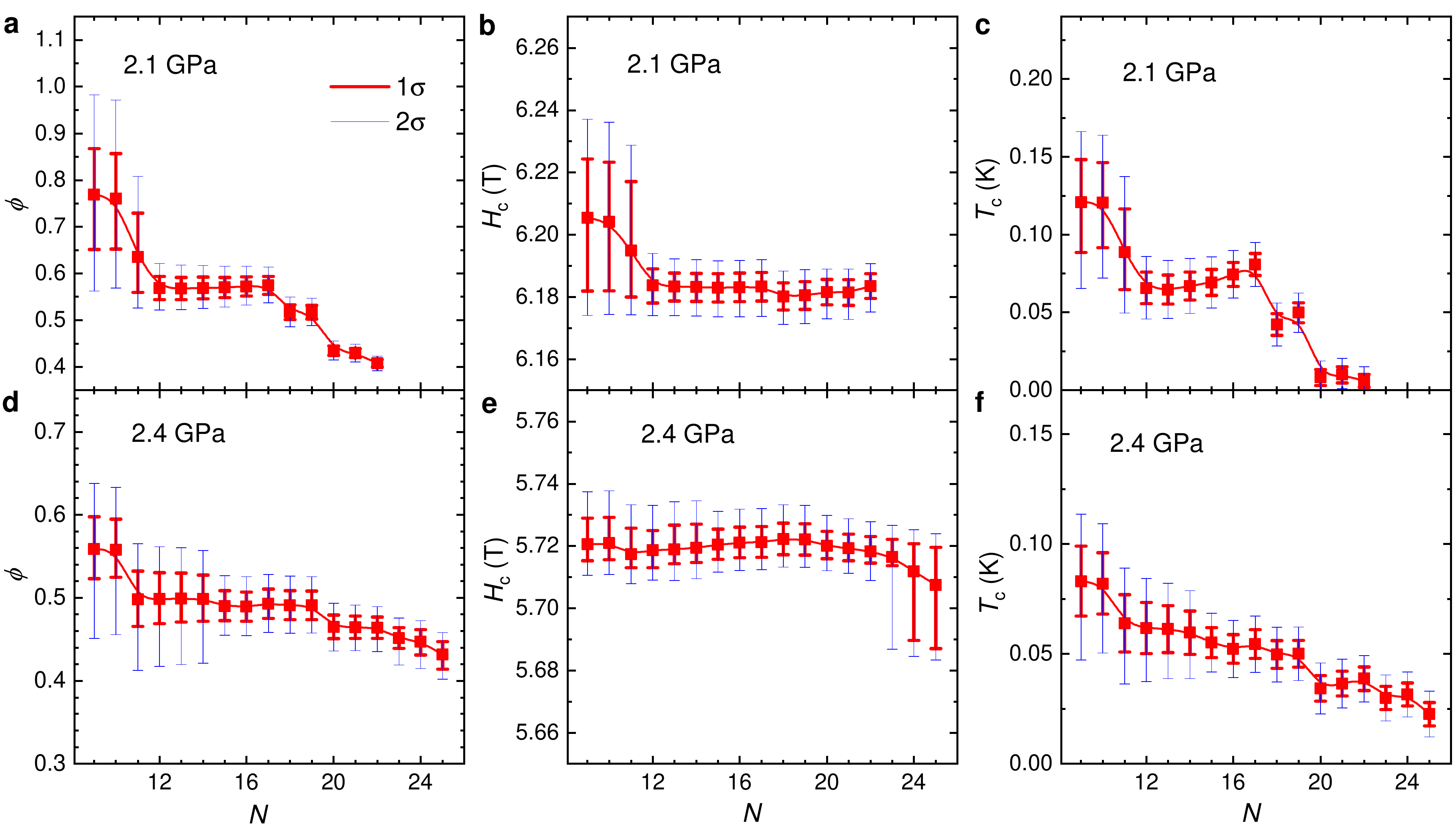}
\caption{\label{fitvsns9} Critical parameters $\phi$, $H_c$, and $T_c$
obtained by fitting the data at 2.1~GPa and 2.4~GPa to the form of
Eq.~(\ref{ebcp}), shown as functions of the total number, $N$, of data
points used in the fitting procedure. Red and blue error bars mark
respectively the $1\sigma$ and $2\sigma$ uncertainties of the fitting
parameters obtained with each $N$. The solid lines are guides to the eye.}
\end{figure*}

As shown above, the values of $H_{\rm c}$, $T_{\rm c}$, and $\phi$
obtained from $T_{\rm P}$ and $T_{\rm N}$ at each pressure are consistent within the errors.
Therefore, it is reasonable to assume the same value of $\nu$ for both order
parameters, which again is motivated in light of the expected duality of the PS
and AFM phases \cite{Wang_PRX_2017,Qin_PRX_2017}. Considering that
the two transitions meet at a bi-critical point $(H_{\rm c},T_{\rm c})$,
a general fitting form
\begin{equation}
T_{\rm P,N} = T_{\rm c} + a_{\rm P,N} |H - H_{\rm c}|^{\phi}
\label{ebcp}
\end{equation}
can be applied. A poor fit to the above form would
indicate a more generic first-order transition.

To use Eq.~(\ref{ebcp}), it is important to establish the width of the critical regime,
i.e., where $H$ is sufficiently close to $H_{\rm c}$ for the scaling form to apply,
while, in the present case where $T_{\rm c}>0$, the transition temperatures also
exceed $T_{\rm c}$. To obtain the five fitting parameters in Eq.~(\ref{ebcp}) describing
the field-dependent transition temperatures measured at each of the two pressure
values, the Bayesian inference procedure \cite{Allenspach_2021}
is again applied.

The output of the process of sampling the parameters is a multidimensional probability
distribution, and we first visualize this in Figs.~\ref{contour}a-\ref{contour}d by
showing projections of the distribution that illustrate its dependence on $\phi$,
$H_{\rm c}$, and $T_{\rm c}$, for the two measurement pressures. It is clear that the
distributions specify a single well defined maximum in the space of each parameter,
with narrow intervals of uncertainty in the parameters that reflect the error bars
inherent to the data.

Before discussing the optimal parameter values determined by the fitting
procedure, it is necessary to establish the unknown width of the critical
scaling regime. For this we performed the analysis using $9 \le N \le 25$
data points, ordered in magnetic field by their separation from $H_{\rm c}$,
and in Fig.~\ref{fitvsns9} we show the evolution of the optimal fitting
parameters $\phi$, $H_{\rm c}$ and $T_{\rm c}$ with $N$. One may anticipate
a deteriorating fit at large $N$, where the points no longer obey critical
scaling, and at small $N$, where there are simply too few points to extract a
reliable functional dependence. Indeed we observe that as $N$ is decreased
below 20, all the fitting parameters for both pressures converge towards
values that are nearly constant over the range $12 \le N \le 16$. However,
for $N < 11$ the fitting becomes less reliable and we neglect these estimates.
Thus we take the parameters to follow for $N = 12$, where the data at both
pressures show good convergence to a reliable fit.

\begin{figure*}[t]
\includegraphics[width=11cm]{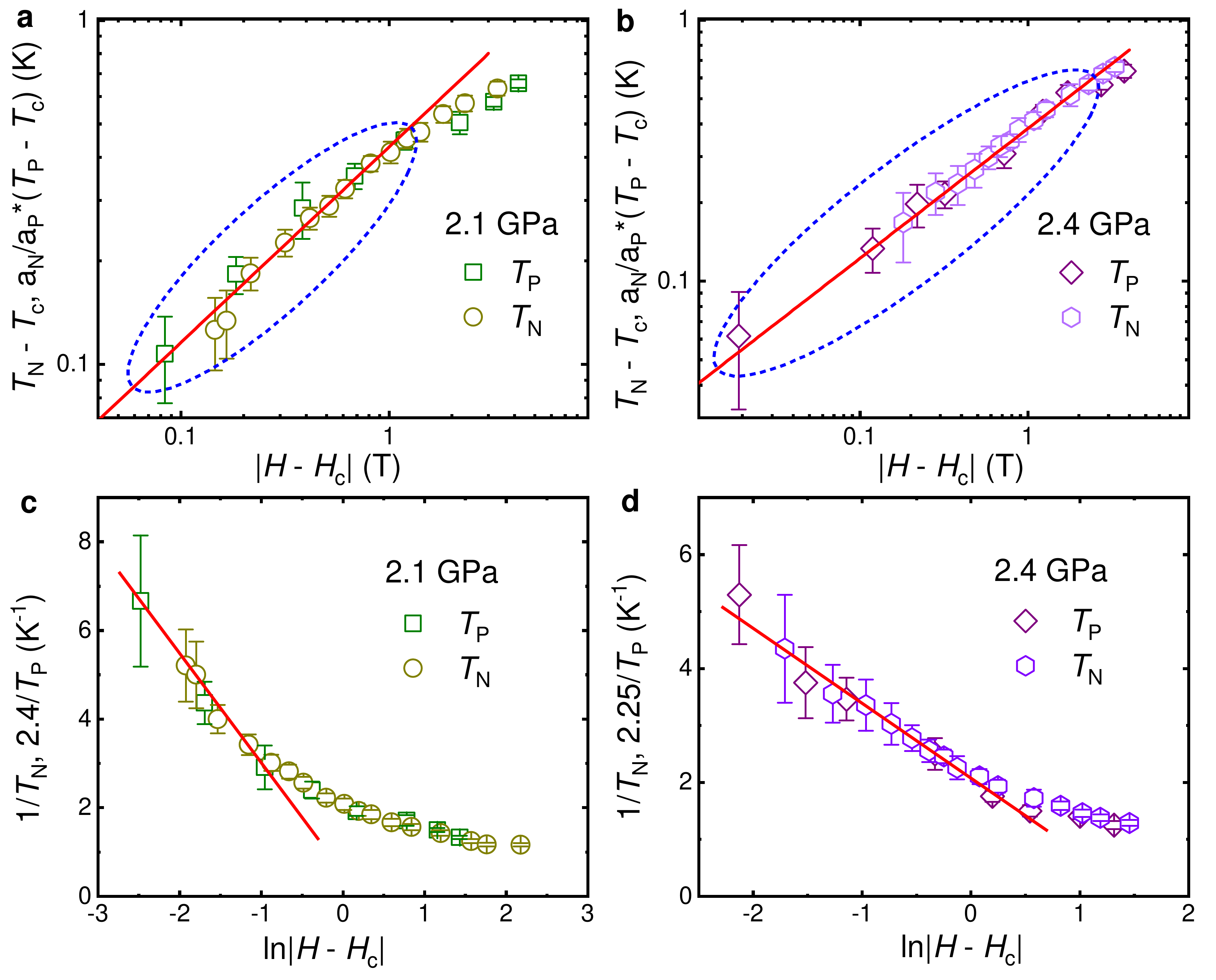}
\caption{\label{scaling} {\bf Fits of the transition temperatures.}
{\bf a,b} Power-law scaling of $T_{\rm P,N} - T_{\rm c}$ with $|H - H_{\rm c}|$
obtained by five-parameter fits to the data using the near-critical form
of Eq.~(\ref{ebcp}), shown for the experimental pressures $2.1$ GPa in {\bf a}
and $2.4$ GPa in {\bf b}. A constant factor $a_{\rm N}/a_{\rm P}$ is used to scale
$T_{\rm P}$ and $T_{\rm N}$ to the same $y$ axis. Although the fitting was performed by
monitoring convergence to the form given by the first 12 data points (straight
red lines), this form applies to all of the data enclosed within the dashed
ellipses, thus specifying the width of the critical scaling regime. {\bf c,d}
Inverse transition temperature shown as a function of $\ln|H - H_{\rm c}|$,
following the logarithmic form of Eq.~(\ref{tplogformtc}). The rescaling
factors were taken from $a_{\rm N}/a_{\rm P}$ in panels {\bf a} and {\bf b},
where the pressures are the same as in {\bf c} and {\bf d}, respectively.}
\end{figure*}

Another means of demonstrating the quality of the near-critical fit and
the width in field of the critical scaling regime is to show the dependence
of $|T_{\rm P,N} - T_{\rm c}|$ on $|H - H_{\rm c}|$ on logarithmic axis, as we do
in Figs.~\ref{scaling}a and \ref{scaling}b. In this form, data in the
critical regime fall on straight lines whose gradient is $\phi$, and scaling
of $T_{\rm P}$ to $T_{\rm N}$ is ensured by the ratio of the prefactors, $a_{\rm N}$
and $a_{\rm P}$, obtained from the Bayesian procedure. Only very close to
$H_{c}$ do some points deviate from the expected fit, which nevertheless lies
well within their error bars (expanded by the logarithmic axes), because of
the difficulty we have in identifying $T_{\rm P,N}$ in this regime. Far from
$H_c$, we find that many more than $N = 12$ data points are well described by
the critical scaling form with a single exponent $\phi$, and the fact that
$T_{\rm P}$ and the $T_{\rm N}$ continue to fall on the same crossover curve out
of the scaling regime underlines the duality between the PS and AFM ordered
phases. From Figs.~\ref{scaling}a and \ref{scaling}b we conclude not only
that the critical fit is fully consistent but also that the width of the
critical regime extends well beyond 1~T in both directions ($N = 17$ at
2.1~GPa and $N = 20$ at 2.4~GPa).

Returning now to the projected probability distributions as functions shown in
Figs.~\ref{contour}a-\ref{contour}d, the optimized fitting parameters, with
uncertainties determined from the the 2$\sigma$ level,
are $H_{\rm c} = 6.184 \pm 0.010$~T, $T_{\rm c} = 0.066 \pm 0.020$~K,
$\phi = 0.569 \pm 0.040$, and $a_{\rm N}/a_{\rm P} = 0.378 \pm 0.034$ for $P = 2.1$
GPa and $H_{\rm c} = 5.719 \pm 0.012$~T, $T_{\rm c} = 0.062 \pm 0.023$~K,
$\phi = 0.498 \pm 0.070$ and $a_{\rm N}/a_{\rm P} = 0.367 \pm 0.030$
for $P = 2.4$~GPa. The errors presented in the main text
are at the $1\sigma$ level ($68\%$ credible interval).
The most important single property of these fits is the
remarkably low value of $T_{\rm c}$ at both pressures. Critical values of order
0.06 K are more than one order of magnitude below the typical (non-critical)
values of both $T_{\rm P}$ and $T_{\rm N}$, which is not a feature of a generic
bicritical or triple point and strongly suggests that the finite $T_{\rm c}$ is a
weak, or residual, effect, for example one arising from 3D perturbations to a
system controlled by 2D physics.

Turning to the critical exponent, $\phi$, we observe at the $2\sigma$ level
that our $2.1$~GPa result, $\phi = 0.569 \pm 0.040$, and our $2.4$~GPa result
(which has a somewhat larger error bar), $\phi = 0.498 \pm 0.062$, are
mutually compatible. Certainly the exponent at 2.4 GPa, a pressure we have
argued appears close to the DQCP of SrCu$_2$(BO$_3$)$_2$, is compatible with
the estimates obtained for the SO($5$) DQCP, $\phi = z \nu$ with generically
$z=1$ and $\nu \approx 0.46$ \cite{Sandvik_CPL_2020,Nahum_PRX_2015}. A very similar
value has been obtained for the O(4) case \cite{Qin_PRX_2017}, which is the
more likely type of DQCP controlling the behavior here.

In the O(4) DQCP case, the model system studied in Ref.~\cite{Qin_PRX_2017}
has anisotropic, planar spin interactions, i.e., O($2$) AFM order parameter, and
four-fold degenerate dimerized state. However, it has been argued \cite{Lee_PRX_2019}
that the DQCP is the same as that obtaining when a scalar PS order parameter
combines with an O($3$) AFM order parameter. It is not clear whether an O($3$)
DQCP exists (to our knowledge this case has not yet been considered theoretically),
but given the extremely small uniform magnetization at the PS-AFM transition in
SrCu$_2$(BO$_3$)$_2$ (Sec.~\ref{safm}), it is possible that the relevant DQCP in this
case is still the O($4$) one. In other words, the uniaxial deformation of the O($4$)
symmetry is insignificant at the temperatures $T \gg T_{\rm c}$ for which we study
critical scaling.

\subsection{Logarithmic analysis}
\label{sstats2}

Given that at least the $2.1$ GPa system has a rather strongly discontinuous behavior of
the AFM order parameter (Fig.~4c of the main text), the applicability of the near-critical
form Eq.~(\ref{ebcp}) of the transition temperatures can be questioned. The very low $T_{\rm c}$
values compared to the transition temperatures when $H$ is not close to $H_{\rm c}$ is then
unusual and calls for a mechanism not requiring very close proximity to a QCP. As we have
discussed in the main text, emergent symmetry induced by a 2D DQCP far into a first-order
line is such a mechanism.

In an ideal 2D system with an O($N$) order parameter and $N > 2$, a perturbation
making one component of the interactions larger pushes the system to the ``Ising side,''
where it has an excitation gap and undergoes a finite-temperature phase transition
with a scalar order parameter. In contrast, a perturbation making one component smaller
still has a continuous order parameter symmetry, O$(N - 1)$, and there can be long-range
order only at $T = 0$ (with a quasi-ordered KT phase at $T>0$ in the special case $N=3$).

A renormalization-group analysis of the Heisenberg model ($N = 3$) with this type of
anisotropy \cite{Irkhin_PRB_1998} deduced a logarithmic form of the transition temperature
on the Ising side; the same analysis should also apply to $N > 3$. In the CBJQM at zero
external field, a first-order PS--AFM transition was found with a coexistence state whose
order-parameter distribution indicated the emergence of O($4$) symmetry up to the largest
system sizes studied, $L \simeq 100$ \cite{Zhao_NP_2019}. Then, moving into the PS side
corresponds to a uniaxial deformation of the order parameter and the logarithmic form is expected;
specifically \cite{Irkhin_PRB_1998}
\begin{equation}
T_{\rm P} (g<g_c) = A\ln^{-1}|C (g_c - g)|,
\label{tplogform1}
\end{equation}
where $g=J/Q$. This form was also confirmed in Ref.~\cite{Zhao_NP_2019} for $g$ close to $g_c$.
In the most likely scenario, the emergent symmetry is violated at some long length scale $\Lambda$,
corresponding to some small energy scale $\epsilon$ (which can be taken as the gap of the Goldstone
mode that corresponds to the Ising direction). The above form should then only be valid down to
$T_{\rm P}$ of order $\epsilon$, but this energy scale was not reached in Ref.~\cite{Zhao_NP_2019}.

In our experiments on SrCu$_2$(BO$_3$)$_2$, we study the phase transitions at fixed pressure versus the magnetic
field, and the emergent symmetry is reduced to O($3$). Also in this case must there be a smooth approach of
$T_{\rm P}$ to $0$ (assuming now momentarily that the symmetry is asymptotically exact) and Eq.~(\ref{tplogform1})
should apply with $g_c - g$ replaced by $(H_{\rm c}-H)^a$, where the exponent $a$ accounts for the different
forms of gap closing versus $g$ and $h$; $(g_c-g)^{1/2}$ \cite{Irkhin_PRB_1998} and $H_{\rm c}-H$, respectively.
However, the value of the exponent can be absorbed into the factor of proportionality and we do not need to
consider it further.

Experimentally, we expect violations of the O($3$) symmetry not only from the fundamental 2D effect related to
the distance to the DQCP, but also because of 3D effects (and potentially other sources, like impurities, that we have
not discussed). Thus, the logarithmic form should break down below some temperature. As in the case of the near-critical
form Eq.~(\ref{ebcp}), we may again add $T_{\rm c}$ as an offset optimized in the fitting procedure. However, our fits to
are actually optimal with $T_{\rm c}=0$, i.e., Eq.~(\ref{tplogform1}) without any added constant, though statistically
acceptable fits are also obtained with $T_{\rm c}$ values up to those obtained in the previous section. The fit shown
in Fig.~5f of the main paper are for zero offset, and the further analysis presented below also
was done with $T_{\rm c}=0$.

It is not immediately clear whether the AFM transition temperature can also be incorporated into a common fitting form in this case.
Interestingly, however, weak inter-layer (3D) couplings $J_\perp$ between 2D Heisenberg AFM layers also leads to a logarithmic form;
$T_{\rm N} (J_\perp) = \ln^{-1}(C/J_\perp)$ \cite{Irkhin_PRB_1998}. In the experiments the microscopic coupling $J_\perp$ is fixed, but it is
still possible that an effective, ``renormalized'' 3D coupling $J_\perp(H)$ can be defined, in light of the fact that 2D AFM order is present only
above $H_{\rm c}$ and increases with $H$. Thus, we assume here that both transition temperatures take the common form
\begin{equation}
T_{\rm P,N} = A_{\rm P,N}\ln^{-1}| C_{P,N} (H - H_{\rm c})|,
\label{tplogformtc}
\end{equation}
though we are less confident in the form of $T_{\rm N}$ than $T_{\rm P}$ and only show results for the latter in Fig.~5f of the main paper.
Here we will discuss both transitions.

At both pressures, $2.1$ and $2.4$~GPa, the optimal $H_{\rm c}$ value is statistically indistinguishable from the value obtained
using the  near-critical form of Eq.~(\ref{ebcp}). To gauge the quality of the logarithmic fit, in Figs.~\ref{scaling}c and \ref{scaling}d
we show the inverse transition temperatures on semi-log axes. The logarithmic fit remains valid over a significantly larger
range of fields at $2.4$ GPa (where it includes 15 data points) than at $2.1$ GPa, which would indeed be expected if $2.4$ GPa
lies closer to the DQCP, because the emergent symmetry would then be better established (i.e., manifested on longer length scales).

Although the near-critical forms in Figs.~\ref{scaling}a,b remain valid somewhat further away from $H_{\rm c}$, corrections are
expected here in both cases, and the non-universal width of the scaling regime is not a criterion for selecting the best critical
form. As mentioned in the main text, each of the fitted forms has its range of validity in principle, but the experiments may be
in a cross-over region where there are corrections to both forms but they still work reasonably well.

\section{Quantum-critical scaling of $1/T_1$}
\label{sscaling}

A quantum-critical point at $T_c=0$, reached at some tuning parameter $g=g_c$, should be associated
with a $T>0$ scaling regime \cite{Sachdev_book}, often referred to as the ``critical fan'' because of its shape extending out
from the $T=0$ point into a wide region in the $(g,T)$ plane. Scaling in the critical fan is one of the most distinctive and important
consequences of quantum criticality, reflecting detectable far-reaching influence of the $T=0$ critical point even when this point
itself cannot be reached.

The most extensively studied example of $T>0$ quantum criticality is in a class of 2D quantum antiferromagnets with
two exchange constants $J_1$ and $J_2$ (reviewed in Ref.~\cite{Sandvik_AIP_2010}), where the $J_2$-coupled spins form
dimers (and all spins belong to a dimer). As a function of the ratio $g=J_2/J_1$, the ground state is a N\'eel AFM
for $g < g_c$ and a unique dimer singlet state for $g>g_c$. Unlike the PS state of the SSM, there is no spontaneous
symmetry breaking in this gapped phase. There is no $T>0$ order in the system for any $g$, but different low-temperature
regimes with distinct physical properties can be defined by two macroscopic energy scales; the spin stiffness $\rho_s$
for $g < g_c$ and the gap $\Delta$ for $g > g_c$. Both these energy scales vanish continuously as $g \to g_c$, and the
critical fan is roughly located at temperatures above $\rho_s$ for $g < g_c$ and above $\Delta$ for $g > g_c$. In other
words, in the critical fan $T$ is the dominant energy scale.

At $g_c$, the quantum-critical scaling regime extends all the way down to $T=0$, while for $g \not = g_c$ the curves
$\rho_s(g)$ and $\Delta(g)$ define smooth cross-over boundaries to different low-temperature behaviors, called renormalized
classical and quantum disordered, respectively \cite{Chubukov_PRB_1994}. Upon increasing $T$, the critical behavior controlled by
the point $(g_c,T_c=0)$ must break down when (or before) the correlation length decreases to order one lattice spacing, where
the continuum description of the system (upon which critical behaviors rely) is no longer valid.

In systems with $T>0$ phase transitions, on one or both sides of $g_c$, the transition temperature(s) vanish as $g \to g_c$ and
a critical fan should still exist, either bordered by two ordered $T>0$ phases or by a single ordered phase and an energy scale
in a disordered phase as discussed above. In the case of SrCu$_2$(BO$_3$)$_2$, we expect a critical fan in the $(H,T)$ plane
(i.e., in the notation above, $g=H$) between the PS and AFM phases. The fan does not necessarily extend down all the way to the phase boundaries
if these boundaries are themselves thermal phase transitions, which are governed by their own critical behaviors. This is true for the AFM
phase of SrCu$_2$(BO$_3$)$_2$, where gapless classical spin fluctuations may mask the 2D quantum critical fluctuations at the same staggered
wave-vector. However, the PS ordering involves singlets, and its critical Ising-type plaquette fluctuations should be better decoupled
from the quantum-critical spin fluctuations residing in a different part of the spin and momentum space. Thus, we expect the quantum-critical
spin fluctuations to reach close to the PS phase boundary in phase diagrams such as Fig.~5f.

Scaling behaviors in the critical fan of the O($3$) transition in the aforementioned Heisenberg spin systems have been derived using field-theory
approaches for a number of physical observables \cite{Chubukov_PRB_1994} and detailed comparisons have been carried out with results of numerical simulation
studies of spin Hamiltonians (with Ref.~\cite{Sen_PRB_2015} being perhaps the most detailed study).

In the case of the spin-lattice relaxation rate, the
expected general behavior in the fan extending from any quantum-critical point of a 2D system should be $1/T_1 \propto T^{\eta/z}$, where $\eta$ is the
standard critical exponent governing the power-law decaying spin correlations and $z$ is the dynamic exponent (see the SI of Ref.~\cite{Hong_PRL_2021}
for an elementary derivation). Assuming $z=1$, which is the case for a DQCP, we have $1/T_1 \propto T^{\eta}$, ideally for arbitrarily low temperatures
exactly at $g=g_c$. Slightly away from $g_c$, the same power law applies, but with a constant correction $b(g)$ \cite{Chubukov_PRB_1994};
\begin{equation}
\frac{1}{T_1} = b(g) + aT^{\eta},
\label{t1gform}
\end{equation}
where $b(g_c)=0$ and the sign of $b(g)$ is negative or positive in the gapped and gapless phase, respectively. The factor $a$ and the functional form
of the  additive contribution $b(g)$ are known (to some approximation) from analytical calculations in some cases \cite{Chubukov_PRB_1994}, but we are not
aware of predictions for $1/T_1$ and a DQCP specifically.

In our experimental results for SrCu$_2$(BO$_3$)$_2$, we observe power-law scaling in what appears to be a quantum-critical region above
the PS phase at 2.4 GPa (Fig.~5c). The exponent $\eta \simeq 0.20$ is much larger than the well known values $\eta\approx 0.03$ for the O($2$) and O($3$)
critical points in 2+1 dimensions. The value is closer to results obtained with models realizing DQCP physics without magnetic fields
\cite{JQ_PRL_2007,Nahum_PRX_2015,Sandvik_CPL_2020}, though we are not aware of results in the presence of a magnetic field. It should be pointed out here
that the existence of DQCPs with various emergent symmetries is an ongoing area of research \cite{Wang_PRX_2017,Zhao_PRL_2020,Lu_PRB_2021}. In the
case at had here, an O($3$) symmetry would be naively expected but, as we discussed in, the very small observed magnetization at the AFM--PS
transition point (\ref{sstats1}) suggests that the DQCP controlling the critical behavior should be one with emergent O($4$) symmetry.
The cross-over from O($4$) to O($3$) symmetry upon moving away from the DQCP is further discussed in Sec.~\ref{cbjqmb}. We here discuss further
our evidence for near-criticality in SrCu$_2$(BO$_3$)$_2$ at the highest pressure studied, reflected in $T>0$ scaling behavior of the form Eq.~(\ref{t1gform})
and presented in the main paper as Fig.~5e.

At first sight, a complication of the critical-fan scenario
is that the common transition point $T_{\rm c}$ is not strictly zero. However, as indicated in Fig.~2b,
we still expect a quantum critical fan above $T_{\rm c}$ if this temperature is much lower than the relevant energy scales of the adjacent phases.
In the present case, those energy scales can be taken as the transition temperatures $T_{\rm P}$ and $T_{\rm N}$, which increase rapidly for $H$
away from $H_c$. In the case of the PS phase, above which our analysis here will be focused, $T_{\rm P}$ safely exceeds the extrapolated $T_{\rm c}$
value for $H$ at and below $5.7$ T at $P=2.4$ GPa.

On the AFM side of the transition, at $5.8$ T and higher field in Fig.~5d, we only have limited data above the peak that signifies the transition to
long-range N\'eel order. This transition would be absent in a single 2D layer, which can undergo O(3) symmetry breaking only at $T=0$. Already
very weak 3D couplings $J_\perp$ can push the transition temperature $T_{\rm N}$ to rather high values, since the dependence on $J_\perp$ is
logarithmic; $T_{\rm N} \propto J_{\rm 2D}/\ln(J_{\rm 2D}/J_{\perp})$ \cite{Irkhin_PRB_1998,Sengupta_PRB_2003}, where $J_{\rm 2D}$ is
the effective magnetic coupling within the layers \cite{J2Dnote}.

As seen in Fig.~5d, the ordering peak is quite broad, and none of the data above the
peak show a reduction in $1/T_1$ with decreasing $T$. Thus, we are not able to analyze any predominantly 2D quantum-critical fluctuations here
(and at higher temperatures, where we do not have data, the scaling behavior would be terminated when the classical paramagnetic regime
above the critical fan is entered, as seen on the PS side). On the AFM side at 2.1 GPa, in Fig.~5b, we do have data at higher temperatures
and the influence of the AFM ordering peak far above $T_{\rm N}$ is very clear. The quantum-critical spin fluctuations are visible in $1/T_1$
on the PS side, in Fig.~5c, because there is no spin ordering and the 3D magnetic couplings do not play a significant role there.

In Fig.~5e we demonstrate consistency with the expected critical scaling form (\ref{t1gform}) with a temperature independent shift. Since the field values
correspond to the gapped phase, we expect $b_H \equiv b(H)$ (where we now replace the generic parameter $g$ by the field strength) to be positive, approaching
zero as $H \to H_{\rm c}$. From general scaling arguments \cite{Chubukov_PRB_1994} we expect the form $b_H \propto (H_c-H)_0^d$ for $H$ close to $H_c$, but we
do not know the value of the exponent $d$. We therefore carry out a fitting procedure for the fields close to $H_{\rm c}$, between $5.4$ T and $5.7$ T,
where the exponent and the factor of proportionality are optimized for the best data collapse to the form Eq.~(\ref{t1gform}) of $1/T_1$, using the optimal
value of the exponent on $T$; $\eta=0.20$. For the smaller field strengths we optimize the values $b_H$ individually. The results for the so obtained
$1/T_1(H)+b_H$ are shown in Fig.~5e, where the optimized $b_H$ values along with the fitted power-law form are shown in the inset. The description of the
experimental data with common power-law over almost a decade of temperature is apparent, along with the faster decays and more rapid increase (in the
one case, $H=5.4$ T, where we have data) at the lower and upper bound, respectively, of the critical scaling regime for given $H$.

Including our estimated error bars from the fitting procedure, our results are $\eta=0.20 \pm 0.02$ and $d=0.77 \pm 0.04$. The exponent $\eta$ estimated
for an O($4$) DQCP from static correlation functions in a planar version of the $J$-$Q$ model is $\eta_{\rm JQ} = 0.13 \pm 0.03$ \cite{Qin_PRX_2017}, thus
marginally in agreement with our experiments.

An important aspect of the quantum critical scaling in SrCu$_2$(BO$_3$)$_2$ is its manifestation in our experiments only at 2.4 GPa, not at 2.1 GPa.
At the lower pressure a substantial peak is instead seen in Fig.~5a, and this peak was interpreted as the PS ordering temperature according to the NMR
line-width analysis in Fig.~3b and also considering the good agreement with the peak observed in the previous specific heat measurements
\cite{Guo_PRL_2020,Larrea_Nature_2021}. In our scenario outlined in Fig.~2, the system moves closer to the DQCP with increasing pressure, thus causing
an increase in the spin fluctuations in the $(T,H)$ regime between $T_{\rm P}(H)$ and $T_{\rm N}(H)$. It is at least plausible that these spin fluctuations
also affect the strength of the PS ordering, i.e., the overall factor in the singlet density modulations below $T_{\rm P}$ [i.e., the factor $a$ in the
critical PS order parameter $m_p = a(T_{\rm P}-T)^\beta$] is reduced when singlets ``break''. Thus, even if the PS transition may have a broad precursor
at higher $T$ also at $2.4$ Gpa, this may not significantly affect $1/T_1$ until the spin excitations become gapped by the long-range PS order. The
observed behavior without PS ordering peak in Fig.~5c supports this scenario. Note also that phase separation is only observed close to $H_{\rm c}$ on
the AFM side of the transition (Figs.~4a,b).

Indeed, at the point $(H_{\rm c},T_{\rm c})$ we know that the coexisting order parameters are weaker at 2.4 GPa---though only observed in the AFM order,
Fig.~4c, a similar weakening of PS order should also be expected---and that trend should persist also for the short-range PS order above $T_{\rm P}$. That
overall lowering of the plaquette strength should then also translate into less contributions to $1/T_1$ from the essentially classical plaquette fluctuations
above $T_{\rm P}$ (which also impact $1/T_1$ less directly than the spin fluctuations), thus suppressing the peak at $T_{\rm P}$. At 2.1 GPa, Fig. 5a, the
depletion of spin fluctuations in the PS liquid phase until the formation of the ordering peak is obvious. While $1/T_1$ also decreases sharply for $T$
above $2$ K at 2.4 GPa (Fig.~5c), the abrupt change of behavior to the common power law points to another dominant source of spin fluctuations, which,
as we have argued above, is the quantum criticality.

\section{Field-driven transitions in quantum spin models}
\label{scbjq}

To study the nature of the PS--AFM transition theoretically, we perform SSE QMC simulations of the CBJQM,
as detailed in the Methods section. This model is related to the SSM [and, we argue, to SrCu$_2$(BO$_3$)$_2$] at
least to the extent of hosting PS and AFM phases with the same order-parameter symmetries. While QMC simulations of
the SSM in the $J/J'$ regime of its QPTs are afflicted by a severe minus-sign problem \cite{Henelius00,Wessel18}, the CBJQM
is fully accessible to QMC simulations and, thus, offers unique opportunities for quantitative studies of the universal
aspects of the PS--AFM transition.

The PS--AFM transition in the CBJQM has been shown at zero field to be of first-order in the sense that the PS
and AFM order parameters coexist with finite values \cite{Zhao_NP_2019}. At a conventional first-order quantum
phase transition, the coexistence state can be understood as an analogy to uniform thermodynamic states separated
by free-energy barriers. The Hilbert space for a large system subdivides by energy barriers into separate parts for the
two different ground states. On a finite lattice at $T=0$, there is a characteristic time for tunneling between
these parts of the Hilbert space that diverges with the system size. In a realistic situation in a material, spatial
domains in different phases would form.

In the CBJQM \cite{Zhao_NP_2019} (and other related models \cite{Serna_PRB_2019,Takahashi_PRR_2020}) the coexistence state at
the PS--AFM transition hosts emergent O($4$) symmetry of the four-dimensional vector $(m_x,m_y,m_z,m_p)$, which combines the three
components of the staggered (AFM) magnetization and the scalar PS order parameter. Here it should be noted that the Hamiltonian does
not have any point in parameter space where such a symmetry is explicit, and the symmetry instead emerges on long
length scales. The emergent continuous symmetry fundamentally changes the coexistence state. In a finite system the
order can be continuously rotated between the AFM and PS states without passing through energy barriers. This type
of coexistence of two ordered phases can also be characterized as a supersolid order, where two non-zero order parameters
can coexist at the same spatial location. In the thermodynamic limit, the O($4$) symmetry is broken (the time scale of
rotations diverges), and any mix of AFM and PS order can be realized in principle.
In practice, domain with different symmetry-breaking will form.

The most likely scenario is that the emergent symmetry is violated above some length scale, thus being reduced to
O($3$)$\times Z_2$ of independently fluctuating AFM and PS order parameters. In cases studied so far, which includes not
only O($4$) symmetry \cite{Zhao_NP_2019,Serna_PRB_2019,Sun_CPB_2021} but also a model with four-fold degenerate PS state
and emergent SO($5$) coexisting order parameter \cite{Takahashi_PRR_2020},
this length scale can be at least hundreds of lattice spacings. The  reduction of the
O($4$) order parameter symmetry into O($3$)$\times Z_2$  has been studied in detail when inter-layer couplings are turned
on \cite{Sun_CPB_2021}. Here we characterize the putative O($3$) symmetry at the PS--AFM transition in the CBJQM when the O($3$)
symmetry of the AFM order parameter is reduced to O($2$) by the external magnetic field. We will demonstrate clear analogies
to behaviors observed experimentally in SrCu$_2$(BO$_3$)$_2$.

In addition to the CBJQM we also here study the anisotropic Heisenberg model, the $S=1/2$ XXZ model defined in Eq.~(\ref{hxxz}),
in which the symmetry of the order parameter changes versus $\lambda$. At zero magnetic field, for $\lambda<0$ the ground-state is
an XY AFM state with an O($2$) order parameter, while for $\lambda>0$ the ground-state changes to an Ising phase with $Z_2$
order parameter. In this case, the transition between XY and Ising orders is clearly not first-order in the conventional sense;
only the direction of the AFM order parameter in spin space flips discontinuously as the isotropic O($3$) point is crossed,
while the magnitude of the order parameter does not change. The PS--AFM transition in the CBJQM at $h=0$ is analogous to  this
``spin-flop'' transition, though in this case the symmetry is not explicit but emergent. The two order parameters of the CBJQM
would of course also not a priori be expected to form components of the same vector, and unless the transition is analyzed
in detail it would appear to be a clear-cut case of a discontinuous jump between two completely different ordered states with
unrelated symmetry breaking

When Ising-anisotropic, the XXZ model undergoes a spin-flop transition  versus an external magnetic field, in this case from
the Ising AFM phase to a  canted XY AFM phase \cite{Kosterlitz_PRB_1976}. This transition is in some respects similar to the
field-driven AFM state in the CBJQM, though there are important differences as we will elucidate below. One important aspect
of these differences is that the discontinuities at the first-order transition are much more pronounced in the XXZ model when
comparing systems with similar sized spin gaps at zero field. The small discontinuities that we find in the CBJQM lend support
to our argument that it is a suitable model for describing the universal aspects of the PS--AFM transition that we have
identified in SrCu$_2$(BO$_3$)$_2$.

We present results for the field-driven phase transitions of both models in Sec.~\ref{cbjqma}.
In Sec.~\ref{cbjqmb} we discuss the evidence for emergent symmetry in the CBJQM provided by the histograms
in Figs.~6d-f of the main text and further by
the dimensionless cross-correlation ratio $\langle m_{xy}^2 m_p^2 \rangle/\langle m_{xy}^2 \rangle \langle m_p^2 \rangle$. In
Sec.~\ref{cbjqmc} we discuss the spin gap, with examples of the field dependence for both the CBJQM and XXZ model.
In Sec.~\ref{scbjqd} we summarize our scenario for the PS--AFM transitions mechanism in the CBJQM.

\subsection{First-order quantum phase transitions}
\label{cbjqma}

\begin{figure}[t]
\includegraphics[width=7.5cm]{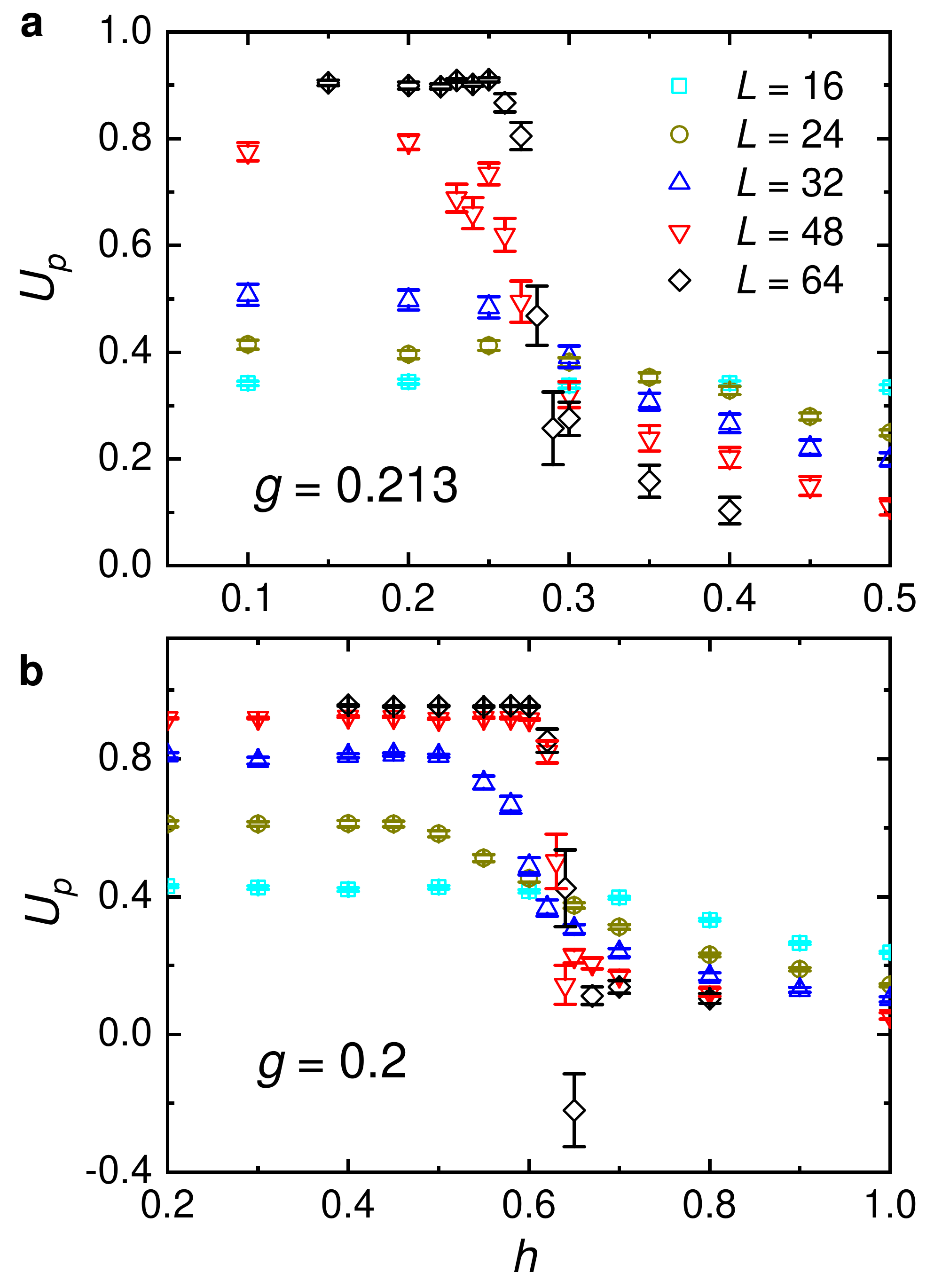}
\caption{\label{numericals1} Binder cumulant of the PS order parameter of
the CBJQM, shown as a function of the reduced field at $g = 1/4.7 \approx 0.213$
(in {\bf a}) and $g = 1/5=0.2$ (in {\bf b}). Simulations were performed for a range of system
sizes up to $L = 64$ and at temperature $T = J/L$. The transition field, $h_c$,
is determined from the crossing point of the $U_p(L)$ curves. The negative
value of $U_p$ close to $h_c$ for the largest system size in panel {\bf b}
indicates an asymptotically conventional first-order transition.}
\end{figure}

As the first characterization of the field-driven phase transition of the CBJQM,
we calculate the Binder cumulant of the PS order parameter,
\begin{equation}\label{Eq.Binder}
U_p = \frac{3}{2} \left( 1 - \frac{\langle m_p^4 \rangle}{3\langle m_p^2
\rangle^2} \right),
\end{equation}
where $m_p$ is the order parameter of the PS phase, which we have defined
in Eq.~(\ref{mpdef}). In the thermodynamic limit, the Binder cumulant obeys the
properties $U_p \rightarrow 1$ in the PS phase and $U_p \rightarrow 0$ in the AFM
phase. For a given ratio $g=J/Q$ the transition field $h_c(g)$ can be
extracted from the crossing point of the different curves, $U_p (L)$, obtained
from calculations on systems of finite size, $L$.

Figure \ref{numericals1} shows such finite-size analysis for two model parameters, $g \approx 0.213$ ($Q=4.7$, $J=1$)
in Fig.~\ref{numericals1}a and $g = 0.2$ ($Q=5$) in Fig.~\ref{numericals1}b, both obtained in simulations
at $T = J/L$ with maximum system size (length) $L = 64$. Here it should be noted that the PS
ground state is a singlet, and a calculation exactly at $T=0$ would deliver a constant value of
$U_p$ up to the point where a level crossing with the lowest magnetized state takes place. This $h$-independent constant value still
depends on the system size, reflecting finite-size fluctuations of the order parameter. Almost
constant values of $U_p$ are indeed observed Fig~\ref{numericals1}a at small values of $h$,
before a decrease that becomes sharper with increasing system size. The rounding before the
transition at $h=h_c$ reflects finite-temperature effects, while for $h > h_c$ there are $T>0$
effects as well as effects of the ground state magnetization evolving with $h$.

\begin{figure}[t]
\includegraphics[width=7.5cm]{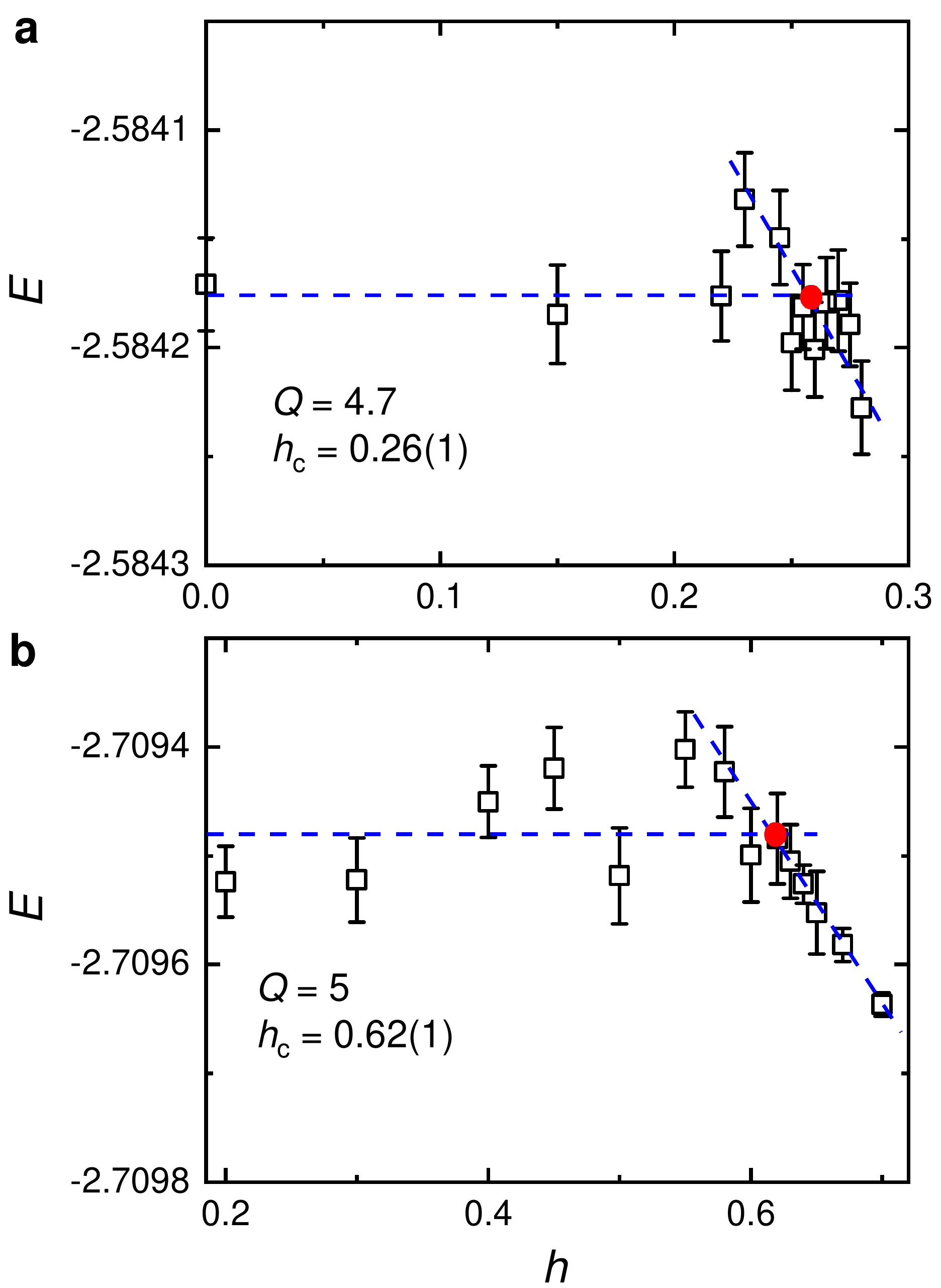}
\caption{\label{Evsh47} {\bf a} Internal energy of the CBJQM at $Q=4.7$ ($g \approx 0.213$) with system
size $L=48$ simulated at $T=1/L$. The crossing of the two dashed lines fitted to the two branches give
the transition field $h_c$ marked with a red circle. {\bf b} Same plot at $Q=5$ ($g=0.2$).}
\end{figure}

From the crossing point of the Binder cumulants for the two largest system sizes we obtain
$h_c = 0.27 \pm 0.01$ for $g = 0.213$ and $h_c = 0.62 \pm 0.01$ for $g = 0.2$. We note that,
at $g = 0.2$, $U_p$ of the $L = 64$ system is not a monotonic function of $h$ but drops to
negative values near the transition in Fig.~\ref{numericals1}b. This behavior is characteristic of
a conventional first-order transition \cite{Sandvik_AIP_2010}. At $g = 0.213$, the Binder cumulant remains
positive up to $L = 64$ in Fig.~\ref{numericals1}a, though eventually, for larger system sizes,
we also expect a negative peak to emerge at this coupling.

\begin{figure}[t]
\includegraphics[width=7cm]{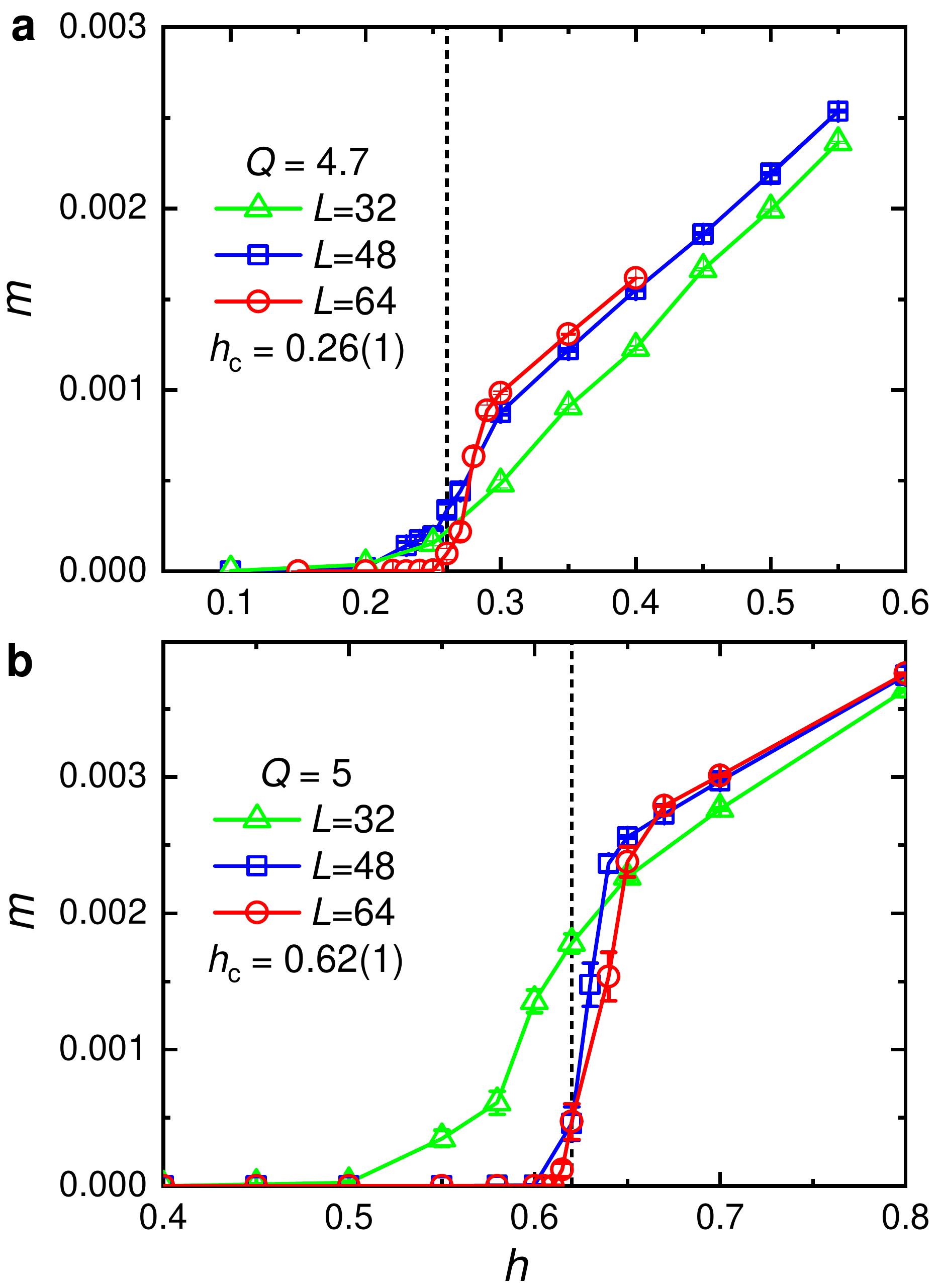}
\caption{\label{MzHQ5} {\bf a}: Magnetization per spin vs field strength in the CBJQM at $Q=4.7$, simulated
  on lattices of size $L=32,48$, and $64$ at temperature $T=1/L$. {\bf b}: The same at $Q=5$.}
\end{figure}

For the CBJQM at $h = 0$, no negative values of the Binder cumulants were observed up to system
size $L = 96$ \cite{Zhao_NP_2019}, which was taken as evidence (along with other signals) of
an emergent continuous symmetry. If the higher symmetry is ultimately violated, we
expect a conventional coexistence state and negative cumulant, as also seen in the CBJQM
with weak inter-layer couplings \cite{Sun_CPB_2021}. Given that we observe a negative peak for
the largest system in Fig.~\ref{numericals1}b, and also the overall sharp drop of the cumulant
at $h_c$, we expect the putative O$(3)$ symmetry to also be violated on a scale of tens
of lattice spacings, and this length scale should grow as $g_c(h)$ approaches its $h=0$ value.
We point out again that $g_c(h=0)$ is not a critical point but a first-order transition
with a very long length scale of emergent O(4) symmetry, presumably induced by a DQCP slightly
outside the parameter space of the CBJQM

\begin{figure}[t]
  \includegraphics[width=7.5cm]{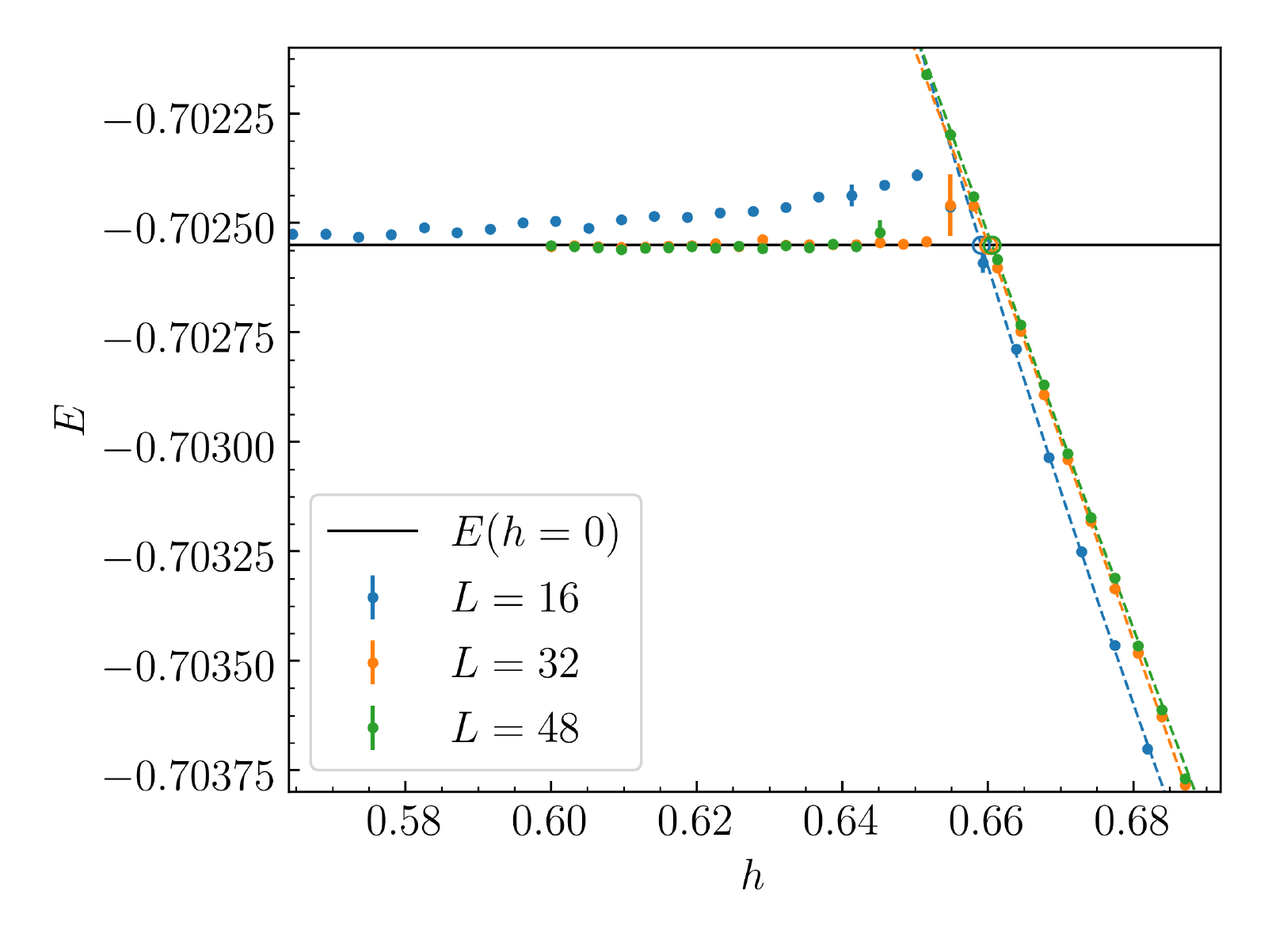}
\caption{\label{xxzh1}
Energy per spin vs the field for the XXZ model at $\lambda =0.1$ for systems of size $L=16$, $32$, and $48$.
The flat portion corresponds to the $h$-independent singlet ground state, which crosses a state with
$m>0$ at $h_c$. We extracted $h_c$ from the crossing points of the fits shown. The horizontal lines
show the $h=0$ energy, which for $L=16$ is slightly below the QMC values because of temperature effects
at $\beta=L$ when the spin gap is small. No such effect is seen for $L=48$. The higher branch for $h < h_c$
reflects metastability in the simulations close to the first-order transition.}
\end{figure}

The level crossing mechanism of the transition can be directly confirmed by studying the low-temperature internal energy versus
the field. In Fig.~\ref{Evsh47} we show results for a system with $L=48$ at two model parameters, $Q=4.7$ ($g=0.213$) in panel a
and $Q=5$ ($g=0.2$) in panel b. In each case, two branches of the energy can be observed (with some effects of meta-stability).
On the $h \le h_c$ branch, the energy should be completely flat, while an almost linear behavior should be expected for $h > h_c$, reflecting the
linear decrease of the energy of a level with non-zero magnetization and the increase of the magnetization with $h$. In Fig.~\ref{Evsh47},
the error bars are large on the relevant small scale of the energy changes and we just fit a line also on this branch.
The crossing point is fully compatible with $h_c$ extracted above from the Binder cumulant. The small slope of the energy
versus $h$ for $h \agt h_c$ reflects the weak nature of the first-order transition in this case.

Fig.~\ref{MzHQ5} shows the field induced magnetization at both $Q=4.7$ and $5$. A temperature rounded small jump (of about $0.25\%$ of the saturation value)
followed by a near-linear increase can be observed at $Q=5$. At $Q=4.7$, the jump is less than half as large and there are relatively larger effects
of finite size and finite temperature. A priori, such small magnetization discontinuity is not expected for a system going through a first-order spin-flop
transition. It should be stressed again that even at the $h=0$ transition point at $Q \approx 4.6$ (Ref.~\cite{Zhao_NP_2019} and Fig.~6b in the main text)
the transition is distinctively first-order, in the sense of hosting substantial coexisting long-range PS and AFM order parameters.

As a bench-mark case further illustrating that the discontinuities observed here in the CBJQM versus the field are indeed small, we next present results
for the field-driven transition out of the Ising phase of the XXZ model at $\lambda=0.1$, where the transition field $h_c$ is almost the same as that of
the CBJQM at $Q=5$. We again set the temperature $T=1/L$ and have studied systems of size up to $L=48$. Figure \ref{xxzh1} shows the internal energy per spin
versus $h$ for three different system sizes. Here we clearly observe meta-stability for the largest system, while the smallest system can still fluctuate enough
to produce an energy slightly above the true ground state energy (which of course is $h$ independent for $h < h_c$). Fitting lines to the two energy branches,
we see that the transition point is well converged at $h_c=0.660$. Though the transition field is comparable to those of the CBJQM at $Q=5$
(which also implies comparable $h=0$ gaps in these two systems), the meta-stability in the XXZ model is much more prominent than in Fig.~\ref{Evsh47}.

\begin{figure}[t]
\includegraphics[width=7.5cm]{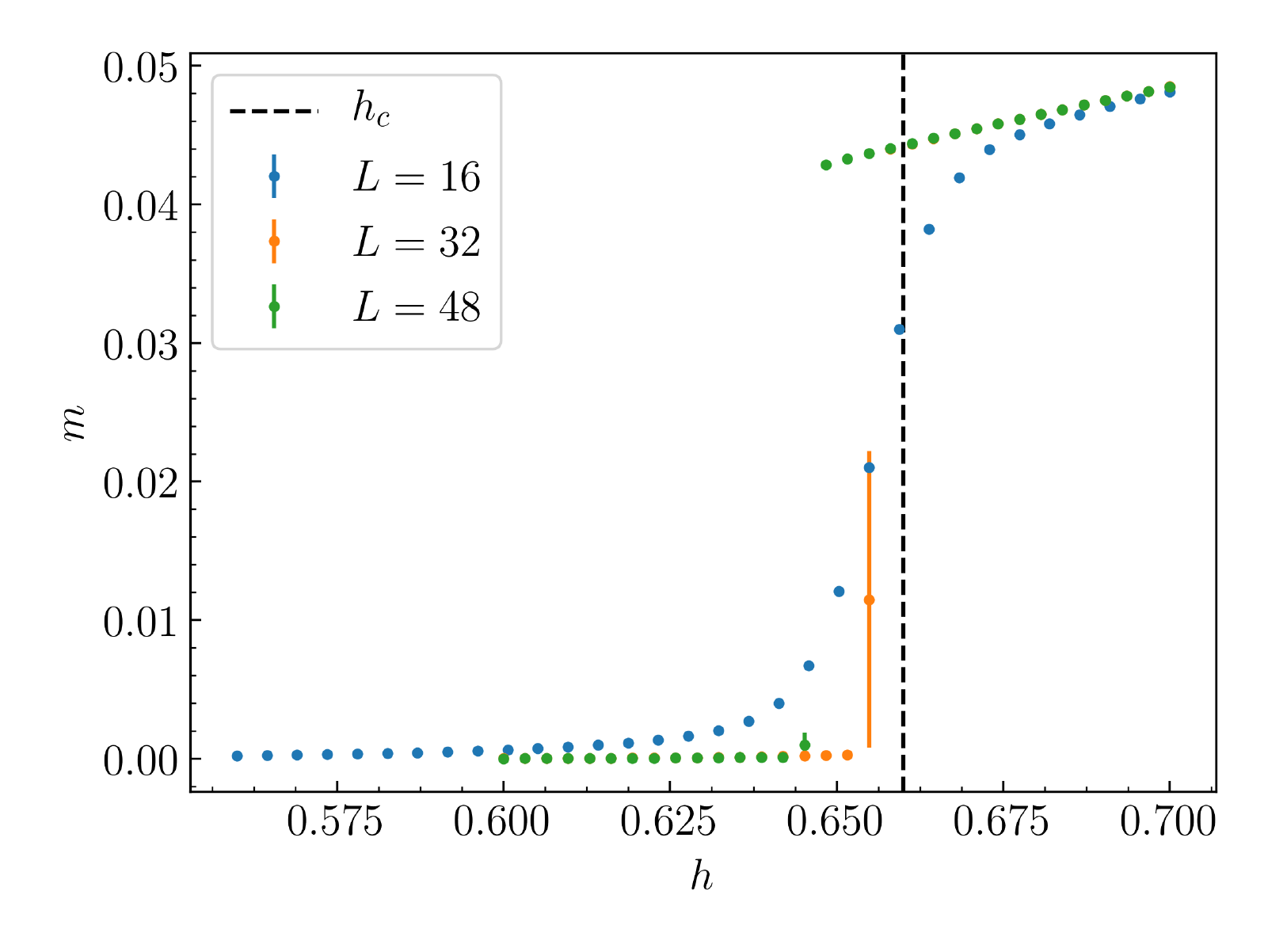}
\caption{\label{xxzh2}
Magnetization per spin in the same XXZ simulations as in Fig.~\ref{xxzh1}. For $L=32$, one very large error bar reflects a
simulation that escaped from the meta-stable state during the simulation. The transition point determined from the results
in Fig.~\ref{xxzh1} is shown with the vertical dashed line.}
\end{figure}

The magnetization versus the field is shown for the same system sizes in Fig.~\ref{xxzh2}. Here the jump in $m$ is more than 20 times larger than that of
the CBJQM at $Q=5$ (Fig.~\ref{MzHQ5}a) even though the values of $h_c$ and $\Delta(h=0)$ are almost the same.
Thus, judging from both the energy and the magnetization, the transition in the XXZ model is much more strongly first-order, even though the relevant energy
scales ($h_c$ or the $h=0$ gap) of the two models are comparable (within $10\%$ of each other), and at first sight the transition mechanism is the same.

An important difference between the conventional spin-flop transition in the XXZ model and the PS--AFM transition is that the magnetic field
does not explicitly couple to the objects (plaquette singlets) forming the PS order parameter. In contrast, in the XXZ model the magnetic field couples
to the uniform Fourier component of the spins, and the Ising order parameter is the staggered component of the same spins. The magnetic field in the
CBJQM explicitly deforms only the AFM order, taking it from O($3$) to O($2$) symmetric while leaving the PS state essentially intact. Thus, with emergent
O($4$) symmetry at the PS--AFM transition at $h=0$, it is not surprising if an O($3$) symmetry survives after one of the components of the AFM order has been
suppressed by $h>0$. Because of the level crossing, the transition is still eventually first-order with no exact O($3$) symmetry, but the influence of the
O($4$) DQCP (existing in an extended parameter space) is much more robust than the influence of the exact O($3$) point on the order parameters at the
spin-flop transition in the XXZ model.

\subsection{Emergent O($3$) symmetry}
\label{cbjqmb}

Our numerical test of an emergent O(3) symmetry in the CBJQM, shown in Figs.~6d-6f of the main text,
is based on the probability distribution function of the PS order parameter, $m_p$, across the QPT. An O(3) symmetry would
combine $m_p$ with the two XY-AFM order parameters to form a three-component supervector,
${\bf n} = (m_x,m_y,m_p)$, which has a uniform probability distribution on
the sphere of the order-parameter space
\begin{equation}
{\bf n}^2 = R^2,
\end{equation}
where $R$ is the radius of the sphere. Specifically, integrating out the
$m_x$ and $m_y$ components gives the distribution of $m_p$ in the presence
of an O(3) symmetry as
\begin{equation}
P(m_p) = \left \{ \begin{array}{ll} 1/(2R) & {\rm for} ~ |m_p| \leq R,
\\ 0 & {\rm for} ~ |m_p| > R, \end{array} \right.
\label{pmpo3sym}
\end{equation}
whence the expectation of a uniform distribution over a region of finite
width centered at $0$ when the system is very close to the transition point,
$h_c$.

\begin{figure*}[t]
  \includegraphics[width=14cm]{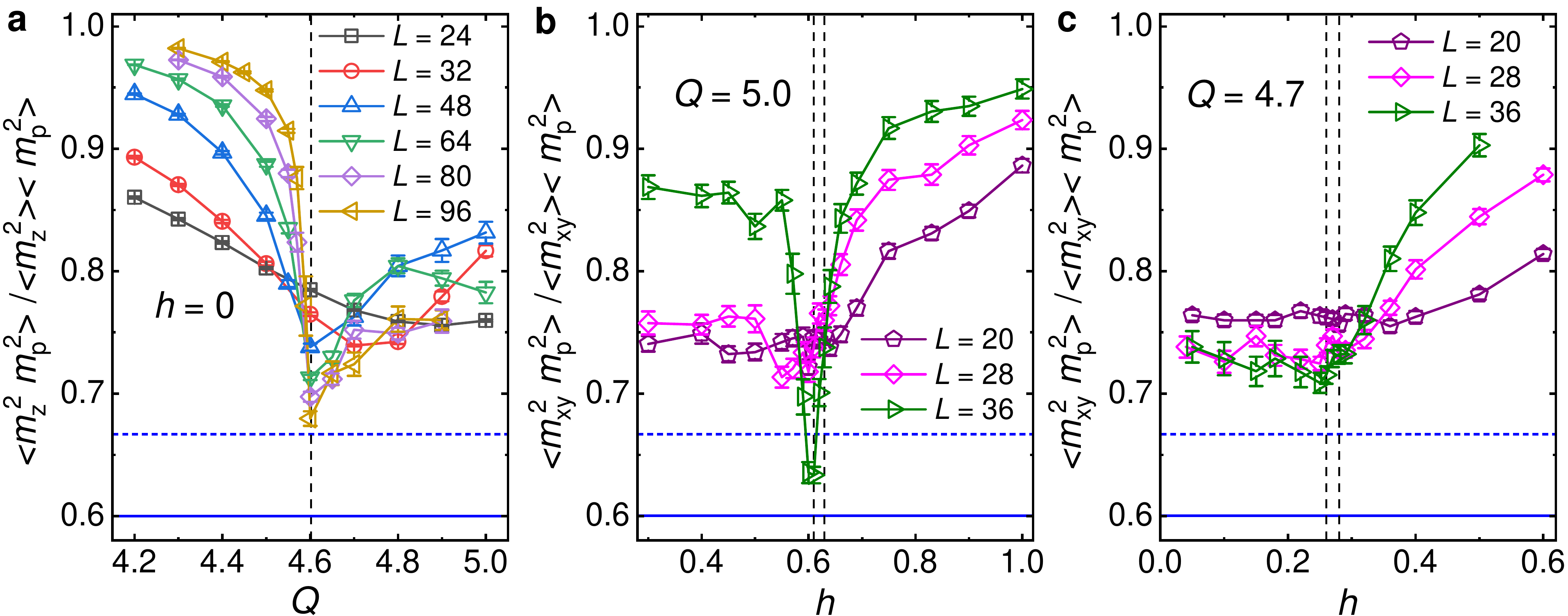}
  \caption{\label{ratio} Field dependence of the cross-correlation ratio, Eq.~(\ref{cxp}), for different system sizes. In a, results are
    shown vs $Q$ at $h=0$, where $m_{xy}^2=(m_x^2+m_y^2)/2$ can be replaced by $m_z^2$ thanks to the O($3$) symmetry of the AFM order parameter.
    In b and c, results are shown vs the field at fixed values of $Q$; $Q=5$ in b and $Q=4.7$ in c. The horizontal lines are drawn at the values
    pertaining to O($3$) (solid lines) and O($4$) (dashed lines) symmetry. The vertical dashed lines indicate the previously
    determined transition points plus and minus their standard errors.}
\end{figure*}

The probability distributions given by our simulations (Figs.~6d-6f of the
main text) do indeed bear out this expectation: the double-peak structure of
$P(m_p)$ in the PS phase reflects a $Z_2$ symmetry, which is broken in the
thermodynamic limit (where the peaks become $\delta$-functions). The AFM
phase is marked by a single peak centered at $m_p = 0$, because its order
and symmetry-breaking take place in a different channel (namely, breaking
of global spin-rotation symmetry). Right at $h_c$, the clear plateau feature
in $P(m_p)$ indicates that the emergent symmetry is O($3$). The rounding of the
edges can naturally be explained by finite-size effects, which correspond to
a fluctuating radius $R$ of the sphere in Eq.~(\ref{pmpo3sym}).

The results in Figs.~6d-6f were generated for a system size $L = 32$. It is
difficult to obtain statistically good  results (smooth histograms) for larger
sizes, because of the long autocorrelation time (which grows rapidly with
the system size) in the simulations. Given the negative cumulant peak in
Fig.~\ref{numericals1}, we would expect to see a three-peak structure
developing in place of the flat portion of the distribution in Fig.~6e for
larger system sizes. However, the observation of the signatures of O($3$)
symmetries for $L = 32$ already demonstrates that this symmetry is emergent.
The symmetry is likely inherited from an O($4$) DQCP when the magnetic field is
turned on, through an extension of the mechanism for first-order transitions with
emergent O($4$) symmetry discussed in Ref.~\cite{Serna_PRB_2019}.
Here we also point out that the sum of the distributions in
Figs.~6d (in the PS state) and Figs.~6f (in the AFM state) does exhibit a
three-peak structure, though not of course with sharp peaks because of the small system
size. The sum distribution roughly reflects what would be expected in a conventional
coexistence state (see also Ref.~\cite{Sun_CPB_2021}).

In the XXZ model very close to the isotropic point, we also expect remnants of the O($3$) symmetry
(which in this case is exact at $\lambda=0$). However, as we saw above in Sec.~\ref{cbjqma}, in the XXZ model the
magnetization jump at $H_c$ is relatively much larger. Therefore, the length scale at which the O($3$) symmetry is
violated should be much larger in the CBJQM for comparable distance from the $h=0$ transition point. This distance
can be taken as the transition field $h_c$ in either model, or, equivalently, the spin gap at $h=0$.

To further justify the emergent O($3$) symmetry in the CBJQM, we calculate the cross-correlation ratio
\begin{equation}
C_{xy,p}  = \frac{\langle m_{xy}^2 m_p^2 \rangle}{\langle m_{xy}^2 \rangle \langle m_p^2 \rangle},
\label{cxp}
\end{equation}
where $m_{xy}^2$ is the squared off-diagonal (XY) AFM order parameter
\begin{eqnarray}
  m_{xy}^2 & = & \frac{1}{2} (m_x^2 + m_y^2) \nonumber \\
  & = & \frac{1}{4N^2} \sum_{i,j} (-1)^{i-j} (S_i^+ S_j^- + S_i^- S_j^+) .
\end{eqnarray}
The ratio (\ref{cxp}) measures the covariance between the AFM and PS order parameters and, thus, it can be used to detect the symmetry
of the joint order parameter vector $(m_x,m_y,m_z,m_p)$. It is obvious that $C_{xy,p}=1$ when $m_{xy}$ and $m_p$ are completely uncorrelated.
Therefore, we expect the ratio to be close to $1$ deep inside either the PS or the AFM phase. On the other hand, if the two order parameters are
connected by an O($3$) symmetry, it is easy to see that the ratio should take the value of $3/5$, while if the symmetry is O($4$) the value is
$2/3$. Similar cross-correlations were previously studied in the context of te DQCP with emergent SO(5) symmetry \cite{Nahum_PRL_2015,Sreejith_PRL_2019}.

For $h=0$, the AFM order parameter is inherently O($3$) symmetric and $m_{xy}^2$ in Eq.~(\ref{cxp}) can be replaced by $m_z^2$, which makes
the entire correlation function diagonal in the $S^z$ basis used in our SSE simulations. However, for $h>0$ the operator $m_{xy}^2$ is off-diagonal
and has to be treated in a different way using a string estimator, as explained in Sec.~\ref{methods:cbjq}. These off-diagonal measurements
are much noisier than the corresponding diagonal ones (also because the simulations overall are more challenging when $h>0$),
and we can therefore not reach as large system sizes for $C_{xy,p}$ as for the corresponding diagonal quantity $C_{z,p}$ at $h=0$.

We first discuss $h=0$, where emergent O($4$) symmetry was discovered in previous work by examining primarily the order-parameter distribution
\cite{Zhao_NP_2019}. Results for the ratio $C_{z,p}$ are shown versus $Q/J$ ($J=1$) for several system sizes in Fig.~\ref{ratio}a. Here we observe values
tending clearly toward $1$ in the AFM phase, $Q < Q_c$. The results are more affected by long autocorrelations in the PS phase, where the error bars, thus,
are larger. The convergence toward $1$ is also overall less obvious---there may be some non-trivial covariance between the long-ranged PS
order parameter and the short-ranged AFM order parameter in the PS phase. The most important aspect of these results is the sharp minimum in $C_{z,p}$,
which flows with increasing $L$ toward a value of $Q$ completely consistent with the known value $Q_c(h=0)=4.600$ \cite{Zhao_NP_2019}. The value of
$C_{z,p}$ shows a significant size dependence, but, indeed, reaches close to the expected O($4$) value $C_{z,p}=2/3$ for the largest system size,
$L=96$.

Next, we consider $Q>Q_c(h=0)$ and scan the ratio $C_{xy,p}$ versus $h$. Fig.~\ref{ratio}b shows results for three system sizes at $Q=5$.
Here we again observe a sharp minimum, especially for the largest system size, $L=36$, where the minimum is located at $h \approx 0.61$.
Considering that the trend with increasing $L$ is a slight drift of the minimum toward larger values of $h$, the results are consistent with the
transition point $h_c(Q=5) = 0.62 \pm 0.01$ that we extracted from other quantities previously. The value of the ratio at the minimum is now below
the O($4$) value for $L=36$, and the trend with increasing $L$ is to lower values. While the very high cost of simulations for larger $L$
prohibit us from confirming that the O($3$) value is reached, the observed trends nevertheless support our assertion of emergent O($3$) symmetry.

Finally, in Fig.~\ref{ratio}c we show results obtained at $Q=4.7$, close to $Q_c(h=0)=4.6$. Here the minimum versus $h$ is less sharp than in
Fig.~\ref{ratio}b and the values of the ratio are overall significantly higher when $h \approx h_c$.
Comparing with the $h=0$ results in Fig.~\ref{ratio}a for $Q$
close to the transition point, the values of $C_{z,p}$ are very similar for sizes $L \simeq 30$ to those of $C_{xy,p}$ when
$h \approx h_c \approx 0.27$. We can explain this behavior as a symmetry cross-over effect: For very small $h>0$, the system should initially,
for moderate system sizes, behave as if O($4$) symmetry is emerging when $Q \to Q_c \approx Q_c(h=0)$, while for larger sizes the necessary effects
of $h>0$ to eventually take the symmetry of the AFM order parameter down from O($3$) to O($2$) will also imply violation of the O($4$) symmetry and
flow of the ratio toward the smaller O($3$) value. Again, our systems are not sufficiently large to follow this behavior in its entirety, but, taken
together, all results in Fig.~\ref{ratio} are certainly supportive of this O($4$) $\to$ O($3$) cross-over scenario.

\begin{figure}[t]
\includegraphics[width=7.75cm]{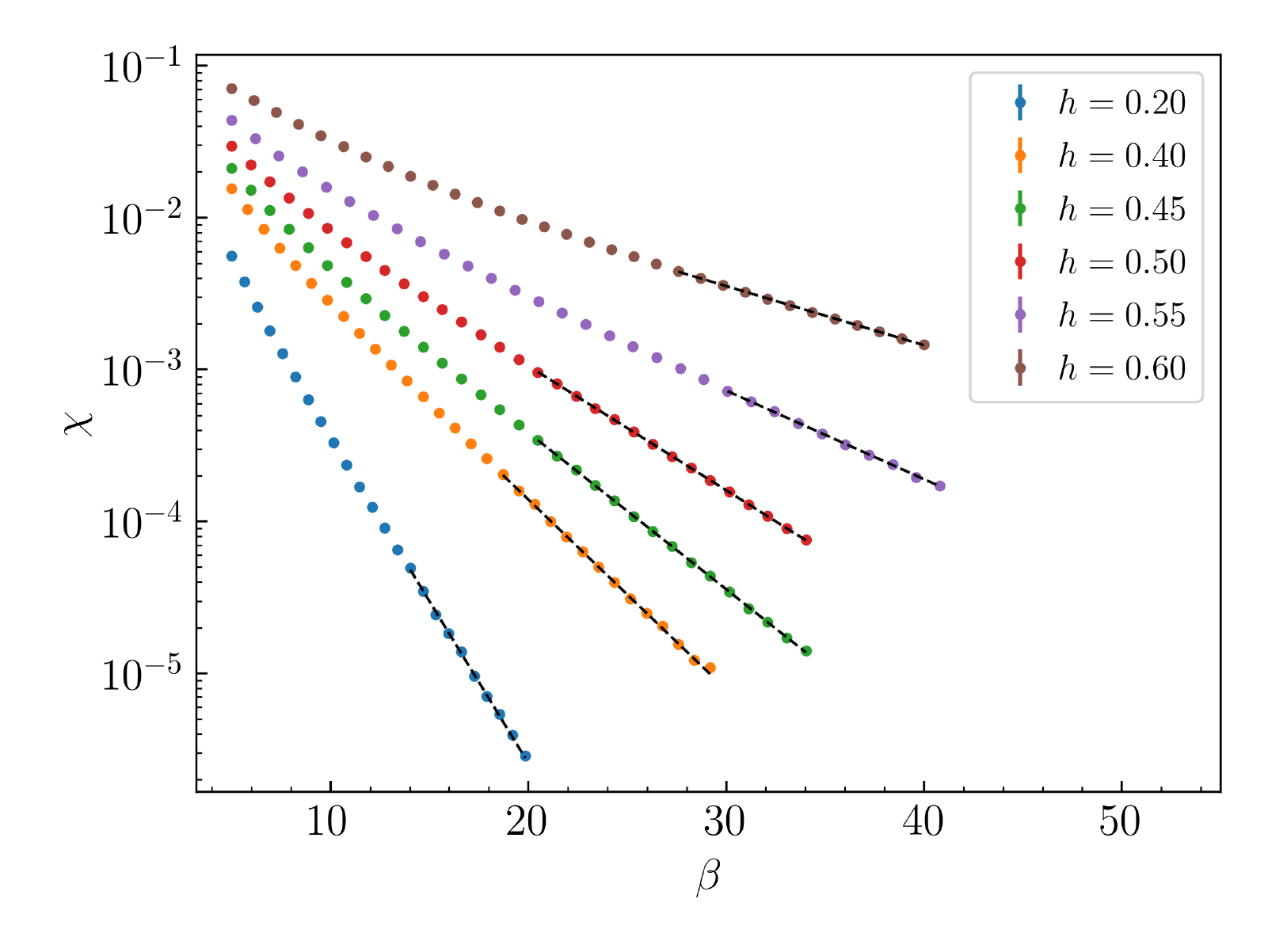}
\caption{\label{xxzh3}
Uniform susceptibility versus the inverse temperature of the XXZ model with anisotropy parameter $\lambda=0.1$
and several field values. The curves (almost straight lines) are fits to the full form in Eq.~(\ref{chiformkhw}),
but the corrections to the leading exponential form are very small.}
\end{figure}

\subsection{Excitation gap}
\label{cbjqmc}

As discussed in the main text, the spin excitation gap closes linearly with the field, with no appreciable discontinuity
at the transition, shown for SrCu$_2$(BO$_3$)$_2$ in Fig.~6a and the CBJQM in Fig.~6c. We here provide details of the
CBJQM gap calculation, and again also study the XXZ model as a benchmark.

We first extract the gap in the Ising phase of the anisotropic Heisenberg model from low-temperature behavior
of the susceptibility, defined in Eq.~(\ref{chidef1}) and computed in the simulations in the standard way in terms
of magnetization fluctuations according to Eq.~(\ref{chiqmc}).

The Ising-like ground state is doubly degenerate in the thermodynamic
limit, with a gap between the quasi--degenerate states closing exponentially
as a function of the system size. When the system is locked in one of
these ground states, which has total $S^z=0$, and assuming a single band of
excitations with $|S^z|=1$ in momentum space with a quadratic minimum above
the gap $\Delta$, the susceptibility at low temperatures takes the form
\begin{equation}
\chi = \frac{A{\rm e}^{-\Delta\beta}}{1+\beta^{-1} A{\rm e}^{-\Delta\beta}} \to
A{\rm e}^{-\Delta\beta} ~~(\beta \to \infty),
\label{chiformkhw}
\end{equation}
where $\beta = J/T$ and $A$ is a model dependent parameter.

\begin{figure}[t]
\includegraphics[width=7.75cm,clip]{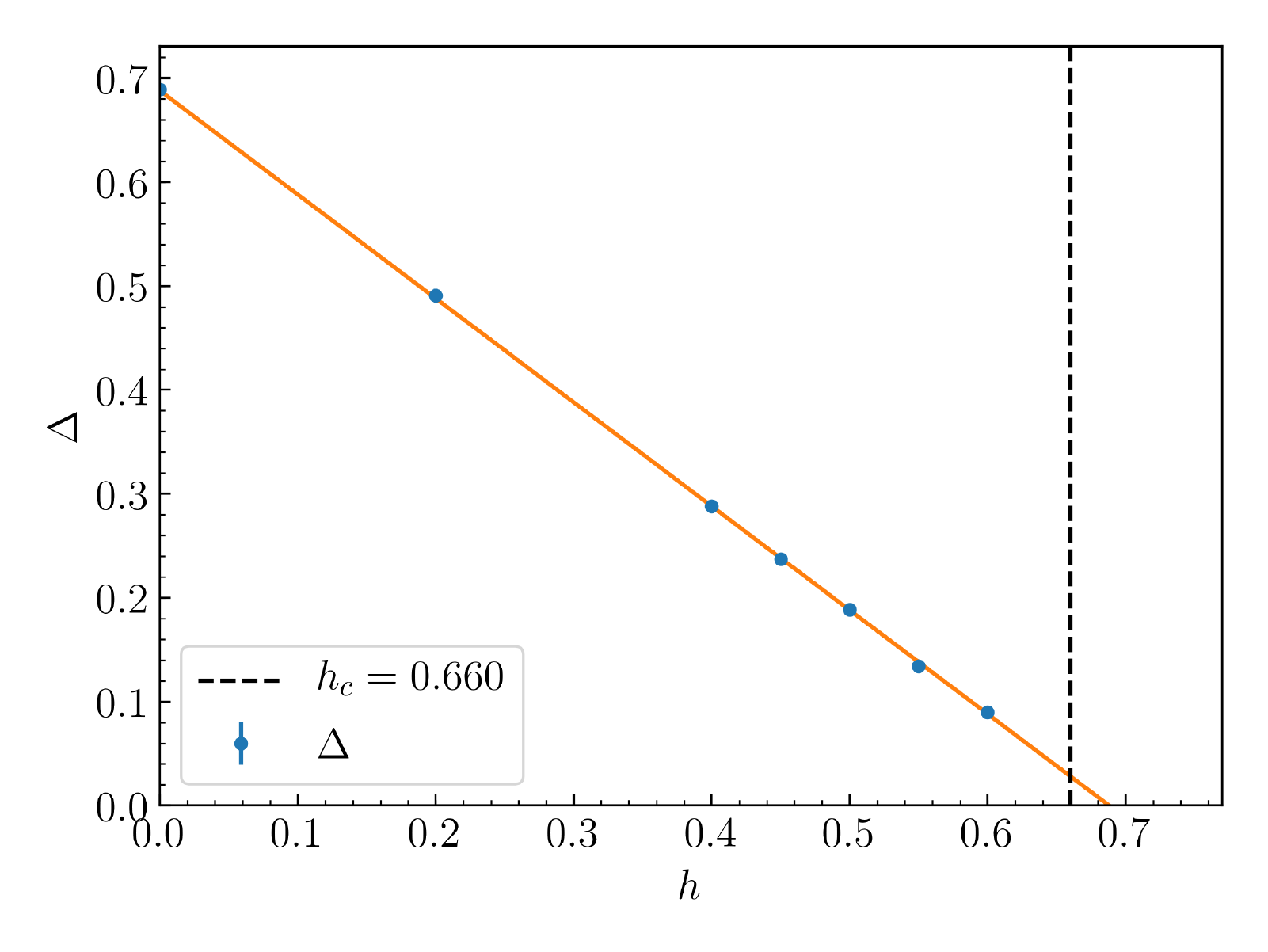}
\caption{\label{xxzh4}
  The spin gap of the XXZ model at $\lambda =0.1$, extracted from the low-temperature uniform susceptibility as illustrated
  in Fig.~\ref{xxzh3}. The vertical dashed line marks the first-order transition value $h_c$ of the field,
  where the ground state changes discontinuously from one with $m=0$ to
  one with $m>0$ (as shown in Fig.~\ref{xxzh2}). The gap to the $S^z=1$ state for $h<h_c$ is of the form $\Delta(h)=\Delta(0)-h$,
  until the discontinuous jump to zero at $h_c$.}
\end{figure}

Typical QMC results for $\chi$, obtained with system sizes $L$ up to
$64$, are graphed versus the inverse temperature $1/T$ ($J=1$) in
Fig.~\ref{xxzh3} on a lin-log plot, where the asymptotic exponential form
corresponds to a straight line. We fit the data at low temperatures to
the full functional form Eq.~(\ref{chiformkhw}). When using only the
asymptotic form (which we do when analyzing the experimental data and
also the CBJQM data below) the fits are still good and the extracted gaps
only change insignificantly (which is reflected in the essentially
straight fitted lines in the plot), as long as only data points at
sufficiently low temperatures are included.

We graph $\Delta$ versus spin anisotropy $\lambda$ in Fig.~\ref{xxzh4}. As expected, the gap decreases linearly with the field, consistent
with the form $\Delta(h)=\Delta(0)-h$ expected when the lowest excitation has $S^z=1$, and extrapolates to a point above the transition field $h_c$
extracted in Fig.~\ref{xxzh2}. Thus, the gap exhibits a discontinuous jump of about $0.04$ in the thermodynamic limit.

\begin{figure}[t]
\includegraphics[width=7.6cm]{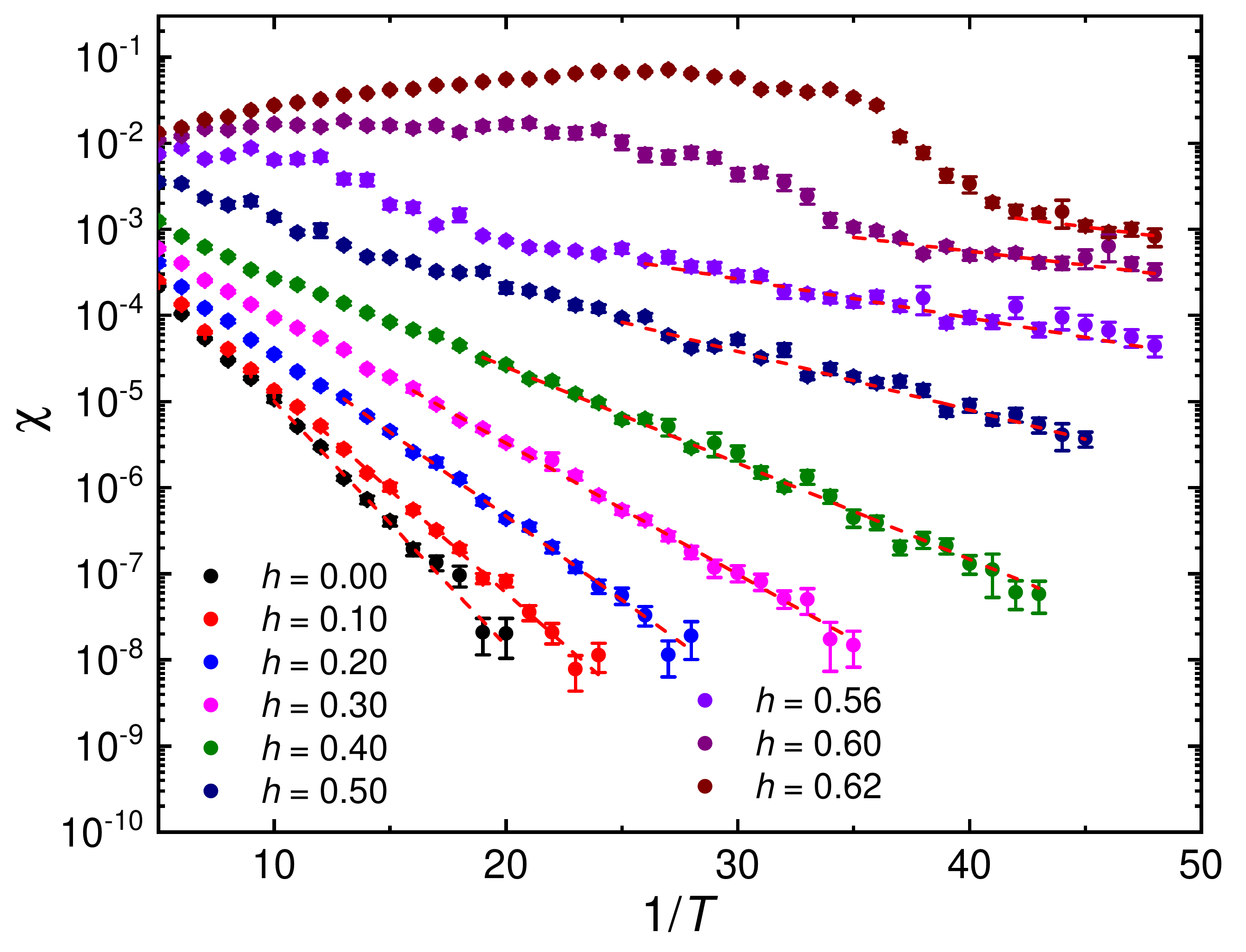}
\vskip-2mm
\caption{\label{numericals4}
Uniform susceptibility of the CBJQM with $Q=5$ computed for $L = 48$ and shown as a function
of the inverse temperature at several different fields. The dashed lines are fits to the form $\chi \propto e^{-\Delta/T}$, which were
used to extract the spin-gap vs $h$ shown in Fig.~6c of the main text.}
\end{figure}

We extract the PS gap of the CBJQM in the same way. The susceptibility computed at various fields for $Q = 5$ and system size $L = 48$
is shown in Fig.~\ref{numericals4}. For fields close to $h_c$ ($0.6\lesssim h \lesssim 0.62$), $\chi(T)$ first slightly increases with
decreasing temperature, then forms a broad peak before decreases at sufficiently low temperatures. The low-temperature $\chi(T)$ data in the
PS phase always follow a clear exponential decay in the PS phase, which we use here without corrections to extract $\Delta(h)$.
The temperature has to be very low (and the system size very large) to reach the asymptotic behavior when $h$ is close to $h_c$.
The intricate behavior reflects a strong competition between the AFM and PS states close to the transition. The system size used
here is sufficiently large for obtaining well-converged results for $h$ up to $0.60$, but we also show $h=0.62$ results for reference.

The resulting gap for $h \le 0.62$ is graphed in Fig.~6c in the main text. The strong fluctuating data and the narrow available temperature range of
temperatures with exponentially decaying $\chi(T)$ prevent an accurate determination of the gap values close to $h_c$, as reflected in the corresponding
large error bars in Fig.~6c. It is nevertheless clear that the gap discontinuity at $h_c$ is small, though the relatively large uncertainty
in the location $h_c$ of the transition field of the CBJQM makes it impossible to obtain a meaningful estimate of the ratio of the gaps
in the CBJQM and XXZ model (Fig.~6c versus Fig.~\ref{xxzh4}). The dramatic difference in the magnetization jump (Fig.~\ref{xxzh2} versus \ref{MzHQ5})
is much more obvious and may be a more prominent hallmark of the PS--AFM transition in the CBJQ model as well as in SrCu$_2$(BO$_3$)$_2$.

Our gap calculations for the CBJQM have focused on $Q=5$, for which it is easier to obtain size-converged results than closer to the transition. Like
the reduction of the magnetization jump when moving from $Q=5$ to $4.7$ in Fig.~\ref{MzHQ5}, the gap discontinuity should also be significantly reduced
and be more similar to the barely discernible gaps observed at $H_c$ in SrCu$_2$(BO$_3$)$_2$ at 2.1 and 2.4 GPa (Fig.~6a).

\subsection{PS--AFM transition mechanism}
\label{scbjqd}

The very different magnetization jumps point to a fundamental difference between the first-order phase transitions in the two models. When the
magnetization per spin at a field immediately above $h_c$ is large, as in the XXZ model at the anisotropy value $\lambda=0.1$ studied here, there are
hardly any remnants of the Ising-type order left in the ground state. The coexistence state can therefore not host any enhanced symmetry. In
contrast, with the very small magnetization jump at $h_c$ in the case of the CBJQM, the coexistence ground state still can have significant PS
features left.

Our microscopic picture of the transition in the CBJQM is that the singlet PS ground state is associated with a tower of spin-$S$ states, corresponding to successive
excitations of plaquette singlets into triplets (triplons).
In a field, an $S^z=S$ state corresponds to a number $n=S^z$ of triplons, or a volume fraction $\rho =S^z/L^2=m$.
Assuming that the interactions between the triplons are weak, states with different $n$ all become almost degenerate at $h_c=\Delta(0)$ on account of
the negative energy shifts $hS^z=hn$ of the $h=0$ energies located at $\Delta_n=n\Delta_1$ above the $n=0$ ground state.

In reality, we know
that the interactions must be effectively weakly attractive, up to a certain density, so that a state with finite $n \propto L^2$ crosses the $S=0$ PS ground state
first, thus leading to the observed magnetization jump. However, given that the triplon (or magnetization)
density is very small, $\rho \approx 0.0025$ in Fig.~\ref{MzHQ5} at $Q=5$
and $\rho \approx 0.001$ at $Q=4.7$, these crossing states still can have significant PS character. Moreover, the ``broken'' triplet plaquettes can form
increasing AFM order as the number of triplets $n$ increases. The degeneracy of states with triplet density from zero to the transition value of $\rho$
(which in our picture form the highly degenerate coexistence state) is consistent with a continuous symmetry of the coexistence state, though we do not at
present know the detailed mechanism by which the AFM order emerges from the broken plaquettes of the PS state and how the required O($3$) Anderson rotor
tower is realized. However, our observation of O($3$) symmetry in the order-parameter histogram (Fig.~6e) demonstrates its existence.

Another way to qualitatively understand the emergent O($3$) symmetry in the CBJQM is from the removal by the field of one of the AFM components of the
emergent O($4$) order-parameter vector at $g_c$ when $h=0$. This process simply corresponds to an O($4$) model with an uniaxial deformation parameter,
similar to $\lambda$ in XXZ model, which takes the symmetry of the coexistence state at $h>0$ down to O($3$) by gapping out one of the AFM order-parameter
components. This picture explicitly demonstrates the difference between the order parameter transformations in the two models: In the three-component XXZ
model, the Ising order has to be explicitly destroyed in order to induce the AFM order. The Ising and XY AFM parts of the Hilbert space then are very
different and must be separated by energy barriers. In the CBJQM, the transition just corresponds to a further infinitesimal deformation of the O($3$) sphere,
with the coexistence state having the full O($3$) symmetry. The XY AFM order can be introduced gradually without complete destruction of the PS state, as described
above, and a set of degenerate states with different $S^z$ can establish a path for continuous changes (rotations) of the order parameter from PS to AFM
without energy barriers. In reality, we know that the transition in the CBJQM also is ultimately conventionally first-order, but the idealized picture
still describes the situation up to a large length scale, with detectable ramifications that we observe both in the CBJQM and
in SrCu$_2$(BO$_3$)$_2$.

A further observation, based on Fig.\ref{ratio} and the smallness of the magnetization jump, is that the coexistence phase also exhibits a cross-over from
O($4$) to O($3$) symmetry even when transition field $h_c$ is not very small. This robustness of O($4$) symmetry (i.e., its persistence up to large length scales)
supports our assertion that the DQCP point
marked in Fig.~2 in the main text should be of the O(4) kind \cite{Lee_PRX_2019} even in the presence of the magnetic field. Strictly speaking, this point
may exist as a true critical point only at $h=0$, but with a very long correlation length and in practice realizing the DQCP phenomenology also at $h>0$.


\begin{thebibliography}{99}
\bibitem{J2Dnote}
In a 2D Heisenberg model with nearest-neighbor couplings $J$, we simply have $J_{\rm 2D}=J$.
However, competing in-plane couplings, frustration in particular, can lead to a smaller effective
value of $J_{\rm 2D}$.
\end{thebibliography}
\end{document}